\RequirePackage[T1]{fontenc}
\documentclass[12pt]{article}

\usepackage[height=8.85in,width=6.45in]{geometry}

\usepackage[utf8]{inputenc}
\usepackage{amsmath}
\usepackage{amssymb}
\usepackage{mathtools}
\usepackage{physics}
\numberwithin{equation}{section}
\usepackage{slashed}
\usepackage{braket}
\usepackage[svgnames]{xcolor}
\usepackage[colorlinks,citecolor=DarkGreen,linkcolor=FireBrick,linktocpage]{hyperref}

\usepackage{acronym}

\newacro{EFT}[EFT]{effective field theory}
\acrodefplural{EFT}{effective field theories}
\newacro{CFT}[CFT]{conformal field theory}
\acrodefplural{CFT}{conformal field theories}
\newacro{IR}[IR]{infrared}
\newacro{UV}[UV]{ultraviolet}
\newacro{ESC}[ESC]{emergent string conjecture}
\newacro{KK}[KK]{Kaluza-Klein}
\newacro{CKN}[CKN]{Cohen-Kaplan-Nelson}
\newacro{CEB}[CEB]{Covariant entropy bound}
\newacro{BPS}[BPS]{Bogomol'nyi-Prasad-Sommerfeld}
\newacro{GSO}[GSO]{Gliozzi-Scherk-Olive}
\newacro{RS}[RS]{Rankin-Selberg}

\usepackage{varioref}       
\usepackage{cleveref}  

\crefname{table}{table}{tables}
\Crefname{table}{Table}{Tables}
\crefname{figure}{figure}{figures}
\Crefname{figure}{Figure}{Figures}

\usepackage{cite}
\usepackage{graphicx}
\usepackage{tikz}
\usepackage{tikz-cd}
\usepackage{times}
\usepackage{courier}
\usepackage{bm}
\usepackage{subfig}
\usepackage{ascmac}
\usepackage{tikzsymbols}
\usepackage{tensor}
\usepackage{comment}

\usepackage{mdframed}

\makeatletter
\AddToHook{cmd/appendix/before}{\def\cref@section@alias{appendix}}
\makeatother

\renewenvironment{figure}[1][]{
  \begin{originalfigure}[#1]
    \begin{mdframed}[linecolor=black!0,backgroundcolor=black!0]
}{
    \end{mdframed}
  \end{originalfigure}
}
\renewenvironment{table}[1][]{
  \begin{originaltable}[#1]
    \begin{mdframed}[linecolor=black!0,backgroundcolor=black!1]
}{
    \end{mdframed}
  \end{originaltable}
}

\usepackage{colortbl}
\definecolor{lightyellow}{rgb}{1.0, 0.95, 0.7}
\definecolor{lightblue}{rgb}{0.7, 0.9, 1.0}
\definecolor{lightpink}{rgb}{1.0, 0.85, 0.95}
\definecolor{lightgreen}{rgb}{0.7, 1.0, 0.4}

\newcommand*{\bZ}{\mathbb{Z}}
\newcommand*{\bR}{\mathbb{R}}

\newcommand{\cF}{\mathcal{F}}

\newcommand{\cN}{\mathcal{N}}
\newcommand{\cO}{\mathcal{O}}
\newcommand{\cR}{\mathcal{R}}

\newcommand{\cI}{\mathcal{I}}
\newcommand{\cQ}{\mathcal{Q}}
\newcommand{\cG}{\mathcal{G}}
\newcommand{\cV}{\mathcal{V}}
\newcommand{\bC}{\mathbb{C}}
\newcommand{\fu}{\mathfrak{u}}
\newcommand{\td}{\mathrm{d}}
\newcommand{\cZ}{\mathcal{Z}}
\newcommand{\cD}{\mathcal{D}}

\def\boo{0.0}

\def\xlattice#1#2#3{
\begin{tikzpicture}[scale=.5]
\filldraw[color=black!5!white](-.5,-.5) rectangle (1.5,1.5);
\draw[->] (-1,0) -- (2,0);
\draw[->] (0,-1) -- (0,2);
\foreach \x in {0,1} {
	\foreach \y in {0,1}{
		\pgfmathsetmacro\a{mod(#1 * \x - #2 * \y,2)}
		\ifx\a\boo
			\filldraw[color=#3] (\x,\y) circle (.5em);
		\else
			\filldraw[fill=white,draw=gray] (\x,\y) circle (.5em);
		\fi
	}
}
\end{tikzpicture}
}

\newenvironment{eqaed}
    {\begin{equation}
    \begin{aligned}
    }
    { 
    \end{aligned}
    \end{equation}
    \ignorespacesafterend
    }

\begin{document}

\begin{titlepage}

\begin{flushright}
{IFT-UAM/CSIC-26-23 \\ MPP-2026-35}
\end{flushright}

\vskip 3cm

\renewcommand*{\thefootnote}{\fnsymbol{footnote}}
\setcounter{footnote}{1}

\begin{center}

{\Large \bfseries UV/IR relations from the worldsheet}

\vskip 1cm
Christian Aoufia$^{1,}$\footnote{\url{christian.aoufia@estudiante.uam.es}}, Ivano Basile$^{2,}$\footnote{\url{ibasile@mpp.mpg.de}}, Giorgio Leone$^{3,}$\footnote{\url{giorgio.leone@unipd.it}} and Matteo Lotito$^{1,}$\footnote{\url{matteo.lotito@ift.csic.es}}
\vskip 1cm

\begin{tabular}{ll}
$^1$ & \emph{Instituto de F\'isica Te\'orica IFT-UAM/CSIC}\\ & \emph{C/ Nicol\'as Cabrera 13-15, Campus de Cantoblanco, 28049 Madrid, Spain} \\
$^2$ & \emph{Max-Planck-Institut f\"{u}r Physik (Werner-Heisenberg-Institut)}\\
& \emph{Boltzmannstraße 8, 85748 Garching, Germany} \\
$^3$ & \emph{Dipartimento di Fisica e Astronomia ``Galileo Galilei'', Università degli Studi di Padova}\\ & \emph{INFN Sezione di Padova, Via F. Marzolo 8, 35131 Padova, Italy
}
\end{tabular}

\end{center}

\begin{abstract}
    \noindent We derive universal scaling relations for the low-energy effective action of string theory, connecting the vacuum energy and gauge couplings to higher-derivative Wilson coefficients. At one-loop in string perturbation theory, these generic parametric relations follow from modular and conformal invariance of the worldsheet, independently of the specific low-energy phase of the theory, and they become non-trivial in species limits. As a result, we substantially strengthen our previous case for the emergent string conjecture and connect UV/IR mixing to swampland principles. We argue that our results persist to higher loops, hinting at a pathway to study strong couplings using dualities. Further accounting for open-string contributions, if any, our results lead to parametric inequalities which reproduce holographic bounds and support the magnetic weak-gravity conjecture and the dark dimension scenario.
\end{abstract}

\end{titlepage}

\setcounter{tocdepth}{2}
\tableofcontents

\renewcommand*{\thefootnote}{\arabic{footnote}}
\setcounter{footnote}{0}

\section{Introduction}\label{sec:introduction}

Any attempt to connect the very high-energy signatures of quantum gravity directly to real-world observations is bound to meet an inescapable obstruction: up to the energy scales that we can access, physics can be reliably described within the framework of \ac{EFT}. This description still accounts for gravity quantum mechanically, and its low-energy features (such as gravitationally induced entanglement or quantum corrections to Newtonian attraction) are calculable and accessible to foreseeable experiments; however, the smoking guns of any given completion of a gravitational \ac{EFT} to the \ac{UV} are inevitably hidden by the renormalization-group flow across several orders of magnitude of separation between energy scales. In other words, the direct signatures of \ac{UV}-complete quantum gravity are decoupled from \ac{IR} physics. While this state of affairs may appear discouraging, there are some potential ways out: \\

\noindent \paragraph{New physics at the \ac{EFT} cutoff.} The \ac{EFT} cutoff signaling the presence of new physics may be lower than expected, and reflect the existence of new particles or interactions\footnote{Note that an unknown light but extremely weakly coupled sector could also be a source of new physics. We will not consider such a scenario in this paper.}. This is not related to quantum gravity \emph{per se}, but it may be the case that the consistency conditions of the \ac{UV}-completion, together with the observed features of our universe, severely constrain the type of new physics that can appear even at the field-theoretic level. One striking example of this is embodied by the dark dimension scenario \cite{Montero:2022prj}, which proposes that the observed smallness of dark energy be related to the presence of one (possibly two \cite{Anchordoqui:2025nmb}) mesoscopic\footnote{In this paper, ``mesoscopic'' refers to extra dimensions whose characteristic size is (parametrically) smaller than the cosmological (possibly infinite) scale of the $d$-dimensional macroscopic universe and much larger than the $d$-dimensional Planck length. The latter hierarchy implies that mesoscopic extra dimensions are parametrically larger than the Planck length in any dimension.} extra dimension(s), whose \ac{KK} scale $m_\text{KK}$ is the \ac{EFT} cutoff. Although in such scenarios the new physics above the cutoff is field-theoretic (in this case, described by a higher-dimensional \ac{EFT}), its specifics are supposedly dictated by quantum-gravitational consistency. This distinguishes them from other rationales for similar beyond-standard-model scenarios \cite{Arkani-Hamed:1998jmv}. \\

\noindent \paragraph{Quantum gravity and the species scale.} Another distinguished scale in gravitational \acp{EFT} is the species scale $\Lambda_\text{sp}$ \cite{Dvali:2001gx, Veneziano:2001ah, Dvali:2007hz, Dvali:2007wp, Dvali:2009ks, Dvali:2010vm, Dvali:2012uq, Caron-Huot:2024lbf, ValeixoBento:2025iqu}, the subject of considerable recent interest \cite{Castellano:2022bvr, vandeHeisteeg:2022btw, Cribiori:2022nke, Cribiori:2023sch, Cribiori:2024qsv, vandeHeisteeg:2023dlw, vandeHeisteeg:2023ubh, Castellano:2023aum, Blumenhagen:2023yws, Basile:2023blg, Castellano:2024bna, Blumenhagen:2024lmo, Blumenhagen:2024ydy, Aoufia:2024awo, Basile:2024dqq, Calderon-Infante:2025ldq} especially in the context of the swampland program. In order to separate $\Lambda_\text{sp}$ from related scales, such as the scale of strong gravitational coupling or upper bounds given in terms of species counting\footnote{In particular, these estimates of (upper bounds to) $\Lambda_\text{sp}$ may be subject to uncertainties for strongly coupled \ac{UV}-completions, due to the contributions from black-hole microstates.}, we define the $\Lambda_\text{sp}$ associated to an \ac{IR} phase\footnote{By ``\ac{IR} phase'' we mean a gravitational \ac{EFT} together with all \acp{EFT} which supersede it going up in energy scales until, above $\Lambda_\text{sp}$, no \ac{EFT} description exists.} of gravity as the minimal scale above which no \ac{EFT}, no matter in which dimension, can reliably describe the physics. This is a parametric definition that suffices to capture the relevant physics, but one can concoct sensible numerical counterparts of $\Lambda_\text{sp}$ by choosing a threshold for the onset of higher-spin resonances in the S-matrix \cite{Caron-Huot:2024lbf} or a specific normalization for some higher-derivative Wilson coefficient in the gravitational sector\footnote{By ``gravitational sector'' we do not refer to operators built from the Riemann curvature and its covariant derivatives; these can originate from field-theoretic mechanisms, as recently highlighted in this context \cite{Aoufia:2024awo, Basile:2024oms, Calderon-Infante:2025ldq}. Specifically, we mean the sector of the effective action which corrects the Einstein-Hilbert term including the Planck-scale prefactor \cite{Calderon-Infante:2025ldq}.}, as for example the genus-one free energy of the topological string \cite{vandeHeisteeg:2022btw} in type II Calabi-Yau compactifications or certain protected $R^4$ terms in maximal supergravity \cite{Castellano:2023aum, vandeHeisteeg:2023dlw}. \\

\noindent For our purposes, the most interesting cases are \emph{species limits}, i.e when $\Lambda_\text{sp} \ll M_{{\text{Pl}};d}$ is parametrically below the Planck scale of our $d$-dimensional \ac{EFT}; if the \ac{EFT} admits decompactifications up to $D > d$ dimensions, by definition one has $\Lambda_\text{sp} \lesssim M_{{\text{Pl}};D}$, and when $\Lambda_\text{sp} \ll M_{{\text{Pl}};D}$ the \ac{UV}-completion is weakly coupled. This means that when the species scale is parametrically sub-Planckian, new physics emerges, in the guise of either mesoscopic extra dimensions (as discussed above) or weakly coupled degrees of freedom that \ac{UV}-complete gravity. This scenario must manifest at tree-level \cite{Cheung:2016wjt} and is strongly constrained by unitarity and causality \cite{Camanho:2014apa, Arkani-Hamed:2020blm, Geiser:2022exp, Bachu:2022gof, Caron-Huot:2022ugt, Caron-Huot:2022jli, Arkani-Hamed:2023jwn, Basile:2023blg, Bedroya:2024ubj, Herraez:2024kux, Cheung:2024uhn, Cheung:2024obl, Albert:2024yap, Cheung:2025nhw, Cheung:2025tbr, Calisto:2025tjo}, along with some non-perturbative results from bootstrap \cite{Guerrieri:2021ivu, Guerrieri:2022sod} and swampland methods \cite{Kim:2019vuc, Kim:2019ths, Montero:2020icj, Bedroya:2021fbu, Hamada:2021bbz, Bedroya:2023tch, Kim:2024eoa, Delgado:2024skw, Kaufmann:2024gqo, Baykara:2025gcc}. As a result, weakly coupled \ac{UV}-complete gravity (at least in $d>3$ spacetime dimensions) seems to behave very much like string theory, and quantum-gravitational effects ought to be visible far below $M_{{\text{Pl}};d}$. This is the physical content of the \emph{\ac{ESC}}, which even within top-down string constructions only works thanks to a beautiful dovetail of (non-)perturbative dualities that make extra dimensions \cite{Aoufia:2024awo, Ooguri:2024ofs} or the fundamental string \cite{Lee:2018urn, Lee:2019wij, Klaewer:2020lfg, Alvarez-Garcia:2021pxo, Hassfeld:2025uoy, Monnee:2025ynn, Monnee:2025msf} emerge. However, \emph{prima facie} this does not seem to support our cause---string theory (as any quantum theory of gravity \cite{Arkani-Hamed:2007ryu}) comes with a landscape: a wealth of \ac{IR} phases, sifting through which is a main goal of string phenomenology \cite{Marchesano:2024gul} (see \cite{McAllister:2025qwq} for recent developments). As we shall explain in detail, the strategy we follow in this paper is to seek \emph{universal low-energy properties} of string theory that do not depend on the specific \ac{IR} phase within the landscape. \\

\noindent \paragraph{\ac{UV}/\ac{IR} mixing and low-energy data.} Even when some signatures of \ac{UV}-complete quantum gravity are amplified by the above factors, the relevant scales may still be out of reach of foreseeable experiments. Indeed, as an example, the dark dimension scenario \cite{Montero:2022prj} purports a species scale of roughly $10^{10}$ GeV; far below the Planck scale $M_{\text{Pl};4} \approx 10^{19}$ GeV, but still rather large relative to accelerator technology\footnote{Of course, unrelated to the point we are making, the main attractive aspect about the dark dimension scenario is the much more accessible micron-sized \ac{KK} scale, pertaining to the titular extra dimension(s).}. However, these \ac{UV} quantities are tied to \ac{IR} quantities, especially those that are \ac{UV}-sensitive such as the vacuum energy or scalar masses. In quantum gravity, these connections are often unexpected from the point of view of \ac{EFT}, because of the inherent \emph{\ac{UV}/\ac{IR} mixing} which can be ascribed to black-hole physics \cite{Cohen:1998zx} and more generally holography \cite{Bousso:1999xy}. String theory exhibits relations of this type in great generality \cite{Abel:2021tyt, Abel:2023hkk, Basile:2024lcz}, connecting low-energy data (scalar potential, gauge couplings) to high-energy data (\ac{EFT} cutoff, species scale). As hinted at above, in this paper we seek generic patterns in the \ac{IR} physics of string theory, which will take the form of \ac{UV}/\ac{IR} relations. \\

\subsection{Toward generic features of low-energy string theory}\label{sec:big-picture}

In light of the above considerations, we can now introduce the program we follow in this paper, which stems from the recent developments of \cite{Abel:2021tyt, Abel:2023hkk, Aoufia:2024awo, Basile:2024lcz} but traces back much earlier \cite{Banks:1987cy, Banks:1988yz, Lust:2008qc, Anchordoqui:2009mm}. As mentioned above, we seek to derive universal properties that string theory exhibits at low energies, namely patterns among \ac{IR} and \ac{UV} quantities that are shared by stringy \acp{EFT}. \\

\noindent \paragraph{Our setup.} In order to make progress toward this ambitious goal, we will restrict our focus in a number of ways. To begin with, we consider extreme long-distance physics, which is dominated by gravitons, abelian gauge fields (if any) and (pseudo-)moduli scalar fields with a shallow potential (if any)\footnote{In principle, the \ac{EFT} could also contain strongly coupled, possibly non-Lagrangian \ac{CFT} sectors. Since such sectors do not arise within the worldsheet framework that we consider in this paper, and tend to be otherwise difficult to handle at the level of generality that we strive for, we shall exclude them.}. Up to kinetic mixings which can be diagonalized, we can restrict to a single gauge field without missing relevant ingredients. Furthermore, we only consider local physics and propagating degrees of freedom, thus excluding any topological sector\footnote{At any rate, any such sector is expected to couple to gravitons, possibly via kinetic terms, as was recently argued in a holographic context \cite{Cummings:2026giw}.}. Our methods will also not have much to say about the details of the (model-dependent) scalar sector, apart from its imprints on the gauge-gravitational sector and the standard dilaton dependence dictated by perturbative string theory. Similarly, we do not discuss open-string sectors in detail, although we will attempt to account for their contributions indirectly. All in all, the \acp{EFT} we are interested in are described by Wilsonian effective actions of the form
\begin{eqaed}\label{eq:wilsonian_action}
    S_\text{eff} = \int \td^dx \, \sqrt{-g} \left(\frac{M_{{\text{Pl}};d}^{d-2}}{2} \,  R - \, \frac{1}{4g^2} F^2 - \, V + \sum_{n} c_n \, \frac{\cO_n(R,F,\nabla)}{M_{{\text{Pl}};d}^{n-d}} \right) ,
\end{eqaed}
where the gauge coupling $g$ and the vacuum energy (density) $V$ can depend on scalar fields (if any), and the $c_n$ are Wilson coefficients for local dimension-$n$ operators $\cO_n$. The potential $V$ (and its derivatives) must be shallow with respect to the \ac{EFT} cutoff\footnote{This scale-separation condition ensures that the $d$-dimensional \ac{EFT} description is valid in a parametrically large window of scales. In reliable string constructions, the constant part of $V$ that remains at infinity in field space is zero. Obtaining a non-zero cosmological constant within a fully reliable \ac{EFT} description is an open problem \cite{Coudarchet:2023mfs} which spurred significant recent interest \cite{Montero:2024qtz, Andriot:2025cyi, Bedroya:2025ltj, Bobev:2025yxp, Apers:2026lgi}.} $m^2 \gg M_{{\text{Pl}};d}^{2-d}V$, which is expected to suppress the Wilson coefficients. As we mentioned above, and will discuss below in detail, in general there are both a field-theoretic suppression, controlled by $m$ and with no appearance of $M_{{\text{Pl}};d}$, and a quantum-gravitational suppression, controlled by $\Lambda_\text{sp}$ \cite{Aoufia:2024awo, Basile:2024oms, Calderon-Infante:2025ldq}. We will investigate both the \ac{IR} data ($V$, $g$) and the \ac{UV} data (the $c_n$), and in order to make universal statements we make two further simplifying restrictions. Firstly, we will mostly consider curvature operators without covariant derivatives. These provide the contact terms with the highest number of legs at fixed number of derivatives, and can be computed in two independent ways to be compared. Moreover, up to field redefinitions, these kinds of operators are the first ones appearing in the derivative expansion \cite{Cano:2021tfs}. We will also comment on other kinds of operators in less detail. Secondly, in order for our statements to be robust across the string landscape, we will essentially only consider \emph{parametric} expressions, such as scaling relations which become non-trivial asymptotically in certain limits. As such, we make heavy use of asymptotic notation: the symbol $\sim$ denotes asymptotic equality up to positive numerical prefactors. The relation $f \ll g$ means that $f/g \to 0$ in whichever limit we consider, and $f \gtrsim g$ is its logical negation. Parametric quantities are defined projectively, i.e. up to positive numerical prefactors. Finally, our results will be \emph{generic} in the sense that they dominate parametric relations unless prefactors are finely tuned to vanish. \\

\noindent \paragraph{The worldsheet framework.} In order to compute the desired quantities from string theory, we will employ the worldsheet framework\footnote{In particular, our approach does not include \acp{EFT} built by flowing to the \ac{IR} from backgrounds with Ramond-Ramond fluxes, orientifolds and/or branes or any non-perturbative ingredient. While in principle these are encoded in the string S-matrix, in practice accounting for background shifts in these settings is typically handled by string field theory methods (see, e.g., \cite{Cho:2023mhw, Kim:2024dnw, Frenkel:2025wko, Kim:2026kex} for recent developments).}. The detailed analysis that we will perform takes place in the closed-string sector up to one-loop order in string perturbation theory. However, we will argue that open-string effects and, more importantly, higher-loop effects do not modify the story qualitatively, except possibly weakening some equalities to inequalities. Borrowing the language of \cite{Tachikawa:2021mvw, Tachikawa:2021mby}, we think of perturbative string theory as a map that associates a $d$-dimensional gravitational \ac{EFT} to a worldsheet \ac{CFT} of a specific type\footnote{Of course, the assignment includes the full perturbative \ac{UV}-completion of the \ac{EFT} as well. In particular, at the perturbative level, it includes any other \ac{EFT} in the same \ac{IR} phase that supersedes it at higher energies.},
\begin{eqaed}\label{eq:string_worldsheet}
    \sigma(\mathcal{B}_d) \otimes \text{CFT}_\text{int}(t) \quad \longmapsto \quad \text{EFT}_d(\mathcal{B}_d, t) \, .
\end{eqaed}
Specifically, $\text{CFT}_\text{int}(t)$ denotes the internal sector of the worldsheet \ac{CFT}, and comprises a unitary (super)conformal theory with critical central charges with a unique vacuum and a discrete spectrum. For RNS-RNS constructions the critical central charges are $c = \overline{c} = 15 - 3d/2$, while for heterotic constructions one of them equals $26-d$. The internal \ac{CFT} can have a conformal manifold, parametrized by local coordinates $t = (t^i)$; these correspond to (pseudo-)moduli of the spacetime \ac{EFT}, namely (classical expectation values of) light scalar fields whose classical potential vanishes\footnote{Strictly speaking, for the arguments presented in this paper we only need to take limits in a family of \acp{CFT}. There is a growing body of evidence that any such (continuous) family comprises a conformal manifold \cite{Komatsu:2025cai}. Therefore, we shall identify these two notions in the following.}. In practice, we shall consider a single coordinate $t$ parametrizing a curve in the conformal manifold. The external spacetime sector is described by a (superconformal) non-linear sigma model on a background including the spacetime geometry and any other field profile, collectively denoted $\mathcal{B}_d$. Insofar as the background is uniformly weakly curved with respect to the string scale $\alpha' = M_s^{-2}$, its dynamics (encoded in the string S-matrix or coherent states of vertex operators) is captured by the \ac{EFT}. Therefore, in practice we restrict to the classical empty Minkowski background, since neighbouring backgrounds describe the same \ac{IR} phase. In other words, it is the internal \ac{CFT} that determines the spacetime \ac{EFT} we wish to study; hence, universal (generic) properties of the latter correspond to universal (generic) properties of the former, which is very constrained. The final ingredient in the definition of the assignment in \cref{eq:string_worldsheet} is the \ac{GSO} projection, which dictates the discrete data on how the sum over superconformal structures is performed in the Polyakov path integral. We shall not assume that the \ac{GSO} projection leads to unbroken spacetime supersymmetry; we will however assume that it does not lead to tachyons in the physical spectrum, since their effects---even if we could follow the instability to its final configuration---generically take place at the string scale which lies beyond the \ac{EFT} regime. (In particular, this means that spacetime will contain fermionic excitations \cite{Kutasov:1990sv, Leone:2023qfd}.) \\

\noindent Within this framework and the limitations that we have discussed, we will compute the parametric \ac{EFT} data as a function of $t$ and the $d$-dimensional string coupling $g_{s,d}$, which is set by the dilaton. To this end, we will employ two complementary technical tools:

\begin{itemize}
    \item The background-field method of \cite{Kiritsis:1994yv, Kiritsis:1994ta, Kiritsis:1995ga, Kiritsis:1995dx, Kiritsis:1996yb, Kiritsis:1997hj}, which involves turning on a constant gauge-gravitational background $(R,F)$ in the spacetime sector of the worldsheet. The technical achievement of this approach is retaining modular invariance while regulating the \ac{IR} spectrum; the price to pay is that the resulting generating function only encodes certain combinations of tensor contractions among the full expansion in \cref{eq:wilsonian_action}. In particular, it cannot probe operators with covariant derivatives. This method, which improved upon the initial efforts in \cite{Dixon:1990pc, Antoniadis:1992rq, Kaplunovsky:1992vs}, led to many computations of threshold corrections in specific settings \cite{Petropoulos:1996rr, Petropoulos:1996xz, Kiritsis:1996dn, Kiritsis:1998en, Antoniadis:1999ge, Florakis:2016aoi, Angelantonj:2016gkz}. In this paper, we strive to keep the internal \ac{CFT} as general as possible, and extract generic universal scalings for the \ac{EFT} data combining this method with the asymptotic differential equations introduced in \cite{Aoufia:2024awo, Basile:2024lcz}. Although the main focus of this method is four-dimensional \acp{EFT}, in principle it can apply to higher dimensions by restricting the support of background-field tensors.

    \item The low-energy expansion of string scattering amplitudes, which has long been a source of insights on stringy \acp{EFT} and dualities \cite{Gross:1986iv, Ellis:1987yx, Ellis:1987dc, Montag:1992dm, Green:1999pv, Obers:1999um, Green:2008uj, Richards:2008jg, Berg:2016wux, Berg:2016wux}. Once again, in the spirit of \cite{Aoufia:2024awo}, we will deviate from the vast literature on this subject in two main respects: keeping the internal \ac{CFT} generic, and studying the parametric dependence of low-energy coefficients. As in the background-field approach, concretely this involves deriving and solving asymptotic differential equations for one-loop amplitudes \cite{Aoufia:2024awo, Basile:2024lcz}. Furthermore, in this paper we significantly extend the scope of our preceding analyses, applying our methods to higher-point amplitudes and providing an explicit treatment of kinematic poles and boundary terms.
\end{itemize}

\noindent In our preceding works, the latter method was used to find the parametric scalings governing the species scale (via a specific proxy) \cite{Aoufia:2024awo} and the vacuum energy \cite{Basile:2024lcz}. In this paper, starting with the other important piece of \ac{IR} data (gauge couplings), we extend the program to cover infinitely many higher-derivative corrections in the gauge-gravitational sector, higher-loop contributions, the resulting \ac{UV}/\ac{IR} relations with \ac{IR} data and, finally, some applications to phenomenology. Beyond a big-picture understanding of the string landscape, one of our long-term goals is to place on firmer theoretical grounds a number of swampland-motivated phenomenological models, showing that the bottom-up considerations related to black-hole physics and holography are embodied by concrete mathematical properties within the worldsheet framework we described. In addition, from the swampland perspective, our findings can be viewed as providing a much stronger support for the \ac{ESC} in perturbative string theory, generalizing our previous results in \cite{Aoufia:2024awo, Basile:2024lcz} to infinitely many higher-curvature gauge-gravitational terms in the effective action. \\

\noindent \paragraph{Summary of contents.} The paper is organized as follows.
In \cref{sec:gauge_couplings_higher_derivative_corrections} we set the stage from an \ac{IR} perspective, recall the background field method of \cite{Kiritsis:1994ta}, and apply it to gauge couplings in four-dimensional \acp{EFT}. 
In \cref{sec:universal_thresholds_modular_invariance} we review the \ac{CFT} machinery introduced in \cite{Aoufia:2024awo, Basile:2024lcz}, using the \ac{UV} rigidity stemming from modular and conformal invariance to study species limits and derive asymptotic differential equations. We then apply these methods to gauge couplings \cref{sec:gauge_couplings_species_limits} (and higher-derivative corrections in \cref{sec:bkg_field_method_all-orders}) using the background-field method, and present a unified treatment of low-energy scattering amplitudes in \cref{sec:higher-derivative_corrections_species_limits} (specifically \cref{sec:scattering_amplitudes_all-orders}).
In \cref{sec:UV/IR_relations_and_swampland}  we describe the interplay between the \ac{UV} and \ac{IR} perspectives, discussing how the asymptotic one-loop scalings are affected by higher-loop corrections and other effects. Our findings provide significant top-down support for swampland constraints, which take the form of (parametric) holographic bounds; applying them to our universe, the dark dimension scenario seems to be favored by genericity. 
We conclude in \cref{sec:conclusions} with a summary and open questions on how to push our results in several directions. Finally, the paper contains four appendices. In \cref{app:eisenstein_series_and_regularization} we present some useful material on modular functions that are heavily employed in the main text. In \cref{app:exp_suppression} we provide some mathematical details and proofs regarding boundary terms and subleading asymptotics used in the derivation of asymptotic differential equations. In \cref{app:additive_towers_anisotropic_limits} we discuss how our derivations apply to more general spectra than the ones presented in the main text, including subleading and/or multiple light towers of species. Finally, in \cref{app:unphysical_tachyons} we address the possibility of off-shell tachyons in the string spectrum, which introduce a number of technical subtleties in the derivations presented in the main text. \\

\section{Gauge couplings and higher-derivative corrections}\label{sec:gauge_couplings_higher_derivative_corrections}

\noindent Gauge couplings play a special role among \ac{EFT} observables especially in four spacetime dimensions: besides the vacuum energy, they are in fact the only other quantities persisting in the deep \ac{IR} of the bosonic sector\footnote{A generic gauge theory can in principle confine or undergo Higgsing while running to the \ac{IR}. Here, we mean gauge couplings for which no such process occurs.}. Consequently, they provide an ideal probe of \ac{UV}/\ac{IR} mixing in quantum gravity. In string theory these couplings---together with the species scale and the \ac{EFT} cutoff---depend on (pseudo-)moduli in a peculiar way. Throughout this work, we will argue that \ac{UV}/\ac{IR} mixing manifests as a universal (asymptotic) scaling of one-loop gauge couplings and higher-derivative corrections in terms of these scales, and derive this result from the worldsheet in the following section. In this section, we set the stage by discussing how the relevant one-loop couplings may be computed from the worldsheet \ac{CFT} through various methods, focusing on the background-field method of \cite{Kiritsis:1994ta} which will allow us to express the result in a modular-covariant fashion. We end the section by applying this method to the specific example of the $\text{E}_8 \times \text{E}_8$ heterotic orbifold $T^4/\bZ_2 \times T^2$. \\

\noindent In what follows, as in \cref{eq:wilsonian_action}, we shall normalize the kinetic term of abelian gauge fields with field strength $F$ according to
\begin{eqaed}\label{eq:gauge_normalization_again}
 S_{\text{gauge}} =  -\frac{1}{4g^2}\int \td^d x\, \sqrt{-g}\,  F^2 \, , 
\end{eqaed}
where the Einstein-Hilbert term is implicitly understood to be written in the Einstein frame. One-loop corrections to gauge couplings in four dimensions can be isolated by providing a normalization with respect to the tree-level piece, namely
\begin{eqaed}\label{eq:gauge_oneloop_normalization}
    \frac{16\pi^2}{g^2} = \left.\frac{16 \pi^2}{g^2}\right|_{\text{tree-level}} + \Delta_g \, .
\end{eqaed}
Although we will mainly concern ourselves with abelian theories, let us mention that when we dealing with non-abelian theories, $g$ is understood as normalized so that covariant derivatives are written only in terms of the generators of the Lie algebra. Additionally, we will be interested in higher-derivative corrections involving both gauge and spacetime curvatures $F$ and $R$, which in the Einstein frame take the form
\begin{eqaed}\label{eq:higherdercorr}
   \Delta S_{n,m} = \int \td^d x \, \sqrt{-g} \, c_{n,m} \, \frac{R^n F^m}{M_{{\text{Pl}};d}^{2n+2m-d}} \, ,
\end{eqaed}
with $M_{{\text{Pl}};d}$ the $d$-dimensional Planck mass. Hereafter, we will leave the structure of the contractions in \cref{eq:higherdercorr} unspecified, since it does not affect our analysis.
It is useful to rewrite the coefficient $c_{n,m}$ above in terms of string-frame quantities, as these will be the ones naturally computed from the worldsheet. Focusing on closed-string amplitudes\footnote{We will comment on possible open-string and higher-loop contributions in \cref{sec:UV/IR_relations_and_swampland}.}, the above coefficient is related to the genus-$h$ effective action term 
\begin{eqaed}\label{eq:genus_h_eff_act}
    \Delta S_{n,m} = \int \td^d x \, \sqrt{-g} \, e^{(2h-2)\phi_d} \, a_{n,m}^{(h)} \, \frac{R^n F^m}{M_s^{2n+2m-d}} \, ,
\end{eqaed}
where $M_s$ is the string scale and $\phi_d$ the $d$-dimensional dilaton, which for geometric compactifications from $D>d$ dimensions can be related to the higher-dimensional dilaton $\phi_D$ according to $e^{2\phi_d}=e^{2\phi_D}/\cV$, with $\cV$ the internal volume in string units\footnote{More generally, the relation between $\phi_d$ and $\phi_D$ is expected to be encoded by the sphere partition function, although care must be exercised in its definition.}. Taking into account that the change to Einstein frame can be performed by $M_{{\text{Pl}};d}^{d-2} = e^{-2\delta\phi_d} M_s^{d-2}$ in terms of the fluctuation of $\phi_d$ from its asymptotic background value, one is led to the one-loop relation \cite{Aoufia:2024awo}
\begin{eqaed}\label{eq:string_vs_einstein_coeff}
    c_{n,m} = \left( \frac{M_{{\text{Pl}};d}}{M_s}\right)^{2m+2n-d} \left( \frac{a^{(0)}_{n,m}}{g_{s,d}^2} + a^{(1)}_{n,m}\right) ,
\end{eqaed}
where $g_{s,d}$ is the  $d$-dimensional string coupling\footnote{More generally, genus-$h$ amplitudes inherit a factor of $g_{s,d}^{2h-2}$ from the asymptotic value of $e^{(2h-2)\phi_d}$.}. \\

\noindent The geometric language of decompactifications we used above turns out not to be a special case: the results of \cite{Aoufia:2024awo, Ooguri:2024ofs, Basile:2024lcz}, which we will further extend and reinforce in this paper, tell us that \emph{all species limits comply with the \ac{ESC}}, which is a non-trivial statement even at the worldsheet level due to the existence of non-geometric settings. As such, we will keep using the geometric language when useful throughout the paper. As we will discuss in more detail in \cref{sec:higher-derivative_corrections_species_limits}, general \ac{EFT} considerations for decompactification limits give two types of behaviors for $a_{n,m}$, which are weighted by the species scale and the \ac{KK} scale respectively \cite{Calderon-Infante:2025ldq}. For us, the species scale is always going to be the string scale, since it is bounded by the higher-dimensional Planck scale in decompactification limits and we never consider M-theory-like limits at strong coupling. Then these two terms predicted by \ac{EFT} scale according to
\begin{eqaed}\label{eq:wilson_coeff_EFT_expectation_planck}
    a_{n,m} \sim a^\text{string}_{n,m} \left(\frac{M_{{\text{Pl}};d}}{M_s}\right)^{2n+2m-2} + a^\text{KK}_{n,m} \left(\frac{M_{{\text{Pl}};d}}{m_\text{KK}}\right)^{2n+2m-d} .
\end{eqaed}
However, since these terms are \ac{UV}-sensitive, it is not \emph{a priori} clear that no other terms in between these will emerge; a decompactification would remove all of them except the first term in \cref{eq:wilson_coeff_EFT_expectation_planck}, which is the dimensional reduction of the corresponding higher-dimensional term. In addition to \ac{UV}-sensitive terms of higher-dimensional origin, there could also be other terms stemming from $d$-dimensional loop effects; at any rate, since they arise from a species counting, no such term would overcome the species-like scaling in \cref{eq:wilson_coeff_EFT_expectation_planck}. When $2n+2m-d=0$, the operator is (classically) marginal and a further logarithmic term is expected; similarly, when the operator is (classically) marginal in the higher-dimensional theory, \cref{eq:wilson_coeff_EFT_expectation_planck} can feature a different logarithmic contribution, as we shall see. Furthermore, it is useful to keep in mind a qualitative difference between gauge fields coming from gauge fields ``upstairs'' and those arising from the gravitational sector upon dimensional reduction (i.e. graviphotons). Let us see this in a concrete example. Consider to compactify a five-dimensional theory on a circle of radius R; the one-loop self-energy for a photon coming from a gauge field involves a sum over \ac{KK} replicas of the fields running in the loop, characterized by the same charge $q$. Schematically, the loop computation involves a sum bounded by the species number $N_{\text{sp}}\equiv \Lambda_{\text{sp}} R$ according to
\begin{eqaed}\label{eq:eft_loop_photon}
    \sum_{|n|=1}^{N_{\text{sp}}} q^2 \log \frac{e^{3k}\Lambda_{\text{sp}}}{m_n} = \sum_{|n|=1}^{N_{\text{sp}}} q^2 \log \frac{e^{3k}N_{\text{sp}}}{|n|}  \, , \quad m_n = \frac{|n|}{R} \, ,
\end{eqaed}
where $k$ parametrizes the \ac{UV}-ambiguity of the cutoff and cannot be fixed by the \ac{EFT}. At parametrically large $N_{\text{sp}}$, we thus get
\begin{eqaed}\label{eq:eft_loop_photon_param}
    \sum_{|n|=1}^{N_{\text{sp}}} q^2 \log \frac{e^{3k}N_{\text{sp}}}{|n|}  \overset{N_{\text{sp}}\gg 1}{=} q^2 N_{\text{sp}} -\frac{q^2}2 \log N_{\text{sp}} + O(1) \, .
\end{eqaed}
Repeating the computation for graviphotons, their charge is instead the \ac{KK}-momentum  $q_n = n$, and the sum now reads
\begin{eqaed}\label{eq:graviphoton_coupling}
     \sum_{|n|=1}^{N_{\text{sp}}} n^2 \log \frac{e^{3k}N_{\text{sp}}}{|n|} \overset{N_{\text{sp}}\gg 1}{=} \frac{2+18k}{9} N_{\text{sp}}^3 + 3k N_{\text{sp}}^2+\frac{6k-1}{6}N_{\text{sp}} + O(1) \, ,
\end{eqaed}
where the $k$ factor can be chosen to cancel some terms. Expressing these corrections in terms of the radius measured in 5d Planck units, $r\equiv R\Lambda_{\text{sp}} = N_{\text{sp}}$, we see that graviphotons acquire a leading scaling of $r^3$, instead of the usual $r$ experienced by other photons; moreover, the logarithmic contribution disappears. We will discuss in greater detail on how this distinction arises from the worldsheet.\\

\noindent In order to extract this physics from the worldsheet, one ought to compute Wilson coefficients involving graviton and gauge bosons insertions, which might produce \ac{IR} divergences if the corresponding operator is relevant or marginal\footnote{Although string scattering amplitudes are inherently finite, extracting their analytic dependence in the Mandelstam variables requires a decomposition in multiple pieces which may individually diverge. It is the analytic pieces that are matched to couplings in the Wilsonian effective action of the \ac{EFT}. \label{footnote:divergencies}}. As reviewed in \cite{Kiritsis:1997hj}, there are a number of methods to get a handle on such divergences:
\begin{itemize}
    \item[1.] The original approach in the computation of gauge couplings in four dimensions, due to \cite{Kaplunovsky:1992vs}, is to manually introduce a gauge-invariant regulator for the massless states in the two-point function, stripped of wave-function factors that would make the on-shell amplitude vanish. This approach is, however, not modular invariant. 

    \item[2.] Another approach \cite{Antoniadis:1992rq} computes derivative of couplings with respect to moduli fields by evaluating e.g. gauge-gauge-scalar amplitudes, which are free of \ac{IR} divergences and modular invariant. However, it cannot detect moduli-independent contributions (which are potentially ambiguous in the presence of logarithmic running). Moreover, wave-function factors still need to be removed by hand.

    \item[3.] The most well-defined approach \cite{Kiritsis:1995dx} is to introduce a background field for the gauge and gravitational curvatures directly in the worldsheet \ac{CFT}, and compute off-shell correlators with the \emph{background-field} method. This is the approach which we will follow until \cref{sec:scattering_amplitudes_all-orders}, and we will describe it in detail in the rest of this section.

    \item[4.] The other main approach that we will undertake in \cref{sec:scattering_amplitudes_all-orders} is to extract \ac{IR} coefficients from the low-energy expansion of suitable scattering amplitudes. In general, these may feature tadpoles and kinematic poles, whose behavior is consistent with \cref{eq:wilson_coeff_EFT_expectation_planck} as we will discuss. 

    \item[5.] Finally, one can start instead from the field-theoretic result and perform a modular completion as pursued in \cite{Abel:2021tyt,Abel:2023hkk}. In particular, we will compare the results of this method with the background-field method above.
\end{itemize}

\noindent In what follows we will describe the background-field method, introduce the language of helicity and flavored partition functions and provide an explicit example of the one-loop computation of four-dimensional gauge couplings in heterotic string theory. 

\subsection{Background-field method}\label{sec:background_field_method}
As we have remarked above, gauge couplings in four spacetime dimensions are classically marginal and generically exhibit logarithmic running at one-loop level. From the worldsheet perspective, this is reflected in an \ac{IR} divergence of the two-point correlation function between gauge fields. Refs. \cite{Antoniadis:1994sr,Kiritsis:1995dx} introduced an \ac{IR} regularization procedure involving the replacement of the flat-space sector of the worldsheet \ac{CFT} with a gapped one, thereby introducing an \ac{IR} cutoff $\mu$ to be removed afterwards. The consistency of the string worldsheet reduces the possiblities to a few choices, insofar as one wants the replacement to preserve as much supersymmetry as possible \cite{Antoniadis:1994sr}. Notably, this approach allows to turn on constant background fields for the spacetime and gauge curvatures $(R,\,F)$ in terms of marginal deformations of the worldsheet sigma model and use the background-field method to compute correlators \cite{Kiritsis:1994ta}. Although this method is mainly applied to four-dimensional \acp{EFT} in the literature, it can in principle be applied in any higher dimension by keeping the marginal deformation supported on the four dimensions described by the gapped spacetime \ac{CFT}. Throughout the rest of the paper, as outlined in \cref{sec:introduction}, we will take the initial \ac{CFT} $\mathsf{T} = \mathsf{F} \otimes \mathsf{I}$ to be a tensor product of an external free theory $\mathsf{F}$ and an internal sector $\mathsf{I}$. \\

\noindent In order to perform the regularization, we follow \cite{Kiritsis:1995dx} and replace the flat-space theory $\mathsf{F}$ with the $\mathsf{W}_k$ theory, whose $N =4$ superconformal algebra is constructed from ${\text{SU}(2)}_{ k} \times {\mathbb{R}}_{q}$ currents and their fermionic partners, which generate an additional ${\text{SO}(4)_1}$ algebra\footnote{We will not be interested in global aspects of the gauge group, such as the spectrum of topological operators or the allowed representations of the Kac-Moody symmetries of the \ac{CFT}, so by abusing notation we will denote e.g. $\fu(1)_k$ simply by ${\text{U}}(1)_k$.}. Note that the charge $q=\sqrt{k+2}$ of the linear-dilaton component ${\bR}_{q}$ is fixed to compensate for the central-charge deficit of the compact ${\text{SU}(2)}_{ k}$ Kac-Moody algebra, ensuring criticality of the background for any level $k$. The limit $k\to \infty$ recovers the flat-space theory, as can be intuitively understood by interpreting the background in the large radius limit as $S^3 \times \bR$. The \ac{IR} regulator is thus identified with the mass gap of the theory $\mu^2 =(k+2)^{-1}$ (in string units). The interested reader can find more details in \cite{Antoniadis:1994sr}. Henceforth, we focus on the \emph{regularized} partition function, defined from the torus partition function of the flat-space theory $Z_{F}(\tau)$ according to
\begin{eqaed}\label{eq:ZW_part_func}
 Z_W (\tau, \mu) = \frac{1}{V(\mu)} \, \Gamma_{{\text{SU}(2)}_k} \, Z_{F}(\tau) \, ,  
\end{eqaed}
where $V(\mu) \equiv (8\pi\mu^3)^{-1}$ is the volume of the compact part of the sigma model and $\Gamma_{{\text{SU}(2)}_k}$ its ratio with the free external bosons, i.e. 
\begin{eqaed}\label{eq:gamma_su2_def}
     \Gamma_{{\text{SU}(2)}_k} = (\sqrt{\tau_2} \eta \overline \eta)^3 \sum_{\alpha, \beta =0}^1e^{-i\pi \beta \ell } \chi_\ell(\tau) \overline \chi_{\ell +\alpha(k-2l)}(\overline \tau) \, .
\end{eqaed}
In \cref{eq:gamma_su2_def}, $\chi_{\ell}$ are the ${\text{SU}(2)}_{ k}$ affine characters of $\ell/2$ spin and $\eta$ the Dedekind eta function. Note that in the $k \to \infty$ 
limit one recovers $Z_{F}(\tau,\overline \tau)$, since $\Gamma_{{\text{SU}(2)}_k}$ reduces to $V(\mu)$. Since we are fixing the external sector to be given by $\mathsf{W}_{k}$, we can rewrite the generic \ac{CFT} partition function in a way which renders the characters of the external light-cone bosons $X^3, X^4$ and fermions $\psi^3,\psi^4$ explicit, namely
\begin{eqaed}\label{eq:Zwexternalexplicit}
     Z_W (\tau, \mu) = \frac{1}{\tau_2\eta^2 \overline\eta^2}\frac{\Gamma_{{\text{SU}(2)}_k}}{V(\mu)} \sum_s r(s) \, \frac{\vartheta(s)}{\eta} \, Z_{\text{int}}^{(s)}(\tau) \, , 
\end{eqaed}
where $\vartheta(s)$ are the Jacobi $\vartheta$-functions (with the argument $\tau$ left implicit) and $s$ collectively denotes possible twists due to the choice of spin structure or twisted sector. More precisely, the $\vartheta$-functions pertaining to the fermions in the external spacetime sector only depend on $s$ via the worldsheet spin structure. The coefficient $r(s)$ denotes the choice of \ac{GSO} projection, while $Z_{\text{int}}^{(s)}$ is the partition function for the internal degrees of freedom at fixed $s$. From the worldsheet perspective, gauge groups are realized as level-$k$ Kac-Moody algebras of the worldsheet \ac{CFT}. In particular, focusing on the abelian case, one can normalize the operator product of ${\text{U}(1)}^p$ (right-moving) Kac-Moody currents $\bar J^a$ as 
\begin{eqaed}\label{eq:JJ_normalization}
    \bar J^a(\overline z) \bar J^b (0) = \frac{\delta^{ab}\ k_a}{2\overline z^2} + \dots \, , \quad a=1,\dots,p
\end{eqaed}
where the levels $k_a$ can be reabsorbed in the normalization of the currents, and will be taken to be $k_a=2 \ \forall a$ without loss of generality for the remainder of the paper. For simplicity, we will assume in what follows that the initial \ac{CFT} $\mathsf{F}$ contains exactly one such $\text{U}(1)$ current $\bar J$.\\

\begin{figure}[!t]
    \centering
    \includegraphics[width=0.9\linewidth]{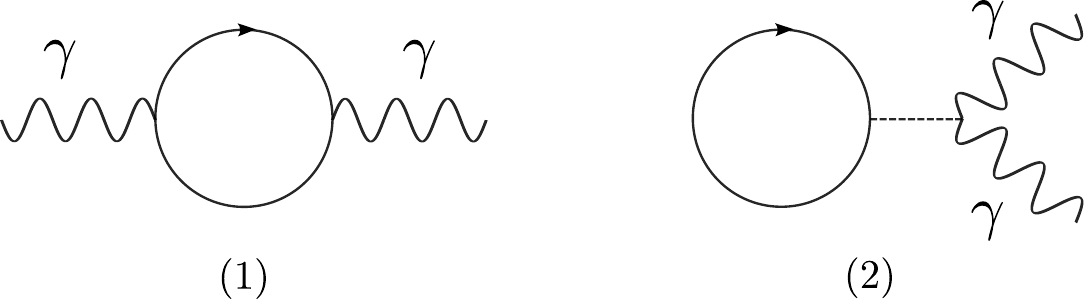}
    \caption{Field-theoretic diagrams contributing to the string-theoretic correlator in \cref{eq:f2correlator}. Dashed lines represent states in the gravitational sector, such as the dilaton. Panel (1) corresponds to the contribution from charged states. Panel (2) instead shows the universal contribution coming from the photon coupling to the gravitational sector, which itself couples to any other state.}
    \label{fig:diagrams}
\end{figure}

\begin{figure}[!t]
    \centering  \includegraphics[width=0.62\linewidth]{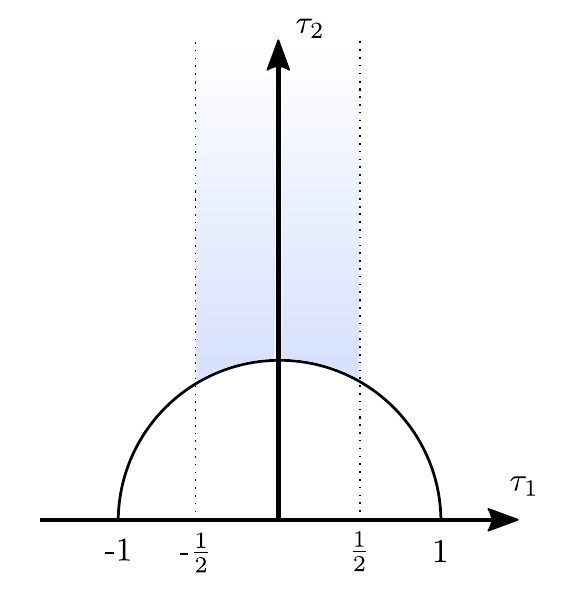}
    \caption{A depiction of the standard ${\text{SL}}(2,\bZ)$ fundamental domain $\cF$, which parametrizes by a complex coordinate $\tau = \tau_1 + i \tau_2$ the moduli space of conformal (equivalently complex) structures of tori.}
    \label{fig:fundamental_domain}
\end{figure}

\noindent With these ingredients, one can turn on marginal deformations corresponding to a purely magnetic constant background field $F$, as well as constant curvature $R$, by adding to the action the following two terms \cite{Antoniadis:1994sr, Kiritsis:1995ga}\footnote{In \cite{Kiritsis:1995dx}, it is remarked that these deformations are not marginal in the original flat theory, so that the background field method can only be employed in the curved theory. This is due to the fact that both curvatures couple to the angular momentum operator $X^{[\mu} \partial X^{\nu]}$, which however is not a proper conformal field. Relatedly, turning on a constant magnetic field induces a non-trivial Weyl anomaly that needs to be compensated by a gravitational term. The situation is different in the case of open strings, for which constant magnetic fields are \emph{bona fide} marginal operators in the flat-space theory on the boundary of the worldsheet.}:
\begin{eqaed}\label{eq:marginal_def_RF}
   & \Delta S_{\text{CFT}}^{(F)} = \int\td^2 \sigma F \left( I^3 + \psi^1 \psi^2 - \psi^3 \psi^4 \right) \bar J \, , \\   & \Delta S_{\text{CFT}}^{(R)} = \int\td^2 \sigma R \left( I^3 + \psi^1 \psi^2 - \psi^3 \psi^4 \right) \bar I^3 \, ,
\end{eqaed}
where $I^3$ and $\psi^1\psi^2-\psi^3\psi^4\equiv i\partial H$ are respectively the Cartan generators of ${\text{SU}(2)}_k$ and ${\text{SU}(2)}_H \subset {\text{SO}(4)_1}$. Denoting with $\overline \cQ$, $\cI$, $\overline \cI$ and $\cQ_H$ the zero-modes of $\bar J$, $I^3$, $\bar I^3$ and $i\partial H$ respectively, one can exactly integrate the deformations inducing a shift in the Virasoro generators \cite{Kiritsis:1995dx}
\begin{eqaed}\label{eq:L0_deformation}
 \delta L_0 = \delta \overline L_0 &\equiv (\cQ_H + \cI)(R\overline{\cI} + F\overline{\cQ}) +\frac{\lambda(F,R)-1 }{2} \left[ \frac{(\cQ_H + \cI)^2}{k + 2} + \frac{(F \overline{\cQ} + R \overline{\cI})^2}{2 F^2 + k R^2} \right] , \\
  \quad\lambda(F,R) &= \sqrt{1 + (k+2)(2F^2 + k R^2)} \, ,
\end{eqaed}
which enter the partition function through the standard expression involving the trace over the ${\mathsf W}_{k}$ Hilbert space. Denoting with $(c,\overline c)$ the central charges of the overall \ac{CFT}, this reads
\begin{eqaed}\label{eq:generating_Z}
    Z_W(\tau, \mu ;F,R) = {\rm Tr} \left[e^{-2\pi\tau_2\left[\left(L_0 + \delta L_0\right)+ (\overline L_0 + \delta \overline L_0)+\frac{c+\overline c}{24}\right]+2\pi i \tau_1 \left(L_0 - \overline L_0 + \frac{c-\overline c}{24}\right)}\right] .
\end{eqaed}
As a result, one can compute off-shell $(n+m)$-point correlators as the coefficients $Z_W^{n,m}$ appearing in power-series expansion with respect to the background fields $Z_W(F,R) = \sum_{n,m}Z_W^{n,m} R^n F^m$. We will in particular be interested in the two-point function involving gauge curvatures, which we can write explicitly in terms of operator insertions as\footnote{We denote by $\langle X\rangle$ the torus partition function with insertion $X$. This is not to be confused with integration over the fundamental domain, since this expectation value is generically a function of $\tau$.} \cite{Kiritsis:1995dx}
\begin{eqaed}\label{eq:f2correlator}
   \langle F^2 \rangle(\tau,\mu) = 8 \pi^2 \tau^2_2 
\left\langle\left(  ({\cal Q}_H+{\cal I})^2  - \frac{k+2}{8 \pi\tau_2} \right)
\left(  \overline{\cal Q}^2  - \frac{1}{4\pi\tau_2} \right) \right\rangle
- \frac{ (k+2)}{4} \, \langle1\rangle \, .
\end{eqaed}
It is worth noting that the last term is proportional to the vacuum energy and vanishes in theories with spacetime supersymmetry. Even in non-supersymmetric settings, however, this contribution is to be matched to a tadpole diagram in the field-theory limit and as such needs to be subtracted from the irreducible correlator (see \cref{fig:diagrams}). More broadly, it is expected to naturally disappear in the computation of one-loop corrections once a suitable renormalization scheme is employed, see \cite{Florakis:2016aoi} for details. Furthermore, notice that differently from similar terms appearing in the product of the two parentheses in \cref{eq:f2correlator}, this term is independently modular invariant, meaning its subtraction is compatible with integration over the fundamental domain $\cF$\footnote{For more details about this subtraction, see \cite{Florakis:2016aoi}.} depicted in \cref{fig:fundamental_domain}. Accordingly, we will omit it in the following. We can finally write the regularized one-loop correction to the gauge coupling as
\begin{eqaed}\label{eq:gaugecoupling}
    \Delta_g(\mu) = -\frac{1}{4\pi^2}\int_{\cF}\frac{\td^2 \tau}{\tau_2^2} \, \langle F^2 \rangle \, .
\end{eqaed}
In the following, we will recast the above expression for the correlator in a manifestly modular-covariant fashion by employing the formalism of flavored partition functions, which we now turn to.

\subsubsection{The flavor and helicity partition functions}\label{sec:flavor_helicity_partition_functions}
In this section we show that the operator insertions in \cref{eq:f2correlator} can be conveniently expressed in terms of \textit{helicity} and \textit{flavored} partition functions\footnote{We refer the reader to \cite{Kiritsis:1997hj, Kraus:2006wn, Dyer:2017rul} for a more thorough discussion.}. Both quantities can be thought of as the insertion of a chemical potential in the defining partition function of the 2d \ac{CFT}. That is, given the zero-mode of a conserved holomorphic\footnote{For illustration purposes, we restrict the discussion to purely (anti-)holomorphic insertions. In the settings at stake, both holomorphic and anti-holomorphic insertions arise.} level-$m$ current $J_0$, one can construct the generating function of the corresponding correlators according to
\begin{eqaed}\label{eq:flavor_partition_func}
 Z(\tau; \zeta) \equiv {\rm Tr}\left( q^{L_0-\frac{ c}{24}} \overline q^{\overline L_0-\frac{\overline c}{24}}e^{2\pi i \zeta J_0}\right) ,
\end{eqaed}
with $(c,\overline c)$ the left- and right-moving central charges and $q=e^{2\pi i \tau}$ the nome of the torus. The resulting partition function is no longer modular invariant, although its modular properties are universal and fixed by the transformation \cite{Kraus:2006wn,Dyer:2017rul,Heidenreich:2024dmr}
\begin{eqaed}\label{eq:modular_trsf_flavor_pf}
    Z\left( \frac{a\tau + b}{c\tau + d}; \frac{\zeta}{c\tau + d}\right) = e^{i\pi m \frac{c \zeta^2}{c\tau + d}}Z(\tau;\zeta)\ , \quad m\in \bZ \, .
\end{eqaed}
This allows one to engineer an ``improved'' modular invariant partition function by attaching a universal prefactor compensating the transformation, which is given by \cite{Schellekens:1986xh, Lerche:1987qk, Kraus:2006wn, Heidenreich:2024dmr}
\begin{eqaed}\label{eq:improvedfpf}
 \hat Z(\tau; \zeta) \equiv e^{\pi m \frac{\zeta^2}{2\tau_2}} \, Z(\tau;\zeta) \, , \quad \hat Z\left( \frac{a\tau + b}{c\tau + d}; \frac{\zeta}{c\tau + d}\right) =  \hat Z(\tau; \zeta) \, .
\end{eqaed}
The fugacity $\zeta$ introduced above is a function of modular weight $-1$, as can be understood by its transformation $\zeta \to \frac{\zeta}{c\tau + d}$ under the modular group. Thus, starting from \cref{eq:improvedfpf}, one can readily read off the modular properties of its derivatives with respect to $\zeta$, resulting in $\partial_\zeta^n$ yielding a modular form of weight $n$. One can then write modular-covariant correlators involving $J_0$ by taking derivatives of the improved generating function. \\

\noindent We now aim to rewrite \cref{eq:f2correlator} in terms of improved flavored partition functions. To begin with, we observe that the charges appearing in the left- and right-moving sectors are respectively $\cQ_H+\cI$ and $\overline \cQ$. The former, in particular, can be understood as the zero-mode of the diagonal combination of $\text{SU}(2)_k$ and $\text{SU}(2)_H$ Cartan generators
\begin{eqaed}\label{eq:hel_op}
    \cN^3 \equiv I^3 + \partial H \, ,
\end{eqaed}
where $i \partial H= \psi^1 \psi^2-\psi^3 \psi^4$ is a bosonized field. Comparing the free-field and $\mathsf{W}_k$ superconformal algebras, one can see that this operator defines a light-cone gauge $\bZ_2$ parity with respect to which the $N=4$ supercurrents $(G,\tilde G)$ and their conjugates have charge $\pm 1/2$ respectively. Denoting by $\cN = \cQ_H + \cI$ the zero-mode of $\cN^3$, one can write this operator as $(-1)^\cN$. The corresponding free-field realization is the helicity operator $\cN_F =N_P + \partial H$, where $N_P$ counts the number of $P=\partial X^3 +i\partial X^4$ minus $P^\dagger$ excitations, while light-cone fermions $\psi^{\pm} = \psi^3\pm i\psi^4$ contribute $\pm1$ to $\partial H$. Thus, $\cN$ can be identified with the ``helicity'' charge in the curved theory \cite{Antoniadis:1994sr}. By direct inspection, we see that the chemical potentials we need to consider take the form
\begin{eqaed}\label{eq:flavpartfunc}
 Z_W(\tau,\mu;v,\overline z) = {\rm Tr}\left( q^{L_0-\frac{\overline c}{24}} \overline q^{\overline L_0-\frac{c}{24}}e^{2\pi i v \cN + 2\pi i \overline z \cQ }\right) .
\end{eqaed}
By the arguments above, in the following we will dub the flavored partition function with respect to $\cN$ the \textit{helicity} partition function. Comparing with other methods, we shall see that correlators involving the helicity charge $\cN$ will indeed be identified with helicity insertions in the flat-space limit. Since the external sector is specified by the $\mathsf{W}_k$ theory, we can straightforwardly turn on the fugacity $v$ by deforming the relevant characters of external light-cone bosons and fermions in \cref{eq:Zwexternalexplicit}. In particular \cite{Florakis:2016aoi} 
\begin{eqaed}\label{eq:affinedeformedsu(2)}
      \chi_\ell (\tau) \to \chi_\ell (v,\tau) \equiv 2\sum_{\gamma \in \Lambda_\ell} q^{(k+2)\langle \gamma, \gamma\rangle} \, \frac{\sin (2\pi (k+2) \langle v,\gamma \rangle)}{\vartheta_1(v)} \, , \ \quad\  \vartheta(s) \to \vartheta(s,v) \, ,
\end{eqaed}
with $\Lambda_\ell = \mathbb{Z}+\frac{2\ell+1}{2(k+2)}$, in terms of the affine lattice of the spin-$\ell$ ${\text{SU}(2)}_k$ representation and the scalar product between lattice vectors $\langle - , - \rangle$. This can be also expressed according to \cite{DiFrancesco:1997nk}
\begin{eqaed}\label{eq:characters_theta}
   \chi_\ell (v,\tau)=\frac{\Theta^{(k+2)}_{2 \ell+1}(v|\tau)-\Theta^{(k+2)}_{-2 \ell-1}(v|\tau)}{\vartheta_1(v|\tau)} \, ,
\end{eqaed}
in terms of the generalized $\vartheta$-functions
\begin{eqaed}\label{eq:generalized_theta}
   \Theta_{m}^{(k)}(v|\tau)=\sum_{n \in \mathbb{Z}+\frac{m}{2k}} q^{kn^2} e^{2 \pi i v kn} \, .
\end{eqaed}
The function $\vartheta_1$ refers to the Jacobi $\vartheta$-function in Jacobi-Erdélyi notation, while $\vartheta (s,v) \equiv \vartheta[s](v|\tau)$ denotes the helicity $\vartheta$-functions. The interested reader can find more details and identities involving these quantities in \cite{Kiritsis:1997hj}. The helicity partition function thus reads
\begin{eqaed}\label{eq:hel_pf}
     Z_W(\tau,\mu;v,\overline z=0)  = \frac{1}{\tau_2\eta^2\overline \eta^2}\frac{\Gamma_{{\text{SU}(2)}_k}(v)}{V(\mu)} \sum_s r(s) \, \frac{\vartheta(s,v)}{\eta} \, Z_{\text{int}}^{(s)}(\tau) \, .
\end{eqaed}
If we now consider the improvement of \cref{eq:flavpartfunc} to the modular-invariant expression\footnote{Consistency of the \ac{GSO} projection in the $\mathsf{W}_k$ theory dictates that $k$ be an even integer \cite{Antoniadis:1994sr,Kiritsis:1995dx}.}
\begin{eqaed}\label{eq:impr_hel_pf}
    \hat Z_W(\tau,\mu;v,\overline z) = e^{\pi \frac{\overline z^2}{2 \tau_2}}e^{\pi \frac{v^2(k+2)}{4 \tau_2}} \, Z_W(\tau,\mu;v,\overline z) \, , 
\end{eqaed}
we readily see that individual insertions in  \cref{eq:f2correlator} can be expressed as
\begin{eqaed}\label{eq:flavorhelicityinsertionscurved}
     &\frac{1}{(2\pi i)^2}\partial_v^2 \hat Z_W |_{v=0} = \left\langle ({\cal Q}_H+{\cal I})^2  - \frac{k+2}{8 \pi\tau_2} \right\rangle \ , \\
     &\frac{1}{(2\pi i)^2}\partial_{\overline z}^2 \hat Z_W |_{\overline z=0} = \left\langle \overline{\cal Q}^2  - \frac{1}{4\pi\tau_2} \right\rangle  ,
\end{eqaed}
while the overall two-point function takes the simple form
\begin{eqaed}\label{eq:f2correlator_withflavor}
    \langle F^2 \rangle = \left.8\pi^2 \tau_2^2 \, \partial^2_v \partial^2_{\overline z} \hat Z_W(\tau,\mu;v,\overline z) \right|_{v=\overline z =0} \, .
\end{eqaed}
In the next section, we will compare this expression for the regularized \ac{CFT} with other approaches to the computation of the gauge coupling, in particular the modular-covariant flat-space approach of \cite{Abel:2023hkk}. This will allow us to further simply the helicity operator insertions by mapping them to derivatives of Jacobi $\vartheta$-functions pertaining to external fermions. 

\subsubsection{Comparison with other methods}\label{sec:comparison_other_methods}
Having established the expression for the one-loop gauge coupling in terms of helicity and flavored partition functions, we now compare the background-field method developed in the previous sections to the flat-space approach of \cite{Abel:2023hkk}. The perspective adopted therein is to motivate a modular-invariant expression for $\Delta_g$ starting from the corresponding field-theoretic computation. In particular, denoting by $\cQ_R$ the charge running in the loop in the representation $R$ of the gauge algebra and by $S$ the associated spin, the strategy of \cite{Abel:2023hkk} is to make sense of an expression of the form
\begin{eqaed}\label{eq:thresh_field_th}
 \Delta_g = -2 \left\langle \left(S^2-\frac{1}{12}\right) \cQ^2_R \right\rangle  
\end{eqaed}
in string theory. The $S^2-1/12$ term comes from the helicity-dependent coefficient multiplying the field-theoretic loop. In the following we specialize to a single $\text{U}(1)$ factor and drop the subscript $R$. The charge bilinear insertion, taken to be in the right-moving sector, can be modular-completed applying the procedure outlined in \cite{Abel:2021tyt} to $\cQ^2 - (4\pi\tau_2)^{-1} $ and matching the insertion in \cref{eq:f2correlator}. The helicity insertion requires a bit more care. Naively, string states possess arbitrarily high spin, however they can be neatly organized in terms of \ac{CFT} data in three sectors: scalars, spinors and vectors. Each sector has ground states of spin $S=0,1/2,1$ respectively, and their descendants contribute to the partition function via linear combination of the $\vartheta$-functions for the external light-cone fermions. Letting $\Theta$ denote such a linear combination, it can be shown that their $q$-expansion reads \cite{Abel:2023hkk}
\begin{eqaed}\label{eq:q_exp_theta}
    \Theta \sim e^{\pi i \tau S^2}(1 + \dots) \, .
\end{eqaed}
The helicity insertion can thus be realized as suitable modular-covariant derivatives acting on the characters pertaining to the external (transverse) fermions. In particular, \cite{Abel:2023hkk} finds that the insertion takes the precise form
\begin{eqaed}\label{eq:cov_d_def}
    D_\tau \Theta(\tau) \equiv \left(\cQ_h - \frac{1}{12}\mathsf{E}_2\right)\Theta(\tau) \, , \quad \cQ_h \equiv \frac{1}{i\pi}\partial_\tau \, ,
\end{eqaed}
with $\mathsf{E}_2$ is the weight-two holomorphic Eisenstein function, which crucially is not modular invariant (see \cref{app:eisenstein_series_and_regularization}) since it compensates the non-modularity of the charge insertion. The full regularized one-loop gauge coupling computed in the flat-space theory is thus written as
\begin{eqaed}\label{eq:abelgauge}
    \Delta_g = -2\tau_2^2\left\langle \left( \cQ_h - \frac{1}{12}\mathsf{E}_2\right)\left( \overline \cQ - \frac{1}{4\pi \tau_2} \right) \cG\right\rangle \, ,
\end{eqaed}
where a modular-invariant regulator $\cG$ has been introduced in the amplitude in order to handle its logarithmic divergences. Importantly, an insertion of $\cQ_h$ is understood as acting only on the relevant combination of $\vartheta$-functions. The approach outlined above is manifestly modular covariant and emphasizes once more that the ``tadpole'' contribution appearing in the modular completion of the charge insertions are required by modular invariance of the amplitude. \\

\noindent We now turn to matching the flat-space insertions to the ones in the curved background. At first glance, the helicity insertion seems qualitatively different from the one found in \cref{eq:f2correlator}. To match the two, we first express the $D_\tau \Theta$ insertion in terms of the string-helicity partition function of the flat-space theory. This is defined as  \cite{Kiritsis:1997hj}
\begin{eqaed}\label{eq:flat_hel_pf}
    Z(v,\overline v) = \left\langle e^{2\pi i v \lambda_{\text{L}} -2\pi i \overline v \lambda_{\text{R}}}\right\rangle ,
\end{eqaed}
with $\lambda_{\text{L},\text{R}}$ the left- and right-moving helicities. In our case, as computed in \cite{Kiritsis:1997hj}, this means deforming $\eta^{-2} \to  \xi(v)\eta^{-2}$ for the two transverse bosons and $\vartheta(s) \to \vartheta(s,v)$ for the two transverse fermions. We consider $\overline v=0$ to match with computations in the previous section, thus arriving at
\begin{eqaed}\label{eq:flat_hel_pf_barv0}
   Z_{F} (\tau, v) =\frac{\xi(v)}{\tau_2 \eta^2 \overline \eta^2}\sum_s r(s) \, \frac{\vartheta(s,v)}{\eta} \, Z_{\text{int}}^{(s)}(\tau) \, ,
\end{eqaed}
which indeed is a flavored partition function with respect to helicity operators. The insertion $\lambda_{\text{L}}^2$ can be evaluated by using the heat equation satisfied by $\vartheta$-functions\footnote{In our normalizations, the differential equation reads $\frac{1}{(4\pi i)^2}\partial_v^2\vartheta = \frac{1}{i\pi}\partial_\tau \vartheta$.\label{foot:heat}} and the properties 
\begin{eqaed}\label{eq:xivi}
     & \xi(v) = \prod_{n=1}^{\infty} \frac{(1 - q^n)^2}{(1 - q^n e^{2\pi i v})(1 - q^n e^{-2\pi i v})} \, , \\
    & \xi(0)= 1 \ , \quad \partial_v\xi(0) =0\ , \quad \partial^2_v\xi (0) = -\frac{\pi^2}{3}(1-\mathsf{E}_2) \, , 
\end{eqaed}
obtaining an \emph{helicity supertrace} coefficient\footnote{Given the string-helicity partition function, one can more generally define $B_n$ helicity supertraces as
\begin{eqaed}\label{eq:b_n_coeff}
    B_n = \frac{1}{(2\pi i)^n} \partial^n_v Z_F |_{v=0} \, ,
\end{eqaed}
and similarly for the right-moving sector.}
\begin{eqaed}\label{eq:b2_hel_pf}
    B_2 \equiv \frac{1}{(2\pi i)^2}\partial_v^2 Z_F|_{v=0} = \left\langle \frac{1}{i\pi }\partial_\tau - \frac{1}{12}(\mathsf{E}_2-1)\right\rangle \, , 
\end{eqaed}
with a $\partial_\tau$ insertion once again understood as acting only on $\vartheta$, as expected. The modular-covariant insertion can be related to the approach of \cite{Abel:2023hkk} by rewriting 
\begin{eqaed}\label{eq:b2flat}
    \left\langle \frac{1}{i\pi}D_\tau\right\rangle = B_2 - \left\langle \frac{1}{12} \right\rangle ,
\end{eqaed}
which indeed looks like a stringy realization of the $(S^2-1/12)$ field-theoretic coefficient. It is worth checking the modular properties of the right-hand side, since the left-hand side transforms as a modular form of weight two, as can be ascertained by performing an S-transformation, 
\begin{eqaed}\label{eq:strasf}
    \tau \to \tau' = -\frac{1}{\tau} \ , \quad D_\tau \to \left( \frac{\partial \tau}{\partial \tau'}\right)D_\tau = \tau^2 D_\tau \, .
\end{eqaed}
One can then perform the following computations:
\begin{eqaed}\label{eq:xivi_manipul}
    &\frac{1}{(2\pi i)^2}\partial_{v'}^2\xi (v')|_{v'=0} = \tau^2\frac{1}{(2\pi i)^2}\partial_{v}^2\xi (0) - \frac{\tau^2}{12}+\frac{i\tau}{2\pi }   + \frac1{12} \, , \\
    &\frac{1}{(2\pi i)^2}\frac{\partial_{v'}^2\vartheta(s,v')|_{v'=0}}{\eta}= \tau^2\frac{1}{i\pi}\frac{\partial_{\tau}\vartheta(s)}{\eta}  - \frac{i\tau}{2\pi \eta} \, ,
\end{eqaed}
where once again we used the heat equation satisfied by the $\vartheta$-functions and the fact that $v$ transforms with weight $-1$ as $ v' = v\tau^{-1}$. Altogether, one finds
\begin{eqaed}\label{eq:b2prime}
    B_2' = \tau^2 B_2 + \frac{1-\tau^2}{12} \, ,
\end{eqaed}
so that by itself $B_2$ does not have a well-defined modular weight. However, $B_2 - \frac1{12}$ is a modular form of weight two, since 
\begin{eqaed}\label{eq:b2prime_trsf}
    B_2'-\frac{1}{12} = \tau^2 \left(B_2 - \frac{1}{12}\right) .
\end{eqaed}
Next, we turn to the background-field computations in \cref{eq:flavorhelicityinsertionscurved,eq:f2correlator_withflavor}. The affine deformed ${\text{SU}(2)}_k$ characters $\chi_\ell(\tau,v) =\sum_{n\in \Lambda_{\ell}} \chi_\ell^{(n)}(\tau, v)$ defined in \cref{eq:affinedeformedsu(2)} satisfy 
\begin{eqaed}\label{eq:partial_v_chi}
     &\partial_v \chi_\ell^{(n)}(0,\tau) = -\frac12 \frac{\vartheta_1''(0)}{\vartheta'(0)^2}\chi_\ell(0)= 0 \, , \\ &\partial_v^2\chi_\ell^{(n)} (0,\tau) = \left[-\frac{4}{3}(2+k)^2 n^2 \pi^2 -\frac{1}{3}\frac{\vartheta_1'''}{\vartheta_1'}\right]\chi_\ell(0) \, .
\end{eqaed}
Notice that the second term can be written as $-\frac{\pi^2}3 \mathsf{E}_2$ using the results of \cite[Appendix F]{Kiritsis:1997hj}, while the first term can be turned into a derivative acting on $\Gamma_{{\text{SU}(2)}_k}$, since in the left-moving sector 
\begin{eqaed}\label{eq:partial_tau_gammasu2}
      \partial_\tau \left[(\sqrt{\tau_2} \eta \overline \eta)^3 \chi_\ell^{(n)}\right]&=\left(\partial_\tau \tau_2^{3/2}\right)\chi_\ell^{(n)} \eta^3\overline \eta^3+(\sqrt{\tau_2}  \overline \eta)^3 (k+2)n \left(\partial_\tau q^{(k+2)n^2} \right)\\&=i\left[-\frac{3}{4\tau_2}+ 2\pi (k+2)n^2\right]  (\sqrt{\tau_2} \eta \overline \eta)^3 \chi_\ell^{(n)} \, ,
\end{eqaed}
so that
\begin{eqaed}\label{eq:partial_v_gammasu2}
     \partial_v^2\Gamma_{{\text{SU}(2)}_k} (v)|_{v=0} = \left[-\frac{2\pi}{3i}(2+k) \partial_\tau + \frac{\pi}{2\tau_2}(2+k) +\frac{\pi^2}{3}\mathsf{E}_2\right]\Gamma_{{\text{SU}(2)}_k}\, .
\end{eqaed}
Using this result and the heat equation satisfied by the $\vartheta$-functions, one can derive the curved-background analogue of $B_2$ according to 
\begin{eqaed}\label{eq:b2w}
    B_2^{W} \equiv \frac{1}{(2\pi i)^2}\partial_v^2 \hat Z_W |_{v=0} = \left\langle \frac{1}{i\pi} D_\tau\right\rangle  + A(\tau) \partial_\tau \Gamma_{{\text{SU}(2)}_k} \, ,
\end{eqaed}
where $A(\tau)$ is a $\tau$-dependent coefficient and once again $D_\tau$ is understood to act only on the $\vartheta$-functions pertaining to the external fermions. Since \cref{eq:b2w} ought to match the modular properties of $D_\tau$, it is also straightforward to confirm that $B_2^W$ is a modular form of weight two, as previously mentioned. The last term in \cref{eq:b2w} vanishes in the flat-space limit, since $\Gamma_{{\text{SU}(2)}_k}$ approaches the $\tau$-independent constant $V(\mu)$\footnote{Note that the limit is uniform in $\tau$ for both the function and its derivative, as argued in \cite{Florakis:2016aoi}.}  \cite{Florakis:2016aoi}. This establishes that the helicity insertions in the flat-space theory, which are realized by the modular-covariant derivative $D_\tau$ acting on a linear combination of $\vartheta$-functions, precisely match the $\cN$ insertion in \cref{eq:flavorhelicityinsertionscurved} in the $k\to \infty$ limit. The $\text{U}(1)$ charge insertion automatically match as well. \\

\paragraph{Gauge couplings in the background-field method.} We have confirmed that the operator insertions agree between the flat-space approach of \cite{Abel:2023hkk} and the background-field method of \cite{Kiritsis:1995dx} in the flat-space limit. This has been done by building the dictionary in terms of helicity and flavor partition functions. In the former approach, the $\cG$ function deforming the one-loop amplitude in \cref{eq:abelgauge} regularizes the logarithmic divergences of the gauge couplings, similarly to how the mass gap $\mu$ regularizes the theory in the latter approach. To simplify the ensuing expressions, and since the difference vanishes in the flat-space limit, we will replace the $\cN$ helicity insertion in the curved-space theory by $D_\tau$ acting on the $\vartheta$-function of the external fermions\footnote{More concretely, using the Legendre representation of $\Gamma_{{\text{SU}(2)}_k}$ \cite{Florakis:2016aoi}, one can bound the modular integral of the difference schematically by
\begin{eqaed}\label{eq:bound_estim}
    \int_{a}^{\infty} \td\tau_2 \, \tau_2^{b}\, k^{c} e^{-d k \tau_2} \, , \quad {a,b,c,d} \text{ positive constants} \, ,
\end{eqaed}
which vanishes exponentially as $k \to \infty$.}. This leads to a refined expression for the gauge coupling in terms of the $\overline \cQ$-flavored partition function
\begin{eqaed}\label{eq:f2correlator_refined}
      \langle F^2 \rangle = \left. \frac{8\pi^2 \tau_2}{\eta^2\overline \eta^2}\frac{\Gamma_{{\text{SU}(2)}_k}}{V(\mu)} \sum_s r(s) \, \frac{D_\tau\vartheta(s,v)}{\eta} \, \partial_{\overline z}^2 Z_{\text{int}}^{(s)}(\tau;\overline z)\right|_{\overline z =0} \, , 
\end{eqaed}
where moreover we used the fact that the $\overline z$-deformation leaves the external sector unaffected, owing to the absence of $\text{U}(1)$ currents. Similar considerations can be made for the higher-derivative corrections in \eqref{eq:higherdercorr}. Specifically, given the expansion in terms of background fields $Z=\sum_{n,m}Z_W^{n,m}R^nF^m$, one is able to obtain the Wilson coefficient of interest by taking a suitable number of derivatives. The interest reader can check explicit expressions in terms of charge and helicity insertions in \cite{Kiritsis:1995dx}. The upshot of the analysis, adapted using the formalism of helicity and flavor partition functions, is that a generic Wilson coefficient can always be expressed as a suitable sum of the type
\begin{eqaed}\label{eq:fmrn_bkg_field}
    \langle R^n F^m\rangle = \tau_2^{n+m}\sum_{p\le n \, , \, q\le m}\partial_{\overline z}^q \partial_v^p \hat Z_W(\tau,\mu;v,\overline z)\bigg|_{v=\overline z=0}
\end{eqaed}
to be integrated over the fundamental domain. To conclude this section, we remark that a background-field method in flat spacetime might be also attainable, at the cost of introducing \ac{IR} divergences which need to be dealt with \cite{Petropoulos:1996xz}. In the following section, we will apply this machinery to work out the form of the one-loop gauge coupling in an explicit example. Note that in the above discussion we left the sum over $s$ generic; however, in order to get a coefficient compatible with the spacetime parity property of $F^2$, the sum over $s$ should be restricted e.g. to the even spin structures. We will see this in detail in the next example. In the subsequent section, instead, we will use the above expression to make statements about the universal scaling of gauge coupling in species limits. 

\subsection{Example: heterotic gauge thresholds}\label{sec:example_heterotic_gauge_thresholds}
The aim of this section is to illustrate in a concrete example the computation of the one-loop gauge coupling via the background-field method and the formalism of flavored partition functions. We take the construction to be the classic heterotic orbifold $T^4/\bZ_2\times T^2$ discussed e.g. in \cite{Kiritsis:1997hj}. The starting point is the heterotic $\text{E}_8 \times \text{E}_8$ theory compactified on a $T^6$, yielding a four-dimensional theory with $N=4$ supersymmetry. Its partition function is given by
\begin{eqaed}\label{eq:het_z_n4}
    Z^{\text{het}}_{N=4} = \frac{\Gamma_{6,6} \, \overline{\Gamma}_{\text{E}_8 }^2}{\tau_2 \, \eta^{8} \, \overline{\eta}^{24}} \frac{1}{2}  \sum_{a,b=0}^1 (-1)^{a+b+ab} \, \frac{\vartheta^4[^a_b]}{\eta^4}   \, ,
\end{eqaed}
where the $\vartheta$-functions for the choice of torus spin structure $[^a_b]$ are given in Riemann notation \cite[Appendix A]{Kiritsis:1997hj}. The $\Gamma$ factors denote the lattice partition functions 
\begin{eqaed}\label{eq:latticesum}
    \Gamma_{n,n} = \sum_{p_{\text{L}},p_{\text{R}}}q^{\frac{p_{\text{L}}^2}{2}}\overline q^{\frac{p_{\text{R}}^2}{2}} \, , \quad  \overline \Gamma_{\text{E}_8}= \sum_{\gamma,\delta=0}^1 {\overline \vartheta^8[^\gamma_\delta]} \, ,
\end{eqaed}
where $p_{\text{L},\text{R}}$ are summed over the torus momentum lattice. The $\text{E}_8$ subscripts denotes that one should sum over the $\text{E}_8$ even, self-dual lattice, for which we directly give the expression in terms of right-moving fermionic characters. In order to break supersymmetry to $N=2$, we choose to quotient by a geometric $\bZ_2$-action on the coordinates $x^i$ of the $T^4 \subset T^6$ as $x^i \to -x^i$. It amounts to the replacement 
\begin{eqaed}\label{eq:repl_torus}
 \frac{\Gamma_{6,6}}{\eta^4 \overline \eta^4} \to \Gamma_{2,2}  Z_{4,4}[^h_g] \, , \quad   Z_{4,4}[^h_g]= \begin{cases}
      Z_{4,4}[^0_0] = \frac{\Gamma_{4,4}}{\eta^4 \overline \eta^4} \, , \quad \\ Z_{4,4}[^h_g] = 2^4\frac{\eta^2 \overline \eta^2}{\vartheta^2[^{1-h}_{1-g}]\overline \vartheta^2[^{1-h}_{1-g}]}\, , \ \ (h,g)\ne(0,0) 
 \end{cases} ,
\end{eqaed}
where $h,g=0,1$ runs on the orbifold blocks of the $(4,4)$ torus lattice, and, because of supersymmetry, a corresponding twist of 4 out of 8 of the left-moving fermions
\begin{eqaed}\label{eq:repl_ferm}
     \frac{\vartheta^2[^a_b]}{\eta^2} \to \ \frac{\vartheta[^{a+h}_{b+g}]\vartheta[^{a-h}_{b-g}]}{\eta^2} \, .
\end{eqaed}
The geometric action should be in general accompanied by an action on one of the $\text{E}_8$ factors to obtain a modular-invariant theory. We choose to adopt the so-called standard embedding which breaks $\text{E}_8$ to ${\text{E}}_7 \times \text{SU}(2)$. In the right-moving sector we thus substitute
\begin{eqaed}\label{eq:repl_e8}
    \frac{\overline \vartheta^2[^\gamma_\delta]}{\overline \eta^2} \to  \frac{\overline \vartheta[^{\gamma+h}_{\delta+g}]\overline \vartheta[^{\gamma-h}_{\delta-g}]}{\overline \eta^2} \, .
\end{eqaed}
Summing over all sectors, we arrive at the partition function for the $N=2$ theory,
\begin{eqaed}\label{eq:hetpartfunc}
Z^{\text{het}}_{N=2} = \frac{1}{2}\sum_{h,g=0}^1\frac{\Gamma_{2,2} \, \overline \Gamma_{\text{E}_8} \, Z_{4,4}[^h_g]}{\tau_2 \eta^4 \overline \eta^{12}} \frac{1}{2}\sum_{\delta,\gamma=0}^1\frac{\overline \vartheta[^\gamma_\delta]^6\overline \vartheta[^{\gamma+h}_{\delta+g}]\overline \vartheta[^{\gamma-h}_{\delta-g}]}{\overline \eta^8} \times \\
\times \frac{1}{2}\sum_{a,b=0}^1 (-1)^{a+b+ab}\frac{\vartheta^2[^a_b]\vartheta[^{a+h}_{b+g}]\vartheta[^{a-h}_{b-g}]}{\eta^4} \, .
\end{eqaed}
To make contact with the notation introduced in the preceding general discussion, notice that in this case the index $s$ appearing e.g. in \cref{eq:Zwexternalexplicit} is associated to spin structure as well as orbifold blocks. \\

\paragraph{Expressing the partition function in terms of characters.} The $T^4/\mathbb{Z}_2 \times T^2$ partition function for ${\text{E}}_8 \times {\text{E}}_8$ with the standard embedding in terms of characters is given by
\begin{eqaed}\label{eq:torus_partition_function_characters}
        Z^{\text{het}}_{N=2} &= \frac{1 }{(\sqrt{\tau_2} \eta \overline \eta )^2} \sum_{m,w \in \mathbb{Z}^2}\chi^{(c=2)}_{m,w} \overline \chi_{m,-w}^{(c=2)}\Bigg \{  \sum_{m>0,w \in \mathbb{Z}^4} \Big (\chi^{(+,c=4)}_{m,w} \overline \chi^{(+,c=4)}_{m,-w}
        \\
        & + \chi^{(-,c=4)}_{m,w} \overline \chi^{(-,c=4)}_{m,-w} \Big) \Big ( \Psi_V + \Psi_{2H} \Big )  \big (\overline \xi_0 \overline \rho_{\bf 133} + \overline \xi_{\frac12} \overline \rho_{\bf 56} \big ) \overline \chi_{\text{E}_8}
        \\
        & + \Big ( |\chi_{e}^{(c=4)} |^2 + |\chi_{o}^{(c=4)}|^2\Big ) \Big ( \Psi_V  \overline \xi_0 \overline \rho_{\bf 133} \overline \chi_{\text{E}_8} + \Psi_{2H} \overline \xi_{\frac12} \overline \rho_{\bf 56} \overline \chi_{\text{E}_8}\Big )
        \\
        & + \Big ( \chi_{e}^{(c=4)} \overline \chi_{o}^{(c=4)} + \chi_{o}^{(c=4)} \overline \chi_{e}^{(c=4)}\Big ) \Big ( \Psi_{2H} \overline \xi_0 \overline \rho_{\bf 133} \overline \chi_{\text{E}_8}+ \Psi_{V}  \overline \xi_{\frac12} \overline \rho_{\bf 56} \overline \chi_{\text{E}_8} \Big )
        \\
        & + \sum_{a=1}^{16} \bigg [ \Big ( |\chi_{\sigma_a}^{(c=4)} |^2 + |\chi_{\tilde{\sigma}_a}^{(c=4)}|^2\Big ) \Big ( \Psi_H  \overline \xi_0 \overline \rho_{\bf 56} \overline \chi_{\text{E}_8} + \Psi_{G}  \overline \xi_{\frac12} \overline \rho_{\bf 133} \overline \chi_{\text{E}_8} \Big )
        \\
        & + \Big ( \chi_{\sigma_a}^{(c=4)} \overline \chi_{\tilde{\sigma}_a}^{(c=4)} + \chi_{\tilde{\sigma}_a}^{(c=4)} \overline \chi_{\sigma_a}^{(c=4)}\Big ) \Big ( \Psi_{G} \overline \xi_0 \overline \rho_{\bf 56} \overline \chi_{\text{E}_8} + \Psi_{H} \overline \xi_{\frac12} \overline \rho_{\bf 133} \overline \chi_{\text{E}_8} \Big ) \bigg ] \Bigg \} \, ,
\end{eqaed}
where $m >0$ means that all the components of $m$ are taken to be positive, defining the orbifold-invariant combination of $\text{U}(1)$ affine characters \cite{DiFrancesco:1997nk}
\begin{eqaed}\label{eq:U1_affine_characters}
    \chi_{m,w}^{(\pm,c=d)}=\frac{\chi_{m,w}^{(c=d)} \pm \chi_{-m,-w}^{(c=d)}}{2} 
\end{eqaed}
with $m,w \neq 0$. 
For $m=w=0$ the character does not correspond to an irreducible representation of the Virasoro algebra, and instead decomposes according to \cite{Dijkgraaf:1989hb}
\begin{eqaed}\label{eq:00_U1_character}
    \chi_{0,0}^{(c=d)}= \Big (\sum_{\ell=0}^{\infty} \chi_\ell^{\text{Vir}} \Big )^{\! d} \, , \quad \chi_\ell^{\text{Vir}} =  \frac{q^{\ell^2}-q^{(\ell+1)^2}}{\eta} \, .
\end{eqaed}
Under the orbifold action $\chi_\ell^{\text{Vir}} \to (-1)^{\ell}\chi_\ell^{\text{Vir}}$, and hence they rearrange themselves into
\begin{eqaed}\label{eq:orbifold_characters}
    &\chi_{e}^{(c=d)}= \sum_{n \in \mathbb{N}^d}^{\infty} \frac{1+(-1)^{\sum_i n_i}}{2}\prod_{i=1}^d \chi_{n_i}^{\text{Vir}} 
    \\
    & \qquad \ \, =\frac12 \bigg ( \sum_{n=0}^{\infty} \frac{q^{n^2}-q^{(n+1)^2}}{\eta}\bigg )^{\! d} + \frac12 \bigg ( \sum_{n=0}^{\infty} (-1)^n \frac{q^{n^2}-q^{(n+1)^2}}{\eta}\bigg )^{\! d} \, ,
    \\
    &\chi_{o}^{(c=d)}=\sum_{n \in \mathbb{N}^d}^{\infty} \frac{1-(-1)^{\sum_i n_i}}{2}\prod_{i=1}^d \chi_{n_i}^{\text{Vir}}
    \\
    & \qquad \ \, =\frac12 \bigg ( \sum_{n=0}^{\infty} \frac{q^{n^2}-q^{(n+1)^2}}{\eta}\bigg )^{\! d} - \frac12 \bigg ( \sum_{n=0}^{\infty} (-1)^n \frac{q^{n^2}-q^{(n+1)^2}}{\eta}\bigg )^{\! d} \, .
\end{eqaed}
One can directly check that these expressions correspond to combinations of the blocks entering the partition function associated with the compact boson in the untwisted sector. Namely, these characters, in terms of the conformal blocks contained in \cref{eq:het_z_n4}, correspond to 
\begin{equation}
\begin{aligned}
    & \chi_e^{(c=4)}= \frac{1}{2} \bigg ( \frac{1}{\eta^4}+\frac{4 \eta^2}{\vartheta^2[^{1}_{0}]} \bigg ) \, ,
    \\
    &\chi_o^{(c=4)}=\frac{1}{2} \bigg ( \frac{1}{\eta^4}-\frac{4 \eta^2}{\vartheta^2[^{1}_{0}]} \bigg ) \, .
    \end{aligned}
\end{equation}
In the twisted sector we have introduced the characters associated with the twist field $\sigma_a$ and the excited twist field $\tilde{\sigma}_a$ for each fixed point $a=1, \ldots ,16$. These can be directly obtained from the combinations of the bosonic piece associated with the compact bosons. Namely, in terms of the conformal blocks contained in \cref{eq:het_z_n4}, we have
\begin{equation}
\begin{aligned}
    & \chi_{\sigma_a}^{(c=4)}= \frac{1}{2} \bigg (\frac{\eta^2}{\vartheta^2[^{0}_{1}]}+\frac{ \eta^2}{\vartheta^2[^{0}_{0}]} \bigg ) \, ,
    \\
    &\chi_{\tilde{\sigma}_a}^{(c=4)}=\frac{1}{2} \bigg ( \frac{\eta^2}{\vartheta^2[^{0}_{1}]}-\frac{ \eta^2}{\vartheta^2[^{0}_{0}]}  \bigg ) \, .
    \end{aligned}
\end{equation}
Moreover, we have introduced the ${\text{E}}_{8,1}$ characters
\begin{eqaed}\label{eq:E8_1_characters}
    \chi_{\text{E}_8}= \frac{\Gamma_{\text{E}_8}}{\eta^8} \, ,
\end{eqaed}
${\text{SU}}(2)_1$ characters (following \cref{eq:affinedeformedsu(2)})
\begin{eqaed}\label{eq:SU2_affine_characters}
    \xi_j=\frac{1}{\eta} \sum_n q^{(n+j)^2}\equiv \frac{\Theta_{2j}^{(1)}}{\eta} \, , 
\end{eqaed}
and those of ${\text{E}}_{7,1}$ associated with the adjoint and fundamental representations,
\begin{eqaed}\label{eq:hat-E_characters}
        &\rho_{\bf 133}= \frac{1}{2 \eta^7} \Big ( \vartheta [\substack{1 \\ 0}](0 |2\tau) \, \vartheta^6 [\substack{1 \\ 0}](0|\tau) + \vartheta [\substack{0 \\ 0}](0|2\tau) \, \vartheta^6 [\substack{0 \\ 1}](0|\tau)  + \vartheta [\substack{0 \\ 0}](0|2\tau)\, \vartheta^6 [\substack{0 \\ 0}](0|\tau)  \Big ) \, ,
        \\
        &\rho_{\bf 56}= \frac{1}{2 \eta^7} \Big ( \vartheta [\substack{0 \\ 0}](0|2\tau) \, \vartheta^6 [\substack{1 \\ 0}](0|\tau) + \vartheta [\substack{1 \\ 0}](0|2\tau) \, \vartheta^6 [\substack{0 \\ 0}](0|\tau)  - \vartheta [\substack{1 \\ 0}](0|2\tau) \, \vartheta^6 [\substack{0 \\ 1}](0|\tau)  \Big ) \, .
\end{eqaed}
Finally, we have introduced the supersymmetric characters \cite{Angelantonj:2002ct}
\begin{eqaed}\label{eq:RNSSUSYZ2}
    &\Psi_V= O_2 V_2 O_4 + V_2 O_2 O_4 - S_2 C_2 C_4 - C_2 S_2 C_4 \, ,
    \\
    &\Psi_{2H}= O_2 O_2 V_4 +V_2 V_2 V_4 -S_2 S_2 S_4 - C_2 C_2 S_4 \, ,
    \\
    &\Psi_{H}= O_2 O_2 C_4 +V_2 V_2 C_4 -S_2 S_2 O_4 - C_2 C_2 O_4 \, ,
    \\
    &\Psi_{G}=  O_2 V_2 S_4 + V_2 O_2 S_4 - S_2 C_2 V_4 - C_2 S_2 V_4 \, .
\end{eqaed}
All in all, the structure of the partition function is thus
\begin{eqaed}\label{eq:structure_partition_function}
    Z^{\text{het}}_{N=2} = \sum_{a, b=0,1} \frac{(-1)^{ a +  b + ab } }{2\eta} \, \vartheta [ \substack{ a \\ b}] \, Z_{\text{int}}[ \substack{ a \\ b}] \, ,
\end{eqaed}
where
\begin{eqaed}\label{eq:internal_het_part_characters_1}
    \frac{Z_{\text{int}} [ \substack{0 \\ 0}]}{O_2 + V_2} = \frac{\Gamma_{2,2}}{\eta^2 \overline \eta^2} & \sum_{m>0, w \in \mathbb{Z}^4} \left( \chi_{m,w}^{(+,c=4)} \overline  \chi_{m,-w}^{(+,c=d)} +  \chi_{m,w}^{(-,c=4)} \overline  \chi_{m,-w}^{(-,c=d)} \right) (O_4 -V_4)\overline \chi_{\text{E}_8} \overline \chi_{\text{E}_8} 
    \\
    &+ ( |\chi_e^{(c=4)} |^2 + | \chi_o^{(c=4)}|^2   ) ( O_4 \overline \rho_{\bf 133} \overline \xi_0 \overline \chi_{\text{E}_8}+ V_4 \overline \rho_{\bf 56} \overline \xi_{\frac12} \overline \chi_{\text{E}_8})   
    \\
    &+  ( \chi_e^{(c=4)} \overline \chi_o^{(c=4)} + \chi_o^{(c=4)} \overline \chi_e^{(c=4)}   ) ( O_4 \overline \rho_{\bf 56} \overline \xi_\frac12 \overline \chi_{\text{E}_8}+ V_4 \overline \rho_{\bf 133} \overline \xi_{0} \overline \chi_{\text{E}_8})   
    \\
    &+ \sum_a   ( |\chi_{\sigma_a}^{(c=4)} |^2 + | \chi_{\tilde{\sigma}_a}^{(c=4)}|^2   ) ( C_4 \overline \rho_{\bf 56} \overline \xi_0 \overline \chi_{\text{E}_8}+ S_4 \overline \rho_{\bf 133} \overline \xi_{\frac12} \overline \chi_{\text{E}_8})   
    \\
    &+  ( \chi_{\sigma_a}^{(c=4)} \overline \chi_{\tilde{\sigma}_a}^{(c=4)} + \chi_{\tilde{\sigma}_a}^{(c=4)} \overline \chi_{\sigma_a}^{(c=4)}   ) ( C_4 \overline \rho_{\bf 133} \overline \xi_\frac12 \overline \chi_{\text{E}_8}+ S_4 \overline \rho_{\bf 56} \overline \xi_{0} \overline \chi_{\text{E}_8}) \, ,
\end{eqaed}
while $Z_{\text{int}} [ \substack{0 \\ 1}]$ is given by
\begin{eqaed}\label{eq:internal_het_part_characters_2}
    \frac{Z_{\text{int}} [ \substack{0 \\ 1}]}{O_2-V_2} = \frac{\Gamma_{2,2}}{\eta^2 \overline \eta^2} & \sum_{m>0, w \in \mathbb{Z}^4} \left( \chi_{m,w}^{(+,c=4)} \overline  \chi_{m,-w}^{(+,c=d)} +  \chi_{m,w}^{(-,c=4)} \overline  \chi_{m,-w}^{(-,c=d)} \right) (O_4 -V_4) \overline \chi_{\text{E}_8} \overline \chi_{\text{E}_8} 
    \\
    &+ ( |\chi_e^{(c=4)} |^2 + | \chi_o^{(c=4)}|^2   ) ( O_4 \overline \rho_{\bf 133} \overline \xi_0 \overline \chi_{\text{E}_8}- V_4 \overline \rho_{\bf 56} \overline \xi_{\frac12} \overline \chi_{\text{E}_8})   
    \\
    &+  ( \chi_e^{(c=4)} \overline \chi_o^{(c=4)} + \chi_o^{(c=4)} \overline \chi_e^{(c=4)}   ) ( O_4 \overline \rho_{\bf 56} \overline \xi_\frac12 \overline \chi_{\text{E}_8}- V_4 \overline \rho_{\bf 133} \overline \xi_{0} \overline \chi_{\text{E}_8})   
    \\
    &+ \sum_a   ( |\chi_{\sigma_a}^{(c=4)} |^2 + | \chi_{\tilde{\sigma}_a}^{(c=4)}|^2   ) ( C_4 \overline \rho_{\bf 56} \overline \xi_0 \overline \chi_{\text{E}_8}- S_4 \overline \rho_{\bf 133} \overline \xi_{\frac12} \overline \chi_{\text{E}_8})   
    \\
    &+  ( \chi_{\sigma_a}^{(c=4)} \overline \chi_{\tilde{\sigma}_a}^{(c=4)} + \chi_{\tilde{\sigma}_a}^{(c=4)} \overline \chi_{\sigma_a}^{(c=4)}   ) ( C_4 \overline \rho_{\bf 133} \overline \xi_\frac12 \overline \chi_{\text{E}_8}- S_4 \overline \rho_{\bf 56} \overline \xi_{0} \overline \chi_{\text{E}_8}) \, .
\end{eqaed}
Moreover, the terms multiplying the spacetime fermions read
\begin{eqaed}\label{eq:internal_het_part_characters_3}
    \frac{Z_{\text{int}} [ \substack{1 \\ 0}]}{S_2+C_2} = \frac{\Gamma_{2,2}}{\eta^2 \overline \eta^2}& \sum_{m>0, w \in \mathbb{Z}^4} \left( \chi_{m,w}^{(+,c=4)} \overline  \chi_{m,-w}^{(+,c=d)} +  \chi_{m,w}^{(-,c=4)} \overline  \chi_{m,-w}^{(-,c=d)} \right) (C_4 +S_4) \overline \chi_{\text{E}_8} \overline \chi_{\text{E}_8} 
    \\
    &+ ( |\chi_e^{(c=4)} |^2 + | \chi_o^{(c=4)}|^2   ) ( C_4 \overline \rho_{\bf 133} \overline \xi_0 \overline \chi_{\text{E}_8}+  S_4 \overline \rho_{\bf 56} \overline \xi_{\frac12} \overline \chi_{\text{E}_8})   
    \\
    &+  ( \chi_e^{(c=4)} \overline \chi_o^{(c=4)} + \chi_o^{(c=4)} \overline \chi_e^{(c=4)}   ) ( C_4 \overline \rho_{\bf 56} \overline \xi_\frac12 \overline \chi_{\text{E}_8} + S_4 \overline \rho_{\bf 133} \overline \xi_{0} \overline \chi_{\text{E}_8})   
    \\
    &+ \sum_a   ( |\chi_{\sigma_a}^{(c=4)} |^2 + | \chi_{\tilde{\sigma}_a}^{(c=4)}|^2   ) ( O_4 \overline \rho_{\bf 56} \overline \xi_0 \overline \chi_{\text{E}_8} + V_4 \overline \rho_{\bf 133} \overline \xi_{\frac12} \overline \chi_{\text{E}_8})   
    \\
    &+  ( \chi_{\sigma_a}^{(c=4)} \overline \chi_{\tilde{\sigma}_a}^{(c=4)} + \chi_{\tilde{\sigma}_a}^{(c=4)} \overline \chi_{\sigma_a}^{(c=4)}   ) ( O_4 \overline \rho_{\bf 133} \overline \xi_\frac12 \overline \chi_{\text{E}_8} + V_4 \overline \rho_{\bf 56} \overline \xi_{0} \overline \chi_{\text{E}_8}) \, ,
\end{eqaed}
and 
\begin{eqaed}\label{eq:internal_het_part_characters_4}
    \frac{iZ_{\text{int}} [ \substack{1 \\ 1}]}{S_2-C_2} =\frac{\Gamma_{2,2}}{\eta^2 \overline \eta^2} & \sum_{m>0, w \in \mathbb{Z}^4} \left( \chi_{m,w}^{(+,c=4)} \overline  \chi_{m,-w}^{(+,c=d)} +  \chi_{m,w}^{(-,c=4)} \overline  \chi_{m,-w}^{(-,c=d)} \right) (C_4 -S_4) \overline \chi_{\text{E}_8}  \overline \chi_{\text{E}_8} 
    \\
    &+ ( |\chi_e^{(c=4)} |^2 + | \chi_o^{(c=4)}|^2   ) ( C_4 \overline \rho_{\bf 133} \overline \xi_0 \overline \chi_{\text{E}_8}-  S_4 \overline \rho_{\bf 56} \overline \xi_{\frac12} \overline \chi_{\text{E}_8})   
    \\
    &+  ( \chi_e^{(c=4)} \overline \chi_o^{(c=4)} + \chi_o^{(c=4)} \overline \chi_e^{(c=4)}   ) ( C_4 \overline \rho_{\bf 56} \overline \xi_\frac12 \overline \chi_{\text{E}_8} - S_4 \overline \rho_{\bf 133} \overline \xi_{0} \overline \chi_{\text{E}_8})   
    \\
    &+ \sum_a   ( |\chi_{\sigma_a}^{(c=4)} |^2 + | \chi_{\tilde{\sigma}_a}^{(c=4)}|^2   ) ( O_4 \overline \rho_{\bf 56} \overline \xi_0 \overline \chi_{\text{E}_8} - V_4 \overline \rho_{\bf 133} \overline \xi_{\frac12} \overline \chi_{\text{E}_8})   
    \\
    &+  ( \chi_{\sigma_a}^{(c=4)} \overline \chi_{\tilde{\sigma}_a}^{(c=4)} + \chi_{\tilde{\sigma}_a}^{(c=4)} \overline \chi_{\sigma_a}^{(c=4)}   ) ( O_4 \overline \rho_{\bf 133} \overline \xi_\frac12 \overline \chi_{\text{E}_8} - V_4 \overline \rho_{\bf 56} \overline \xi_{0} \overline \chi_{\text{E}_8}) \, .
\end{eqaed} \\

\noindent \paragraph{Regularization.} We can now begin the calculation of the one-loop gauge coupling. We first perform the regularization introduced in the previous section, by replacing the external sector by the $\mathsf{W}_k$ theory and writing its partition function as
\begin{eqaed}\label{eq:het_z_regul}
    Z_W^{\text{het}} = \frac{1}{\tau_2 \eta^2 \overline \eta^2}\frac{\Gamma_{{\text{SU}(2)}_k}}{V(\mu)}\sum_{a,b=0}^1 (-1)^{a+b+ab} \, \frac{\vartheta[^a_b]}{\eta} \, Z_{\text{int}}[^a_b] \, ,
\end{eqaed}
where the sum over orbifold blocks is moved inside $Z_{\text{int}}[^a_b]$. \\

\noindent \paragraph{Flavored partition function.} We can now choose which $\text{U}(1)$ gauge factor we should compute the gauge coupling of by turning on a chemical potential associated with its charge, as illustrated in \cref{sec:background_field_method}. In particular, we can construct the flavored partition function for each Cartan generator of $\text{E}_8$. At the level of characters, this means deforming each of the $\vartheta$-functions in $\overline \Gamma_{\text{E}_8}$ as
\begin{eqaed}\label{eq:het_z_flavor}
\overline \Gamma_{\text{E}_8}(\overline z_i) = \frac12 \sum_{\gamma,\delta=0}^1 \prod_{i=1}^8 \, \overline \vartheta[^\gamma_\delta](\overline z_i) \, .
\end{eqaed}
Let us consider only a $\text{U}(1)$ subgroup given by one deformed $\vartheta$-function. One can improve the lattice sum as in \cref{eq:improvedfpf} by adding the prefactor
\begin{eqaed}\label{eq:e8_flavor}
   \hat{ \overline \Gamma}_{\text{E}_8}(\overline z) = \frac12 \, e^{\frac{\overline z^2}{2\tau_2}}\sum_{\gamma,\delta=0}^1 \overline \vartheta[^\gamma_\delta](\overline z)\overline \vartheta^7[^\gamma_\delta] \, ,
\end{eqaed}
so that taking two derivatives and setting $\overline z =0$ one obtains 
\begin{eqaed}\label{eq:e8_flavor_d}
    \frac{1}{(2\pi i)^2}\partial_{\overline z}^2  \hat{\overline \Gamma}_{\text{E}_8} |_{\overline z =0}= \left[\frac{1}{(2\pi i)^2}\partial_{\overline z}^2-\frac{1}{4\pi \tau_2}\right]{\overline \Gamma}_{\text{E}_8} \, ,
\end{eqaed}
which is indeed the appropriate charge insertion of \cref{eq:flavorhelicityinsertionscurved}, upon using the heat equation and identifying $\partial_\tau \vartheta$ with the charge insertion $\overline \cQ$ as in \cite{Kiritsis:1995dx,Kiritsis:1997hj}. \\ 

\noindent \paragraph{Computation of the gauge coupling.} As shown in the preceding section, trading the helicity insertion in \cref{eq:f2correlator} with modular-covariant derivatives of $\vartheta$-functions is only sensitive to the external sector and is thus model-independent. This allows us to begin immediately from the expression in \cref{eq:f2correlator_refined} to compute one-loop gauge couplings. We will see that the resulting expression matches the ones found in the literature, particularly those in \cite{Kiritsis:1997hj}. To this end, we rewrite
\begin{eqaed}\label{eq:f2_het}
      \langle F^2 \rangle_{\text{het}} = \left. \frac{8\pi^2 \tau_2}{\eta^2\overline \eta^2}\frac{\Gamma_{{\text{SU}(2)}_k}}{V(\mu)} \sum_{a,b=0}^1  (-1)^{a+b+ab} \, \frac{D_\tau\vartheta[^a_b]}{\eta} \, \partial_{\overline z}^2 Z_{\text{int}}[^a_b](\tau;\overline z)\right|_{\overline z =0}  \, , 
\end{eqaed}
with the second derivative reading explicitly
\begin{eqaed}\label{eq:het_p2z_int}
    \left.\partial_{\overline z}^2 Z_{\text{int}}[^a_b](\tau;\overline z)\right|_{\overline z =0} = \frac{1}{2}\sum_{h,g=0}^1\frac{\left.\partial_{\overline z}^2\overline \Gamma_{\text{E}_8}(\overline z)\right|_{\overline z = 0} \, \Gamma_{2,2}\, Z_{4,4}[^h_g]}{ \eta^2 \overline \eta^{10}} \frac{1}{2}\sum_{\delta,\gamma=0}^1\frac{\overline \vartheta[^\gamma_\delta]^6\overline \vartheta[^{\gamma+h}_{\delta+g}]\overline \vartheta[^{\gamma-h}_{\delta-g}]}{\overline \eta^8} \\ \times \frac{1}{2}\sum_{a,b=0}^1 (-1)^{a+b+ab}\frac{\vartheta^2[^a_b]\vartheta[^{a+h}_{b+g}]\vartheta[^{a-h}_{b-g}]}{\eta^4} \, .
\end{eqaed}
Moreover, one can further simplify the covariant derivative using 
\begin{eqaed}\label{eq:dtau_vs_partau}
    \frac{D_\tau \vartheta[^a_b]}{\eta} = \partial_\tau\left( \frac{\vartheta[^a_b]}{\eta}\right) ,
\end{eqaed}
as well as $\vartheta$-identities which can be found in \cite[Appendix A]{Kiritsis:1997hj}, to rewrite the charge insertion as
\begin{eqaed}\label{eq:charge_insertion_eisenst}
    \left[\frac{1}{(2\pi i)^2}\partial_{\overline z}^2-\frac{1}{4\pi \tau_2}\right]{\overline \Gamma}_{\text{E}_8}|_{\overline z =0}  = \frac{1}{12}(\hat {\overline{\mathsf{E}}}_2 \overline{\mathsf{E}}_4 - \overline{\mathsf{E}}_6) \, ,
\end{eqaed}
where the $\mathsf{E}_s$ are the holomorphic Eisenstein series defined in \cite{DHoker:2022dxx}, while $\hat{\mathsf{E}}_2$ refers to the modular-invariant, non-holomorphic Eisenstein series (c.f. \cref{app:eisenstein_series_and_regularization}). Thus, the one-loop correction can be compactly repackaged as
\begin{eqaed}\label{eq:f2_het_2}
     \langle F^2\rangle_{\text{het}}  =\frac{8\pi^2 \tau_2}{\eta^2 \overline \eta^2}\frac{\Gamma_{{\text{SU}(2)}_k}}{V(\mu)}\sum_{\text{even}} (-1)^{a+b+ab} \, \partial_\tau\left(\frac{\vartheta[^a_b]}{\eta}\right) C_{\text{int}}[^a_b] \, , 
\end{eqaed}
with 
\begin{eqaed}\label{eq:c_int_het}
C_{\text{int}}[^a_b]  =  \frac{1}{24}(\hat {\overline{\mathsf{E}}}_2 \overline{\mathsf{E}}_4 - \overline{\mathsf{E}}_6) \sum_{h,g=0}^1\frac{\Gamma_{2,2}  \, Z_{4,4}[^h_g]}{ \eta^2 \overline \eta^{10}} \,\frac{\vartheta[^a_b]\vartheta[^{a+h}_{b+g}]\vartheta[^{a-h}_{b-g}]}{\eta^3} \\
\times  \ \frac{1}{2}\sum_{\delta,\gamma=0}^1\frac{\overline \vartheta[^\gamma_\delta]^6\overline \vartheta[^{\gamma+h}_{\delta+g}]\overline \vartheta[^{\gamma-h}_{\delta-g}]}{\overline \eta^8}\ .
\end{eqaed}
To compute this, we can use lattice identities found in \cite{Kiritsis:1997hj}. In particular, we will use
\begin{eqaed}\label{eq:p_thetaovereta_computation}
    \frac{i}{4\pi}\sum_{\text{even}} \partial_\tau \left( \frac{\vartheta[^a_b]}{\eta}\right)\frac{\vartheta[^a_b]\vartheta[^{a+h}_{b+g}]\vartheta[^{a-h}_{b-g}]}{\eta^3} \frac{Z_{4,4}[^h_g]}{|\eta|^4} = \left\{ \begin{aligned}
        &4\frac{\eta^2}{\overline \vartheta[^{1+h}_{1+g}]\overline \vartheta[^{1-h}_{1-g}]} \quad \text{if} \quad  (h,g) \ne (0,0) \, ,\\
        &0 \quad \text{if} \quad  (h,g)=(0,0) \, ,
    \end{aligned} \right.
\end{eqaed}
together with
\begin{eqaed}\label{eq:e6_theta}
    \frac{1}{2}\sum_{(h,g)\ne(0,0)}\sum_{\gamma,\delta}\frac{\overline \vartheta[^\gamma_\delta]^6\overline \vartheta[^{\gamma+h}_{\delta+g}]\overline \vartheta[^{\gamma-h}_{\delta-g}]}{\overline \vartheta[^{1+h}_{1+g}]\overline \vartheta[^{1-h}_{1-g}]} = -\frac14 \frac{\overline{\mathsf{E}}_6}{\overline \eta^6} \, ,
\end{eqaed}
to rewrite the two-point correlator as
\begin{eqaed}\label{eq:f2_het_final}
    \langle F^2\rangle_{\text{het}}  =  8i\pi^2 \tau_2 \, \frac{\Gamma_{{\text{SU}(2)}_k}}{V(\mu)}\left[ \frac{1}{12}\Gamma_{2,2} \frac{\hat{\overline{\mathsf{E}}}_2\overline{\mathsf{E}}_4 \overline{\mathsf{E}}_6 - \overline{\mathsf{E}}_6^2}{\overline \eta^{24}}\right] .
\end{eqaed}
In the normalization discussed above, the gauge coupling is thus given by
\begin{eqaed}\label{eq:het_thr_final}
 \left.\frac{16\pi^2}{g^2}\right|_{\text{one-loop}} = -\frac{1}{4\pi^2} \int_{\cF}\frac{\td^2 \tau}{\tau_2^2} \langle F^2 \rangle = - \frac{2 i}{V(\mu)}\int_{\cF}\frac{\td^2 \tau}{\tau_2} \, \Gamma_{{\text{SU}(2)}_k}\left[ \frac{1}{12} \, \Gamma_{2,2} \, \frac{\hat{\overline{\mathsf{E}}}_2\overline{\mathsf{E}}_4 \overline{\mathsf{E}}_6 - \overline{\mathsf{E}}_6^2}{\overline \eta^{24}}\right] ,
\end{eqaed}
matching the results of \cite{Kiritsis:1995dx}. It is interesting to note that the term $\Gamma_{{\text{SU}(2)}_k}$ in the modular integral plays the role of the modular-invariant regulator $\cG$ in \cref{eq:abelgauge}, and was indeed used in \cite{Abel:2023hkk}. One can manipulate the above expression in order to extract a universal logarithmic contribution due to massless modes, and separate it from the moduli-dependent threshold corrections due to massive modes. We remark that, due to the choice of $\text{U}(1)$ we have made, the gauge coupling scales like the lattice sum $\Gamma_{2,2}$ as a function of the $T^2$ untwisted moduli $T,U$. Concretely, we report the final result for the massive threshold corrections \cite{Kiritsis:1995dx,Dixon:1990pc}, which read
\begin{eqaed}\label{eq:het_thr_log}
 \Delta^{\text{threshold}}_g =  -\,A\log\left( \mathrm{Im} \, T \, \mathrm{Im} \, U \, \abs{\eta^4(T) \eta^4(U)} \right) + (T,U)\text{-independent\ terms} \, ,
\end{eqaed}
with $A$ constant. The above quantity scales with the K\"ahler modulus $\mathrm{Im} \,T$ as $\Delta^{\text{threshold}}_g \sim \mathrm{Im} \, T$, where $\mathrm{Im} \, T$ is (proportional to) the volume of the torus measured in string units. \\

\noindent We have thus seen that the general form of the one-loop gauge coupling in \cref{eq:f2correlator_refined} in terms of the flavor and helicity partition functions correctly reproduces known results in the literature, namely the well-known example of the exceptional heterotic string compactified on $T^4/\bZ_2 \times T^2$. In the following section, we will show that, under conservative assumptions, the string-frame asymptotic behaviour of the gauge coupling computed from the worldsheet in this explicit example is actually universal, and follows from the general properties of the internal \ac{CFT}. 

\section{Universal one-loop asymptotics from modular invariance}\label{sec:universal_thresholds_modular_invariance}
In the preceding discussion, we have introduced a method to compute gauge couplings and higher-derivative corrections of the type of \cref{eq:higherdercorr} from the worldsheet given any internal \ac{CFT}. Even though our computations were specific to $d=4$, similar considerations can be made for arbitrary $d$. In this paper we are interested in deriving universal asymptotic behaviors of the one-loop corrections to these couplings. The main technical tool we will use are asymptotic differential equations, in the spirit of \cite{Aoufia:2024awo}. Specifically, we will assume that the \emph{reduced} partition function of the internal theory $\sf I$---i.e. the internal partition function at fixed twisted sector stripped of the fermionic characters---factorizes into two pieces: one of them, denoted $\cZ_{\text{c}}^{(s)}$ and defined more in detail below, will depend on a marginal parameter $t$. Concretely, echoing \cite{Aoufia:2024awo}, our working assumptions on the \ac{CFT} with (bosonic) partition function $\cZ_{\text{c}}^{(s)}$ are the following (although some of them follow from more physical requirements):
\begin{itemize}
    \item[1.] It contains a marginal parameter $t>0$, related to the geodesic length of a path towards the infinite-distance boundary $t\to \infty$ on its conformal manifold. The fact that $t \to \infty$ lies at infinite Zamolodchikov distance follows from modular invariance when $t \to \infty$ is a species limit (i.e. when the species scale vanishes in $d$-dimensional Planck units), since in that case the spectral gap $\Delta_\text{gap}$ vanishes \cite{Aoufia:2024awo} and the results of \cite{Ooguri:2024ofs} apply.

    \item[2.] In the $t \to \infty$ limit, the spectrum of conformal primaries either contains uniformly \emph{light} states, or \emph{heavy} states with conformal weights $\Delta(t)\gg1$. For convenience, we normalize $t$ relative to the spectral gap, such that asymptotically $\Delta_{\text{gap}} \sim t^{-1}$, while light states are characterized by $\Delta(t) \sim \Delta_* t^{-1}$. As shown in \cite{Aoufia:2024awo}, the existence of an infinite tower of light states actually follows from modular invariance whenever $t \to \infty$ is a species limit. We relax this simplicity assumption in \cref{app:additive_towers_anisotropic_limits}, where we analyze more general spectra including subleading (``additive'' \cite{Castellano:2021mmx, Castellano:2022bvr}) and multiple (``multiplicative'') towers.

    \item[3.] The (integer) spins $j$ of the states are constant along the limit. This is essentially guaranteed by modular invariance, up to pathological discontinuous variations.

    \item[4.] The heavy conformal dimensions have derivatives which at most grow polynomially in the conformal weights themselves, i.e. $\partial_t \Delta(t) \lesssim \Delta^k(t)$ as $t \gg1$, for finite $k$. 
\end{itemize}
Assumption 2 implies the emergence of a light tower \cite{Aoufia:2024awo} whose exponential fall-off complies with the swampland distance conjecture \cite{Ooguri:2006in}, as shown in \cite{Ooguri:2024ofs}. Since the relevant physics is captured by it, we relegate the analysis of more general \ac{CFT} spectra to \cref{app:additive_towers_anisotropic_limits}, which covers subleading and multiple towers (corresponding to more general decompactifications). Assumptions 3 and 4 are merely technical, stemming from the need to define a \ac{CFT} devoid of pathologies, such as extremely rapid oscillations. As a final remark, while the physical string spectrum contains no tachyons for any $t$ (by assumption), states that are not level-matched may have negative conformal weight; we refer to these as \emph{unphysical tachyons} (or ``protogravitons'', in the language of \cite{Dienes:1990ij, Abel:2015oxa}). Since the \ac{CFT} is unitary and $\cZ_{\text{c}}$ contains primary states, unphysical tachyons can arise in the remaining sector of the partition functions we shall examine. We defer their treatment to \cref{app:unphysical_tachyons}, and anticipate that they will not affect the ensuing discussions. Thus, we will not consider them in the following analysis. \\

\noindent Since the \ac{CFT} will contain charged states under the $\text{U}(1)$ symmetry under scrutiny, some assumptions are needed for their spectrum. The unitarity bound $\Delta\ge Q^2$, which holds for every state of the \ac{CFT} \cite{Heidenreich:2024dmr}, implies that light states can only have asymptotically vanishing charges, but heavy states can have arbitrary charges. In particular, we assume that:
\begin{itemize}
    \item[5.] The charge $Q(t)$ of a light state scales asymptotically like $Q^2(t) \sim Q_*^2 t^{-k} ,  \ k \ge1 $. More generally, there could be cases where their charge is exponentially suppressed. They would lead to better-behaved asymptotics, which we will not discuss.

    \item[6.] The derivative of the charges of the heavy states grow at most as fast as some power of their conformal weight, i.e. $\partial_t Q^2(t) \lesssim \Delta^q(t)$ for fixed $q$.
    
\end{itemize}
The case $k=1$ of assumption 5 comprises the case of simultaneous Virasoro and Kac-Moody primaries, for which $\overline Q^2 = \Delta$. Once again, assumption 6 is technical and ensures that operator insertions in the partition function do not diverge pathologically in the limit. Lastly, we remark here some of the bootstrap results that follow from \cite{Afkhami-Jeddi:2020hde} and which will be used throughout the section. Decomposing the \ac{CFT} stress-energy tensor into a Kac-Moody piece and a remainder, one can estimate the density of states as $\rho(Q,\Delta)=\rho(\Delta_{\text{KM}})\rho(\Delta-\Delta_{\text{KM}})$. Since from modular invariance \cite{Afkhami-Jeddi:2020hde} deduced that $\rho(\Delta) \sim \Delta^{c-1}$ for $\Delta\gg1$, $\rho(Q,\Delta)$ can be taken to be at most polynomial in $Q$ and $\Delta$. More in general, the density of states of the total \ac{CFT} bounds the above refinement in terms of the decomposed stress-energy tensor, and the above conclusion applies. \\

\noindent In the subsequent discussion, we review the results of \cite{Aoufia:2024awo} in light of the above assumptions, and how species limits imply some of them. Then, we extend these results to gauge couplings and the higher-derivative corrections discussed above. We do this in two ways: firstly, we do it in detail for the gauge coupling computation in $d=4$ using the background-field method of \cref{sec:background_field_method}, subsequently including higher-derivative operators. Secondly, we generalize the discussion for any $d$ extracting the scalings from a general low-energy analysis of scattering amplitudes.

\subsection{Species limits and modular differential equations}\label{sec:species_limits_modular_differential_equations}
In this section, we briefly review the results of \cite{Aoufia:2024awo}, where a universal scaling for $R^4$ Wilson coefficients was derived in the context of RNS-RNS\footnote{Here the term ``RNS-RNS'' is meant to emphasize that we are not merely considering compactifications of type II superstrings, but in principle more general---possibly non-geometric and/or non-supersymmetric---settings.} closed-string amplitudes involving a generic internal \ac{CFT}. In particular, this was done by virtue of an asymptotic differential equation satisfied by \cite{Aoufia:2024awo}
\begin{eqaed}\label{eq:ar4_coeff}
    a_{4,0}(t) \sim \frac{2 \zeta(3)}{g_{s,d}^2} + 2\pi \sum_sb_s\int_{\cF} \td \mu \, \widetilde \cZ_{\text{int}}^{(s)} \, ,
\end{eqaed}
where $\td \mu \equiv \tau_2^{-2} \td^2 \tau$ is the standard modular-invariant measure and the $b_s$ are suitable constants determined by the \ac{GSO} projection. To be precise, in settings with non-maximal or even no supersymmetry, \cref{eq:ar4_coeff} may not provide the full set of Wilson coefficients for quartic-Riemann operators, since they can feature different kinematic structures and (external) worldsheet integrands. As a numerical equality, \cref{eq:ar4_coeff} only captures specific kinematic contractions, such as the unique one surviving maximal supersymmetry. However, as an asymptotic equality, \cref{eq:ar4_coeff} captures all quartic-Riemann coefficients, as we shall discuss in detail in \cref{sec:scattering_amplitudes_all-orders}. Here $\cZ_{\text{int}}^{(s)}$ is the \emph{reduced} torus partition function for the internal theory $\sf I$ at fixed spin structure (or twisted sector) $s$, which is obtained from the partition function $Z_{\text{int}}$ by stripping away the fermionic characters and multiplying and dividing by $(\sqrt{\tau_2}\eta \overline \eta)^c$, with $c=\overline c=10-d$ the central charge of the \ac{CFT}. In terms of modular properties, the reduced partition function defined above is a (vector-valued\footnote{Intuitively, this means the transformation properties due to the $s$-dependence resemble those of the indices $a,b$ of $\vartheta$-functions $\vartheta[^a_b]$; more precisely, they can be organized into a vector whose components mix under modular transformations with a (generically complex) matrix action accompanied by an overall modular weight.}) modular form of weight zero. The tilde denotes Eisenstein regularization due to possible \ac{IR} divergences for $c>1$ (see \cref{app:eisenstein_series_and_regularization}); it amounts to isolating the slow-growth contribution of $\cZ_{\text{int}}^{(s)}$ scaling as $\tau_2^{c/2}$ at the cusp, and subtracting off an Eisenstein series $E_{c/2}$ of the same growth in order to render the function integrable in a modular-invariant fashion. All in all, we can write $\widetilde \cZ_{\text{int}}$ by decomposing $\cZ_{\text{int}}$ according to
\begin{eqaed}\label{eq:intsumreview}
    \widetilde \cZ_{\text{int}}^{(s)}(\tau)=\tau_2^{\frac{c}2} \sum_{j,\Delta}d_{j,\Delta}^{(s)}\, e^{2\pi i j \tau_1}e^{-2\pi\Delta \tau_2}  - k^{(s)} E_{\frac{c}{2}} \, , 
\end{eqaed}
with $\Delta=h+\overline h$ the (primary) conformal weight, $j=h-\overline h \in \bZ$ the corresponding spin, and $k^{(s)}$ a constant depending on the $s$-sector (if $c=1$ there is no need for the Eisenstein regularization). The degeneracies in \cref{eq:intsumreview} are non-negative, as befits a bosonic partition function. For Narain \acp{CFT}, the reduced partition function can be written in terms of the lattice sum of ${\text{U}(1)}^n\times {\text{U}(1)}^n$ primaries---i.e. \ac{KK} and winding modes---in \cref{eq:latticesum} as $\cZ_{n,n} = \tau_2^{n/2} \, \Gamma_{n,n}$, since the fermionic characters factorize and the $s$-dependence due to the spin structures drops out. We assume that $\widetilde \cZ_{\text{int}}$ satisfies the hypotheses discussed at the beginning of the section. The case where only a factor $\cZ_{\text{c}}$ inside $\cZ_{\text{int}}$ satisfies the assumptions is also discussed in \cite{Aoufia:2024awo}. We shall discuss this case extensively later on in the paper. In the above discussion, we have assumed that the decompactifying sector of the internal \ac{CFT} be non-chiral, in the sense that its left-moving and right-moving central charges $c = \overline{c}$. We now prove that this must be the case on the grounds of modular and conformal invariance. \\

\noindent \paragraph{The decompactifying central charges must be equal.} To begin with, recall that mutual locality of vertex operators with weights $h, \overline{h}$ implies that
\begin{eqaed}\label{eq:mutual_locality}
   h - \overline{h} \in \frac{1}{2} \, \mathbb{Z} \, . 
\end{eqaed}
Moreover, invariance of the torus partition function under modular T-transformations leads to
\begin{eqaed}\label{eq:T-invariance}
    h - \overline{h} - \, \frac{c - \overline{c}}{24} \in \mathbb{Z} \, .
\end{eqaed}
The decompactifying sector of the \ac{CFT} contains light states for which $\Delta(t) = h(t) + \overline{h}(t) \to 0$, which requires $h(t) = \overline{h}(t)$ according to \cref{eq:mutual_locality}. Thus we learn that
\begin{eqaed}\label{eq:c-cbar_24Z}
    c - \overline{c} \in 24 \mathbb{Z} \, .
\end{eqaed}
For the internal \ac{CFT} of a classical string vacuum in $d>3$ non-compact dimensions we have $c , \overline{c} \geq 0$ and $\abs{c - \overline{c}} \leq 26 - d \leq 22$. Hence, \cref{eq:c-cbar_24Z} finally implies
\begin{eqaed}\label{eq:c_cbar_equality}
    c = \overline{c} \, .
\end{eqaed}
In particular, the internal sector of heterotic strings (including the one that produces the gauge sector) cannot decompactify more than the maximal number of dimensions. For example, in toroidal compactifications, decompactifying sector should be described by a sublattice $\Gamma_{c,c} \subset \Gamma_{d,d+16}$. \\

\noindent \paragraph{Modular differential equations.} We now briefly present the results of \cite{Aoufia:2024awo}. The main technical upshot is an asymptotic generalization of the remarkable differential equation satisfied by Narain partition functions \cref{eq:latticesum}, which reads
\begin{eqaed}\label{eq:narain_asym_eq}
    \left(\Delta_{\text{Narain}}+ w_c -\Delta_{\tau}\right) \cZ_{c,c} = 0 \ , \quad w_c \equiv \frac{c}{2}\left(1-\frac{c}{2}\right) \, ,
\end{eqaed}
where $\Delta_{\text{Narain}}$  and $\Delta_{\tau}$ are respectively the Laplacian on the Narain conformal manifold (computed with the Zamolodchikov metric) and the Laplacian over the upper-half plane. Starting from the assumptions fleshed out at the beginning of the section, one can try to look for an analogous differential equation satisfied in the limit $t\gg1$. To do this, \cite{Aoufia:2024awo} split the contribution of light and heavy states and looked for a second-order differential operator in the marginal parameter. The procedure has been refined and generalized in \cref{app:exp_suppression}, the upshot being that the light(est) states dominate the limit while the contributions from heavy states are exponentially suppressed. Since we will go through the derivation once again in \cref{par:graviphotons}, we omit it here and refer the reader to that section or to \cref{app:exp_suppression}. The resulting asymptotic differential equation is 
\begin{eqaed}\label{eq:asymptotic_diff_eq}
    \left( \cD_c + w_c \right) \widetilde \cZ_{\text{int} }(\tau;t) \overset{t \to \infty}{\sim} \Delta_\tau \widetilde \cZ_{\text{int} }(\tau;t) \ , \quad \cD_{c} \equiv -t^2\partial_t^2 -(2-c)t \partial_t \, .
\end{eqaed}
This establishes that the Wilson coefficient for the quartic contact interaction between gravitons satisfies the corresponding equation in $t$ once integrated. Its general solution reads
\begin{eqaed}\label{eq:ar4_solution}
    a_{4,0}(t) \sim t^{\frac{c}{2}}\left( a+ \frac{b}{t}\right) ,
\end{eqaed}
where $a\ne0$ follows from a \ac{RS}-transform argument \cite{Aoufia:2024awo}. For $c=2$, Eisenstein regularization via $\hat E_1$ adds a constant to the right-hand side of the equation, in light of the ``anomalous'' differential equation \cref{eq:eigen_e1}, which translates into the known logarithmic running of the marginal coupling in eight dimensions. Notice that nowhere we assumed the internal \ac{CFT} is geometric or supersymmetric; however, the results of \cite{Ooguri:2024ofs} show that the $t \to \infty$ is in fact a decompactified sigma model; in order to show that \cref{eq:ar4_solution} is telling us the same thing asymptotically, we employ geometric notation pertaining to a decompactification to $D = d+c$ dimensions. Specifically, since $m_{\text{KK}}/M_s =\cV^{-1/c} = \sqrt{\Delta_{\text{gap}}}=t^{-1/2}$, the string-frame coefficient scales with the volume in string units as
\begin{eqaed}\label{eq:ar4_sol_vol}
    a_{4,0}(t) \sim \cV \, .
\end{eqaed}
Using the dictionary between string-frame and Einstein-frame coefficients in \eqref{eq:string_vs_einstein_coeff}, and the fact that $M_s^{d-2} = \cV^{-1} g_{s,D}^2 \, M_{{\text{Pl}};d}^{d-2}$ in terms of the $D$-dimensional string coupling $g_{s,D}$, one confirms the EFT expectation \eqref{eq:wilson_coeff_EFT_expectation_planck} that the Einstein-frame coefficient $c_{4,0} \sim \Lambda_{\text{sp}}^{-6} M_{{\text{Pl}};d}^{6}$, since we are taking the limit at fixed $g_{s,D}$ and thus the species scale is always the string scale. A more refined computation which encompasses partial decompactifications is also able to reproduce the second term in \eqref{eq:wilson_coeff_EFT_expectation_planck} depending on whether the Wilson coefficient is classically irrelevant in the decompactified theory \cite{Aoufia:2024awo}, namely when $D=8$. \\

\paragraph{Species limits and towers of states.} Another important point highlighted in \cite{Aoufia:2024awo} is that in species limits the spectral gap of the \ac{CFT} vanishes, together with the conformal weights of an infinite tower of states. As outlined in \cref{sec:introduction}, we define \emph{species limit} any limit (in the worldsheet conformal manifold\footnote{Of course, the limit in which the string coupling vanishes is also a species limit. In \cref{sec:higher_loops} we will discuss some aspects of combining these possibilities.}) where the species scale, as computed from the (suitable terms of the) Wilson coefficients of the \ac{EFT}, vanishes in Planck units. For our purposes, this occurs precisely when e.g \cref{eq:ar4_coeff} diverges; as we will show later on, the precise choice of modular integral is irrelevant, since they share the same properties. Here we generalize these results for the case where the internal partition function is at fixed $s$-sector (i.e. at fixed spin structure or twisted sector). As in \cite{Aoufia:2024awo} we proceed by contradiction, assuming that the gap does not vanish, and following \cite{Hartman:2014oaa}. To begin with, if $a_{4,0}(t)$ diverges in the limit, the only source of divergence is the integral of the reduced partition function, which we denote $I(t)$ in the following. In the limit, this means that at least one of the integrands diverge. Let us then bound the relevant reduced partition function by its canonical counterpart
\begin{eqaed}\label{eq:bound_redpf}
    |\mathcal{Z}^{(s)}_{\text{int}}| \le Z \equiv \tau_2^{\frac{c}2} \sum_{j,\Delta>0} d_{\Delta, j}^{(s)} \, e^{-2\pi \Delta \tau_2} + \tau_2^{\frac{c}{2}} \equiv Z_{>} + Z_0 \, ,
\end{eqaed}
and its (regularized) integral by the integral over the strip $S=\{\tau_2\ge \sqrt{3}/2 \}$. In the previous expression, we separated the zero-mode $Z_0 = d_0 \, \tau_2^{c/2}$ from the sum $Z_>$ over positive weights, and we used the fact that degeneracies of bosonic partition functions are positive. All the above quantities are vector-valued modular forms of weight zero. Using the dot-product notation, the Wilson coefficient in \cref{eq:ar4_coeff} is thus defined as the integral of $b\cdot \cZ_{\text{int}}$, where $b=(b_s)$ is a vector in the space indexed by $s$. Under an S-transformation $\tau_2 \to 1/\tau_2$, the modular vector $Z$ transforms as
\begin{eqaed}\label{eq:pf_s_transf}
    Z'\equiv Z\left( \frac{1}{\tau_2}\right) = S Z \, ,
\end{eqaed}
where we defined the matrix $S$ implementing the S-transformation on the $s$-blocks such that ${S^2=\mathbb{I}}$, where $\mathbb{I}$ denotes the identity matrix\footnote{There is an analogous matrix implementing T-transformations via phases. Generically, this means that each component of a vector-valued partition function is not T-invariant by itself; however, these phases cancel out in the relevant modular-invariant combinations thereof, as is the case for S-transformations.}. More precisely, $S = S_\text{L} \otimes S_\text{R}^\dagger$ embodies the modular S-transformation on both left-moving and right-moving characters. Splitting $Z$ as in \cref{eq:bound_redpf}, invariance under S-transformations implies that
\begin{eqaed}\label{eq:pf_s_invariance}
   Z_>+  Z_0 = S(Z_>' +  Z_0') \, , \quad Z_>' =  \sum_{j,\Delta>0} d_{\Delta, j}^{(s)} \, e^{-2\pi \frac{\Delta}{\tau_2} } \, , \quad Z_0' = (S d_0) \, \tau_2^{-\frac{c}{2}} \, .
\end{eqaed}
For notational convenience, we will leave implicit the degeneracies $d_{j,\Delta}$ in what follows; thus, sums range over possibly repeated elements. Since $\Delta\ge \Delta_{\text{gap}}$ except for the vacuum, for $\tau_2 >1$ one can write the component-wise inequality
\begin{eqaed}\label{eq:pf_dominance}
    Z_> = \sum_{j,\Delta>0}e^{-2\pi \tau_2 \Delta} e^{-2\pi \frac{\Delta}{\tau_2} } e^{+2\pi \frac{\Delta}{\tau_2} }  \le e^{-2\pi \left(\tau_2 -\frac{1}{\tau_2}\right)\Delta_{\text{gap}}} \, Z'_> \, ,
\end{eqaed}
and therefore\footnote{Although the matrix $S$ generically has complex entries, its action on canonical partition functions must be real. Indeed, $S Z_> + S Z_0$ is manifestly real for all $\tau_2$, and hence---taking the large-$\tau_2$ asymptotics---both $S Z_0$ and $S Z_>$ are real.}
\begin{eqaed}\label{eq:pf_dominance_2}
     S Z_>'- Z_> \ge \left( S  - e^{-2\pi \left(\tau_2 -\frac{1}{\tau_2}\right)\Delta_{\text{gap}}} \, \mathbb{I}\right) Z_>' \, .
\end{eqaed}
Thus, taking the norm of \cref{eq:pf_dominance_2} and using modular invariance we obtain
\begin{eqaed}\label{eq:pf_dominance_modinv}
    \left(1 - e^{-2\pi \left(\tau_2 -\frac{1}{\tau_2}\right)\Delta_{\text{gap}}} \right) \norm{Z_>'} & \le \norm{\left(S - e^{-2\pi \left(\tau_2 -\frac{1}{\tau_2}\right)\Delta_{\text{gap}}} \, \mathbb{I}\right)Z'_>} \\
    &\le \norm{S Z_>'- Z_>} = \norm{Z_0 - S Z_0'} \, .
\end{eqaed}
Finally, using the positivity of $Z_0,\,Z_0'$ and $\tau_2>1$, \cref{eq:pf_dominance,eq:pf_dominance_modinv} lead to
\begin{eqaed}\label{eq:bounded_argument_1}
    \left\lVert Z_> \right\rVert\le \norm{\left(\mathbb{I} - \tau_2^{-c} S\right)d_0} \frac{e^{-2\pi \left(\tau_2 -\frac{1}{\tau_2}\right)\Delta_{\text{gap}}}}{1 - e^{-2\pi \left(\tau_2 -\frac{1}{\tau_2}\right)\Delta_{\text{gap}}} } \, \tau_2^{\frac{c}{2}} \leq \text{const.} \times \frac{e^{-2\pi \left(\tau_2 -\frac{1}{\tau_2}\right)\Delta_{\text{gap}}}}{1 - e^{-2\pi \left(\tau_2 -\frac{1}{\tau_2}\right)\Delta_{\text{gap}}} } \, \tau_2^{\frac{c}{2}} \, .
\end{eqaed}
In particular, all components of $Z_>$ are bounded according to \cref{eq:bounded_argument_1}. For $\tau_2<1$, an analogous argument entails that
\begin{eqaed}\label{eq:bounded_argument_2}
    \left\lVert Z_>\right \rVert\le \frac{\text{const.}}{1 - e^{2\pi \left(\tau_2 -\frac{1}{\tau_2}\right)\Delta_{\text{gap}}}} \, \tau_2^{-\frac{c}{2}} \, .
\end{eqaed}
Hence, we see that assuming the gap not to vanish leads to a contradiction if $a_{4,0}(t)$ diverges. Indeed, the norm of the integral over the strip $S = \{ \tau_2 \ge \sqrt3/2\}$ can be bounded by a function of $\Delta_{\text{gap}}$ which remains finite by virtue of \cref{eq:bounded_argument_1,eq:bounded_argument_2}. Replacing $\Delta_\text{gap}$ with any threshold above it, we can repeat the same argument as above to see that the vanishing spectral gap must be accompanied by an infinite tower of states whose conformal weights also vanish. Indeed, splitting the sum into $Z_>$ and $Z_<$ at any finite conformal weight $\hat \Delta \ge \Delta_{\text{gap}}$, one has the same inequalities with $Z_0$ replaced by the sum $Z_<$ over states with $\Delta \leq \hat \Delta$: thus, by way of contradiction, one has $Z_< =\tau_2^{c/2}\sum_{\Delta\le \hat \Delta}e^{-2\pi \Delta \tau_2} \le N\tau_2^{c/2}$, with $N$ the total (assumed finite) number of states below $\hat \Delta$. This allows to bound the norm $\norm{Z_<' - S Z_<} \leq \text{const.} \times N \left(\tau_2^{c/2} + \tau_2^{-c/2}\right)$, which can be further bound according to \cref{eq:bounded_argument_1,eq:bounded_argument_2} depending on whether $\tau_2 > 1$. Once again, this leads to a contradiction: there must exist an infinite tower of states with vanishing conformal weights whenever one of the sectors $\cZ_{\text{int}}^{(s)}$ develops a divergence. Finally, notice that the above argument goes through even when the vector $b_s$ is replaced by the external sector of a \ac{CFT}---say $B^{(s)}(\tau)$---as it is $t$-independent and cannot develop a divergence. It is worth noting that from a bottom-up perspective the emergence of a tower of light species in species limits was framed in information-theoretic terms in \cite{Stout:2022phm}, where the (asymptotic) factorization of observables in the presence of gravity requires such a mechanism and no other options are known.

\subsection{Gauge couplings in species limits}\label{sec:gauge_couplings_species_limits}
We now turn our attention to the universal scaling of gauge couplings in species limits. Similarly to above, we start our analysis by rewriting the partition function for the curved-background deformation tensored with a generic internal theory $\sf I$ in terms of a reduced partition function ${\cZ}_{\text{int}}$. To do so, first we make the overall moduli dependence of the partition function explicit; namely, we recast \cref{eq:Zwexternalexplicit} as
\begin{eqaed}\label{eq:zwexternalexplict_recast}
   Z_W (\tau, \mu; t) = \frac{1}{\tau_2\eta^2 \overline\eta^2}\frac{\Gamma_{{\text{SU}(2)}_k}}{V(\mu)} \sum_s r(s) \, \frac{\vartheta(s)}{\eta} \, Z_{\text{int}}^{(s)}(\tau;t) \, , 
\end{eqaed}
with $t$ the overall modulus that parametrizes the path we consider within the worldsheet conformal manifold. Then, we move all fermionic characters\footnote{To distinguish them invariantly, one can define the fermionic characters to be the super-Virasoro partners of the primary states which enter the reduced partition function. As such, they cannot depend on moduli due to conformal invariance; all the moduli-dependence features in the primary weights.} $\cZ_{\text{f}}^{(s)}$ out of $Z_{\text{int}}$ and multiply and divide by $(\sqrt{\tau_2} \eta \overline \eta)^6$ to get to 
\begin{eqaed}\label{eq:Zw_t-explicit}
     Z_W (\tau, \mu; t) = \frac{1}{\tau_2^4\eta^8 \overline\eta^8}\frac{\Gamma_{{\text{SU}(2)}_k}}{V(\mu)} \sum_s r(s) \, \frac{\vartheta(s)}{\eta} \cZ_{\text{f}}^{(s)}(\tau) \, \cZ_{\text{int}}^{(s)}(\tau;t) \, .
\end{eqaed}
Notice that $\cZ_{\text{int}}^{(s)}$ behaves as the partition function of a \ac{CFT} of bosons with possibly different central charges $c , \, \overline{c}$\footnote{For comparison, in the heterotic example discussed in \cref{sec:example_heterotic_gauge_thresholds} the central charges would be $c=9,\, \overline{c}=22$ due to the presence of the $\text{E}_8\times \text{E}_8$ lattice.} and with fixed spin structure and twist insertions $s$. Following the discussion in \cref{sec:species_limits_modular_differential_equations}, we assume the internal reduced partition function to factorize into two pieces. The first one, which we denote $\cZ_{\text{c}}^{(s)}$ as mentioned above our list of assumptions, contains the (bosonic) characters of the (necessarily non-chiral) sector which contains the light states. One can think of this contribution as coming from the smallest \ac{CFT} containing the collection of primary states with $t$-dependent conformal weights $\{\Delta(t)\}$ and their modular orbit. The second one, which we simply denote $\cZ^{(s)}$, contains the remaining $t$-independent contributions. Therefore, up to the possible addition of extra sums-over-sectors which we shall not include in the notation, we take 
\begin{eqaed}\label{eq:z_int_split}
    \cZ_{\text{int}}^{(s)}(\tau;t) = \cZ_{\text{c}}^{(s)} (\tau; t) \, \cZ^{(s)}(\tau)
\end{eqaed}
from here onward. In the above expression, we include an extra factor of $\tau_2^{c/2}$ in $\cZ_{\text{c}}$ and $\tau_2^{3-c/2}$ in $\cZ$ to match with the definitional behavior of the reduced partition function given above.
When computing the gauge coupling in terms of (improved) flavored partition functions, we will include the improvement factor $e^{\overline z^2/(2\tau_2)}$ related to the gauge-charge insertion introduced in \cref{eq:improvedfpf} either in $\cZ_{\text{c}}$ or outside of it, depending on whether the conformal weights of the Kac-Moody ${\text{U}(1)}$ algebra depend on $t$ or not. This will allow us to distinguish photons from graviphotons from our worldsheet perspective. \\

\paragraph{Gauge couplings for photons.} Let us first consider the case where the ${\text{U}(1)}$ insertions fall outside of $\cZ_{\text{c}}^{(s)}$. The structure of the correlator in \cref{eq:f2correlator_refined} thus reads 
\begin{eqaed}\label{eq:f2_zc_split}
    \langle F^2\rangle  =  \left. \frac{8\pi^2 }{\tau_2^2\eta^8\overline \eta^8}\frac{\Gamma_{{\text{SU}(2)}_k}}{V(\mu)} \sum_s r(s) \, \frac{D_\tau\vartheta(s,v)}{\eta} \, \cZ_{\text{c}}^{(s)}(\tau;t)\,\partial_{\overline z}^2 \cZ_{\text{rest}}^{(s)}(\tau;\overline z)\right|_{\overline z =0} \, ,
\end{eqaed}
where we collectively denoted $\cZ_{\text{rest}} = Z_{\text{f}} \cZ$ the fermion characters and the $t$-independent terms in $\cZ_{\text{int}}$. In order to evaluate the modular integral of this quantity, we make use of the ${\text{SL}}(2,\bZ)$-invariant spectral decomposition described in \cref{app:eisenstein_series_and_regularization}, and similarly employed in \cite{Aoufia:2024awo}. In particular, let us comment on the $t$-dependence of the spectral coefficients of the regularized (vector-valued) modular function $\tilde \cZ_{\text{c}}^{(s)}(\tau;t) \equiv \cZ_{\text{c}}^{(s)}(\tau;t)-k^{(s)}E_{c/2}$, with $k^{(s)}$ a suitable constant. This reads
\begin{eqaed}\label{eq:fourierexpcz}
    \tilde{\cZ}_{\text{c}}^{(s)}(\tau;t) = \frac3\pi \,  I_{\text{c}}^{(s)}(t) + \sum_m a^{(s)}_m(t) \nu_m (\tau) + \int_{\Re{\ell}=\frac{1}{2}}\td \ell \ \alpha^{(s)}_\ell (t) E_\ell(\tau)  \, .
\end{eqaed}
As was shown in \cite{Aoufia:2024awo}, translating \cref{eq:asymptotic_diff_eq} in terms of these Fourier coefficients allows one to argue that the dominant scaling comes exactly from $I^{(s)}_{\text{c}}(t) \sim t^{c/2}$, while all other coefficients scale as $t^{\frac{c}2-\frac12 \pm i\lambda}$, with $\lambda$ depending on the eigenvalue of the ${\text{SL}}(2, \bZ)$-Laplacian $\Delta_\tau$ as described in \cref{app:eisenstein_series_and_regularization}. \\

\noindent With this result in hand, we are now ready to derive the differential equation obeyed by the gauge coupling by specializing the discussion in \cref{app:exp_suppression} to the case at hand. Let us denote the combination of all other contributions besides $\cZ_{\text{c}}^{(s)}$ in $\langle F^2\rangle$ as $B^{(s)}(\tau)$, so that we simply have
\begin{eqaed}\label{eq:f2_btimesz}
    \langle F^2 \rangle = \sum_s B^{(s)}(\tau) \cZ_{\text{c}}^{(s)}(\tau;t) \, .
\end{eqaed}
Using the $q$-expansion of the $\vartheta$-functions as in \cref{eq:q_exp_theta}, one can see that for $\tau_2 \gg 1$ the $B$-terms scale as $\tau_2^{1-\frac{c}2}$ due to the factor $\tau_2^{3-\frac{c}2}$ in $\cZ_{\text{rest}}$. We now regularize the product $B \cZ_{\text{c}}$, and rewrite it in terms of the regularized $\tilde \cZ_{\text{c}}$. In detail,
\begin{eqaed}\label{eq:widetildebz}
    \widetilde{B^{(s)} \cZ_{\text{c}}^{(s)}} &\equiv B^{(s)}(\tau) \cZ_{\text{c}}^{(s)}(\tau;t) - j_0^{(s)}\hat E_{1}(\tau) \\ &= \left(\hat B(\tau) + j_1^{(s)} \tau_2^{1-\frac{c}2}(\tau) \right) \left( \tilde \cZ_{\text{c}}^{(s)}(\tau;t) + k^{(s)}E_{\frac{c}2}(\tau)\right) - j_0^{(s)}\hat E_1(\tau) \, ,
\end{eqaed}
where $\hat E_1$ denotes ``anomalous'' Eisenstein series obeying \cref{eq:eigen_e1}, which arises because gauge couplings are classically marginal in four dimensions. In the above equation, we furthermore isolated the zero-mode of $B^{(s)}$ as to define a rapid-decay function $\hat B^{(s)}$\footnote{We remark that, since $c\ge1$, we do not subtract any Eisenstein series from $B^{(s)}$, since it is already integrable as explained in \cref{app:eisenstein_series_and_regularization}. }. Choosing the additional constant $j_0^{(s)}=k^{(s)}j_1^{(s)}$, the regularized product is by definition integrable, and thus its Laplacian integrated over the fundamental domain vanishes. This allows us to write\footnote{As described in \cref{app:eisenstein_series_and_regularization}, we normalize the volume of the fundamental domain as $\mathrm{Vol}(\cF)= 3/\pi$.}
\begin{eqaed}\label{eq:sum_b_z_integral}
    \sum_s\int_{\cF} \td \mu \, \Delta_\tau\left(B^{(s)} \cZ_{\text{c}}^{(s)}\right) = -\sum_sj_0^{(s)} \equiv j \, ,
\end{eqaed}
since this compensates the constant contribution coming from $\hat E_1(\tau)$. Expanding the Laplacian inside the integral and evaluating the boundary term at the cusp, we obtain
\begin{eqaed}\label{eq:integral-1}
    -j = \sum_s\int_{\cF}\td \mu \left( \Delta_\tau B^{(s)} \cZ_{\text{c}}^{(s)} -  B^{(s)}\Delta_\tau\cZ_{\text{c}}^{(s)}\right) - 2\lim_{L\to \infty}\sum_s\int_{-\frac12}^\frac12 \td \tau_1\,  \left(B^{(s)} \partial_{\tau_2} \cZ_{\text{c}}^{(s)}\right)_{\tau_2=L} \, .
\end{eqaed}
Expressing this in terms of $\tilde \cZ^{(s)}, \hat B^{(s)}$, we see that the only contribution comes from the Eisenstein and the zero-mode of and $B^{(s)}$; in particular the boundary term evaluates to
\begin{eqaed}\label{eq:boundary_int}
    2j_1^{(s)}k^{(s)}\lim_{L\to \infty}\int_{-\frac12}^\frac12 \td \tau_1 \,  \left( \tau_2^{1-\frac{c}2}\partial_{\tau_2} E_{\frac{c}2}\right)_{\tau_2=L}= j_0^{(s)}  c \, .
\end{eqaed}
In the last equality, we used that the integral over $x$ extracts the constant term of the integrand (the only one that is not exponentially suppressed). We can now apply the operator $\cD_c$ to the (integrated) regularized product and use the fact that $\cZ_{\text{c}}^{(s)}$ satisfies \cref{eq:asymptotic_diff_eq}. We find
\begin{eqaed}\label{eq:dc_reg_prod}
    \cD_c \int_{\cF}\td \mu\, \sum_s\widetilde{B^{(s)} \cZ_{\text{c}}^{(s)}} \sim \sum_s\int_{\cF}\td \mu\, B^{(s)} (\Delta_\tau -w_c)\cZ_{\text{c}}^{(s)} \, ,
\end{eqaed}
so that \cref{eq:integral-1} leads us to 
\begin{eqaed}\label{eq:dc_reg_prod_2}
   \cD_c \int_{\cF}\td \mu\, \sum_s\widetilde{B^{(s)} \cZ_{\text{c}}^{(s)}} \sim j (1-c) + \sum_s\int_{\cF}\td \mu \, (\Delta_\tau -w_c) B^{(s)} \cZ_{\text{c}}^{(s)} \, .
\end{eqaed}
Expanding $B^{(s)}$ and considering that the zero-mode is annihilated by $(\Delta_\tau -w_c)$, we have
\begin{eqaed}\label{eq:B_partial_integr}
    \sum_s\int_{\cF}\td \mu \, (\Delta_\tau -w_c) B^{(s)} \cZ_{\text{c}}^{(s)}  = \sum_s\int_{\cF}\td \mu \, (\Delta_\tau -w_c)\hat B^{(s)} \cZ_{\text{c}}^{(s)}  \sim A \, t^{c/2} \, ,
\end{eqaed}
where we used that $\hat B^{(s)} $ is integrable (even better, exponentially suppressed at the cusp). Thus, one can safely replace $\tilde \cZ_{\text{c}}^{(s)}$ by \eqref{eq:fourierexpcz} and extract the zero-mode directly, arriving at the scaling $A \, t^{c/2}$ with $A$ constant. Finally, summing over sectors and integrating over the fundamental domain, the one-loop correction to the gauge coupling $\Delta_g(t) \equiv c_{0,2}(t) = a_{0,2}(t)$ satisfies the asymptotic differential equation
\begin{eqaed}\label{eq:dc_gauge_coupling}
    \cD_c \, \Delta_g(t) \sim A \, t^{\frac{c}{2}} + j(c-1) \ \implies \ \Delta_g(t) \sim f_1 \, t^{\frac{c}2} + f_2 \, \log t + f_3 \, t^{c-1}
\end{eqaed}
with the $f_i$ constants of integration. The term proportional to $f_3$ is always either subleading or dominant; we now show that when it is dominant, its coefficient must vanish. To do this, let us consider the integral defining $\Delta_g$ and apply the Cauchy-Schwarz inequality,
\begin{eqaed}\label{eq:cauchy_sc_integral}
    |\Delta_g| \le \frac{1}{4\pi^2}\sum_s\left|\int_{\cF} \td \mu \,  \widetilde{B^{(s)} \cZ_{\text{c}}^{(s)}}\right| \lesssim \sum_s\left| I_{\text{c}}^{(s)}(t)\right| \int_{\cF} \td \mu \, \abs{B^{(s)}}^2 = O\left(t^{\frac{c}{2}}\right) ,
\end{eqaed}
where we used the fact that the integral of $\hat B$ is a $t$-independent constant. Consistency of the differential equation thus implies that $f_3=0$ when $c>1$ or that the term is subleading, and the solution becomes
\begin{eqaed}\label{eq:gauge_sol_eq}
    \Delta_g \sim f_1 \, t^{\frac{c}{2}}+f_2 \, \log t \, .
\end{eqaed}
This result is physically consistent: in the case of geometric compactifications, a coefficient $f_3\ne 0$ would have led to a divergent gauge coupling in the decompactified theory, by virtue of its scaling discussed in \cref{sec:gauge_couplings_higher_derivative_corrections}. \\

\paragraph{Gauge couplings for graviphotons.}\label{par:graviphotons} Let us now discuss the case where the ${\text{U}(1)}$ insertion affects $\cZ_{\text{c}}^{(s)}$. To give some intuition on how this might happen, consider the partition function of a compact boson of radius $R$, and let the charge insertion be related to one of the ${\text{U}(1)}$ factors in the ${\text{U}(1)}\times {\text{U}(1)}$ Kac-Moody algebra. Physically, these two conserved charges correspond to a combination of winding states with number $w$ and \ac{KK}-momentum states $n$, namely
\begin{eqaed}\label{eq:compact_boson_kk_momentum}
 p_{\text{L}} = \frac{n}{R} +w R\ , \quad p_{\text{R}} = \frac{n}{R}-wR \, . 
\end{eqaed}
The novelty is that an insertion of $p_{\text{L}}$ or $p_{\text{R}}$ carries a modulus dependence in terms of the radius $R$ of the circle. From the \ac{EFT} perspective, computing the gauge coupling for this ${\text{U}(1)}$ factor would mean to consider the coupling to the graviphoton coming from the gravitational sector of the decompactified theory. We reviewed in \cref{sec:gauge_couplings_higher_derivative_corrections} how these qualitatively differ from the previous case. In the following, we will see how this difference manifests itself in the differential equation satisfied by the coupling. We now derive it in detail following and generalizing the procedure introduced in \cite{Aoufia:2024awo} to derive \cref{eq:asymptotic_diff_eq}. \\

\noindent Consider a generic improved flavoured (reduced) partition function $\hat \cZ_{\text{c}}^{(s)}$. We take it to satisfy the assumptions laid out at the beginning of  \cref{sec:universal_thresholds_modular_invariance}. Decomposing it in Virasoro and ${\text{U}(1)}$ Kac-Moody characters and neglecting the vacuum characters (which were moved inside $\cZ_{\text{rest}}$\footnote{More precisely, Virasoro characters $\chi_h$ contain a universal $q^{-c/24}$ factor due to the vacuum, whose contribution factorizes and is moved outside of $\cZ_{\text{c}}^{(s)}$.}), one obtains \cite{Dyer:2017rul}
\begin{eqaed}\label{eq:virasoro_pf}
  \hat{\cZ}_{c}^{(s)}(\tau; t,\overline z) &= e^{\frac{\pi \overline z^2}{2\tau_2}}(\sqrt{\tau_2}\eta\overline \eta)^c \sum_{\overline Q,h,\overline h}d_{Q,h,\overline h}^{(s)} \, e^{2\pi i \overline z \overline Q} \chi_{h}(q)\overline \chi_{\overline h}(\overline q) \\
    &=e^{\frac{\pi \overline z^2}{2\tau_2}}{\tau_2^{\frac c2}}\sum_{\overline Q, h,\overline h}d_{Q,h,\overline h}^{(s)} e^{2\pi i \overline z \overline Q(t)} q^{h(t)} q^{\overline h(t)} \, ,
\end{eqaed}
where $d^{(s)}_{Q,h,\overline{h}}$ indicates the multiplicity of the level at fixed $h,\overline h,\overline Q$ for each $s$-sector. As remarked at the beginning of this section, due to the unitarity bound each state has $\overline h\ge \overline Q^2$ \cite{Heidenreich:2024dmr}. In particular, since we are summing over primaries, the unitarity bound is saturated for simultaneous charge and Virasoro primaries for which $\overline h=\overline Q^2$. For these states, every assumption on $\Delta(t) = h(t)+\overline h(t)$ transfers to $\overline Q(t)$. More generally, considering light and heavy states, one sees that light states can only have light charges, i.e. $\overline Q(t)^2 \sim \overline Q_*^2t^{-k} ,\ k\ge1$, while heavy states can have arbitrary charges. Rewriting the above in terms of conformal weights and spins, one gets
\begin{eqaed}\label{eq:virasoro_pf_conf_spin}
  \hat{\cal Z}_{\text{int}}^{(s)}(\tau; t,\overline z) =\, e^{\frac{\pi \overline z^2}{2\tau_2}}\,  \tau_2^{\frac c2} \sum_{\overline Q, j,\Delta}d_{Q,j,\Delta}^{(s)} \, e^{2\pi i \overline z \overline Q(t)} e^{2\pi ij \tau_1} e^{-2\pi \tau_2\Delta(t)} \, .
\end{eqaed}
We are interested in evaluating the charge insertion generated by acting with two derivatives with respect to $\overline z$, that is 
\begin{eqaed}\label{eq:pf_conf_spin_charge}
\cF_{c}^{(s)}\equiv\left.\partial_{\overline z} ^2\hat{\cal Z}_{\text{int}}^{(s)}(\tau; t,\overline z)\right|_{\overline z=0} = \tau_2^{\frac c2} \sum_{\overline Q, j,\Delta}d_{Q,j,\Delta}^{(s)} \left(\overline Q^2 - \frac{1}{4\pi \tau_2} \right)e^{2\pi ij \tau_1} e^{-2\pi \tau_2\Delta(t)} \, .
\end{eqaed}
From \cref{eq:pf_conf_spin_charge} one can observe that the charge-independent insertion gives a result proportional to $\cZ_{\text{c}}^{(s)}/\tau_2$. This suggests that $\tau_2\cF_{c}^{(s)}$ could solve the same asymptotic equation as $\cZ_{\text{c}}^{(s)}$. We now prove this, extending the arguments of \cite{Aoufia:2024awo} to the case where the \ac{CFT} contains a ${\text{U}(1)}$ global symmetry. In particular, let us focus on the charge-dependent term, and let us separate the sum into light and heavy states. Applying the operator $(\cD_c+w_c-\Delta_\tau)$ on $\tau_2 \cF_c^{(s)}$ termwise, one gets 
\begin{eqaed}\label{eq:termbytermoperator}
  &(\cD_c+w_c-\Delta_\tau)\left(\tau_2 \,\cF_{c}^{(s)}\right) = 4\pi^2 \tau_2^{1+\frac{c}2} \sum_{\overline Q, j,\Delta}d_{Q,j,\Delta}^{(s)} e^{2\pi ij \tau_1} e^{-2\pi \tau_2\Delta(t)} \\ &\times \left[  2 t^2 (\partial_t \overline Q)^2 + K(t) \overline Q^2 + 2 t \overline Q \Big( t \partial_t^2 \overline Q - \big(c - 2 + 4 \pi t \tau_2 \partial_t \Delta\big) \partial_t \overline Q \Big)   \right] , 
\end{eqaed}
where we have defined
\begin{eqaed}\label{eq:k_factor}
    K(t)& \equiv -c + 4 j^2 \pi^2 \tau_2^2 + 2(c+2)\pi \tau_2 \Delta - 4\pi^2 \tau_2^2 \Delta^2 \\
    &\quad + 2(c-2)\pi t \tau_2 \partial_t \Delta + 4\pi^2 t^2 \tau_2^2 (\partial_t \Delta)^2 - 2\pi t^2 \tau_2 \partial_t^2 \Delta \, .
\end{eqaed}
One can check that for light states, with $\Delta \sim \Delta_*/t$,  $\overline Q(t) \sim \overline Q_*/\sqrt t $, and $j = 0$, the right-hand side vanishes term by term. For light states with $\overline Q \sim \overline Q_* t^{-k/2}$, and $k>1$, the right hand side instead scales like $t^{-k}$. As discussed in \cite{Aoufia:2024awo}, one can interchange this asymptotic result with the sum by virtue of dominated convergence. Indeed, for the case of the partition function, the sum at fixed $t$ can be bounded by a $t$-independent contribution by bounding density of states according to $\rho(\Delta) \stackrel{\Delta \gg 1}\sim \Delta^{c-1}$, as required by modular invariance \cite{Afkhami-Jeddi:2020hde}. Similarly, here the absolute value of the sum can be bounded by the integral over the spectrum 
\begin{eqaed}\label{eq:I_z_bound}
    I_Z=\tau_2^{1+\frac{c}{2}}\int_{0}^\infty \td {\overline Q}\,\int_{\overline Q^2}^\infty \td \Delta  \,  \rho(\Delta,\overline Q) \Delta^p \overline Q^{2q} e^{-2\pi \tau_2 \Delta} \, , \quad p,q \in \bZ_{\ge 0} \, ,
\end{eqaed}
with $\rho(\Delta, \overline Q)$ the density of states; here we used the technical assumption that derivatives of charges and conformal weights grow at most polynomially in themselves. Due to the bootstrap bounds on the spectrum, the integral is bounded according to
\begin{eqaed}\label{eq:I_z_bound2}
 I_Z \le \frac{\Gamma(n)}{(2\pi)^n \tau_2^{n-\frac{c}2}} \, , \quad n =3+2q+p+c \, ,  
\end{eqaed}
which is $t$-independent and finite. The same reasoning as above can also be used for the heavy spectrum, since its contribution vanishes term by term in \cref{eq:termbytermoperator} due to the exponential suppression $e^{-2\pi \Delta}$, independently of the charges insofar as they are not exponentially growing in the conformal weights. We can thus claim that
\begin{eqaed}\label{eq:gauge_eq_gravi}
    (\cD_c+w_c)\left(\tau_2 \,\cF_{c}^{(s)}\right) \sim \Delta_\tau \left(\tau_2 \,\cF_{c}^{(s)}\right) \, .
\end{eqaed}
The subleading corrections to \cref{eq:gauge_eq_gravi} are mainly due to heavy states or light, parametrically non-extremal states, i.e. for which $\Delta \gg \overline Q$; this also establishes that if the spectrum features (additive) towers which conformal dimensions scale in a subleading fashion, they will contribute to the right hand side as $t^{-a}$,\, $0<a<1$. For a discussion of more general spectra, we defer the reader to \cref{app:additive_towers_anisotropic_limits}. To see how the above discussion might apply to an explicit example, consider the compact boson described above, with $\overline Q=p_{\text{R}}$. The (reduced) partition function satisfies the Narain differential equation \cite{Maloney:2020nni}
\begin{eqaed}\label{eq:narain_eq_circle}
    \left( \tau_2^2 \left( \frac{\partial^2}{\partial \tau_1^2} + \frac{\partial^2}{\partial \tau_2^2} \right) + \tau_2 \frac{\partial}{\partial \tau_2} - \frac{1}{4} \left( R \frac{\partial}{\partial R} \right)^2 \right) \Gamma_{1,1} = 0 \, .
\end{eqaed}
Given the explicit form of the partition function, namely 
\begin{eqaed}\label{eq:Z_circle}
   \Gamma_{1,1}(\tau;R)= \sum_{n,w \in \bZ} e^{2\pi i \tau_1 n w-\pi\tau_2 \left( \frac{n^2}{R^2}+w^2R^2\right)} \, , 
\end{eqaed}
we can apply the same differential operator to $ \langle  \tau_2p_{\text{R}}^2 - 1/4\pi\rangle$, arriving at
\begin{eqaed}\label{eq:Z_circle_charge_eq}
    \left( \tau_2^2 \left( \frac{\partial^2}{\partial \tau_1^2} + \frac{\partial^2}{\partial \tau_2^2} \right) + \tau_2 \frac{\partial}{\partial \tau_2} - \frac{1}{4} \left( R \frac{\partial}{\partial R} \right)^2 \right) \left\langle \tau_2 p_{\text{R}}^2 - \frac{1}{4\pi}\right\rangle = \sum_{n,w \in \bZ} A \, w \, e^{-2\pi w^2 R^2} \, ,
\end{eqaed}
where $A=A(\tau; n, w)$ is a function of winding numbers and momenta. Notice that for states that are light at large $R$, which have $w=0$, the right hand side indeed vanishes; in this case all such states are extremal. The contribution of heavy states is instead exponentially suppressed as expected. Thus, the right hand-side asymptotically vanishes by dominated convergence. As we have remarked at the beginning of the section, one can follow the same procedure with $\overline Q=0$ in order to derive \cref{eq:asymptotic_diff_eq}. \\

\noindent We are now able to follow the procedure described in \cref{app:exp_suppression} and derive an asymptotic differential equation for the string-frame one-loop graviphoton gauge coupling. Specifically, the equation will be satisfied by $A_c^{(s)}(\tau;t) \equiv \tau_2 \cF_c^{(s)}$, whose regularization requires to subtract off $E_{1+c/2}$ due to the extra $\tau_2$ factor. Accordingly, the factor $B^{(s)}$ (analogous to that of \cref{eq:f2_btimesz}) now scales like $\tau_2^{-c/2}$ at the cusp. Once again we decompose the regularized product similarly to \cref{eq:widetildebz}, namely
\begin{eqaed}\label{eq:wide_bz_graviph}
 \widetilde{B^{(s)} A_c^{(s)}} &=   B^{(s)}(\tau) A(\tau;t) - \hat E_1(\tau)\\
 &=\left(\tilde B(\tau) + j_1^{(s)} \tau_2^{-\frac{c}2}(\tau) \right) \left( \tilde A_c^{(s)}(\tau;t) + k^{(s)}E_{1+\frac{c}2}(\tau)\right) - j_0^{(s)}\hat E_1(\tau) \, , 
\end{eqaed}
where we still need to subtract $\hat E_1$ since the overall scaling of the product at the cusp does not change. Computations proceed as before, with the crucial difference that the boundary term in \cref{eq:integral-1} now reads
\begin{eqaed}\label{eq:boundary_term_grph}
    2j\lim_{L\to\infty} \int_{-\frac12}^\frac12 \td \tau_1 \left( \tau_2^{-\frac{c}2}\partial_{\tau_2}E_{1+\frac{c}{2}}\right)_{\tau_2=L} = j(c+2) \, .
\end{eqaed}
Repeating the same steps, one is thus led to 
\begin{eqaed}\label{eq:dc_widebz_grph}
   \cD_c \int_{\cF}\td \mu\, \sum_s\widetilde{B^{(s)} A_c^{(s)}} \sim -j (1+c) + \sum_s\int_{\cF}\td \mu \, (\Delta_\tau -w_c) B^{(s)} A_c^{(s)} \, .
\end{eqaed}
The last term can be expanded by rewriting it as $(\Delta_\tau -w_c+c)\hat B^{(s)} - cB^{(s)}$, where the extra contribution comes from the zero-mode which now scales as $\tau_2^{-c/2}$. All in all, using the fact that the term involving $\hat B$ scales like $t^{c/2}$, we have 
\begin{eqaed}\label{eq:dc_gauge_grph}
    (\cD_c+c) \, \Delta_g \sim At^{\frac{c}{2}} -j(1+c) \, ,
\end{eqaed}
where $\Delta_g$ is defined by the appropriate regularization we used in the preceding steps. The relevant solution is
\begin{eqaed}\label{eq:gauge_grph_sol}
    \Delta_g(t) = f_1 \, t^{\frac{c}{2}} + f_2 + \frac{f_3}{t} \, ,
\end{eqaed}
where the homogeneous solution proportional to $t^c$ is excluded by an argument using the Cauchy-Schwarz inequality, similarly to the above paragraph. \\

\paragraph{Comparision with the \ac{EFT}.} We are now in the position to compare our results, derived for a generic internal worldsheet \ac{CFT} and without assuming a geometric compactification (or supersymmetry), with the \ac{EFT} expectations described in \cref{sec:gauge_couplings_higher_derivative_corrections}. For four-dimensional gauge couplings, due to their marginality, the one-loop coefficient \cref{eq:string_vs_einstein_coeff} in the Einstein frame simply reads
\begin{eqaed}\label{eq:c_02}
    c_{0,2}=\frac{a_{0,2}^{(0)}}{g_{s,d}^2} + a_{0,2}^{(1)} \, .
\end{eqaed}
Using the fact that at fixed higher-dimensional string coupling $g_{s,D}$ the tree-level term scales like a volume factor, together with $t^{c/2}=\cV$, one has that asymptotically
\begin{eqaed}\label{eq:gauge_sol_all}
    c_{0,2} \sim \begin{cases}
        f_1 \cV + f_2 \log \cV \quad \text{for photons} \, ,\\
        f_1 \, \cV + f_2+f_3 \, \cV^{-\frac2c} \quad \text{for graviphotons} \, . 
    \end{cases}
\end{eqaed}
In the case of graviphotons, a further rescaling of the field is needed in order to make the charge of each \ac{KK}-mode equal to its level (c.f. \cref{eq:compact_boson_kk_momentum} for the case of the compact boson) and give the field mass dimensions befitting a gauge (local) one-form. Specifically, performing the rescaling $A \to M_s^{-1} \cV^{1/c} \, A $, the suitably normalized gauge coupling for graviphotons reads
\begin{eqaed}\label{eq:c_02_grph}
    c_{0,2}^{\text{graviphoton}} \sim f_1 \cV^{1+\frac2c} + f_2 \cV^{\frac2c}+ f_3 \, .
\end{eqaed}
Comparing with the case of circle compactifications in \cref{eq:graviphoton_coupling}, we see that the subleading terms precisely match the \ac{EFT} computation provided $k=1/6$. More generally, the leading scaling in \cref{eq:c_02_grph} matches the \ac{EFT} for any $c$, since it comes from summing charged-state contributions of the form
\begin{eqaed}\label{eq:EFT_graviphoton_c}
    \sum_{\mathbf{n} \in \bZ^c \, : \, \abs{\mathbf{n}} \lesssim N_\text{sp}^{1/c}} \abs{\mathbf{n}}^2 \overset{N_\text{sp} \gg 1}{\sim} \sum_{n \lesssim N_\text{sp}^{1/c}} n^{c-1} \, n^2 \sim N_\text{sp}^{\frac{c+2}{c}} \sim \cV^{1+\frac{2}{c}} \, .
\end{eqaed} \\

\noindent Having explored in detail the case of gauge coupling in four dimensions using the background-field method, we shall now turn to more general higher-derivative corrections of the type of \cref{eq:higherdercorr}. We first apply the background-field method to the computation of correlators of the type $\langle R^n F^m \rangle$, and then we present another derivation by extracting the (scalings of the) coefficients from string scattering amplitudes without background fields.

\subsection{Higher-derivative corrections in species limits}\label{sec:higher-derivative_corrections_species_limits}

The above discussion shows that the one-loop correction to gauge couplings in infinite-distance limits is universal, and behaves as in decompactification limits. This further supports the \ac{ESC} and provides a connection between \ac{IR} and \ac{UV} data, which we will explore in the next section. This result is analogous to the universal behavior of the vacuum energy found in \cite{Basile:2024lcz}, which relates it to the species scale \cite{Aoufia:2024awo} computed from $R^4$ terms. This prompts the question of whether these generic scaling relations persist throughout the whole low-energy effective action. In this section we answer this question in the affirmative for infinitely many higher-curvature operators involving the Riemann and Maxwell field strengths, namely those of the form $R^n F^m$. The specific tensor contractions are irrelevant for our derivation; however, \emph{a priori} it is possible that Wilson coefficients pertaining to some special linear combinations of tensor contractions behave differently, since the generic leading scalings may in principle cancel out. We expect such terms, if any, to comprise a negligible fraction of all the $R^n F^m$ terms in the effective action.

\subsubsection{Scalings from the background-field method}\label{sec:bkg_field_method_all-orders}

For generic curvature terms without covariant derivatives, we can expand upon the analysis in \cref{sec:gauge_couplings_species_limits} using the background-field method. Insofar as the gauge sector is not part of the decompactifying \ac{CFT}, the full generating function one-loop Wilson coefficients of the form $a_{n,m} = \langle R^n F^m \rangle$ takes the schematic form
\begin{eqaed}\label{eq:bkg_field_method_Z}
    \sum_{n, m} a_{n,m}(t) R^n F^m = \int_{\cF} \frac{\td^2\tau}{\tau_2^2} \, Z_W(\tau,\mu; F, R;t) \, ,
\end{eqaed}
where the partition function $Z_W$ in \cref{eq:generating_Z} introduces the constant gauge-gravitational background in the moduli-dependent partition function of \cref{eq:Zw_t-explicit}. Strictly speaking, this background occupies four of the $d \geq 4$ spacetime dimensions that are present in general; once more, we do not expect this to modify the generic scaling of Wilson coefficients. Since the internal sector and the background dependence factorize, we can apply once again the asymptotic differential equation discussed in the preceding section. The leading-order scaling $\mathcal{Z}_\text{int}^{(s)} \sim t^{c/2}$ then leads to the same universal behavior for all the Wilson coefficients encoded in \cref{eq:bkg_field_method_Z}, possibly up to some vanishing prefactors. The same result can be reached by taking the expression for the correlator in \cref{eq:fmrn_bkg_field} and following the procedure laid out in \cref{sec:gauge_couplings_species_limits} above. As we have discussed in \cref{sec:gauge_couplings_higher_derivative_corrections}, this volume-like scaling arises from the dimensional reduction of higher-dimensional terms, while the other contributions to $a_{n,m}(t)$ will depend on $n,m$. Up to threshold logarithms in $d$ or $D=d+c$ dimensions, which we have discussed in \cite{Aoufia:2024awo} and for gauge couplings, the \ac{EFT} expectation is that \cite{Calderon-Infante:2025ldq}
\begin{eqaed}\label{eq:wilson_coeff_EFT_expectation}
    a_{n,m}(t) \sim a_{n,m}^\text{string} \, t^{\frac{c}{2}} + a_{n,m}^\text{KK} \, t^{n+m-\frac{d}{2}} \, .
\end{eqaed}
In order to derive this more complete scaling for all higher-curvature terms of this form, we now turn to the individual one-loop scattering amplitudes.

\subsubsection{Scalings from scattering amplitudes}\label{sec:scattering_amplitudes_all-orders}

Our strategy is to extract $R^n F^m$ Wilson coefficients as contact terms in the $(n+m)$-point scattering amplitude with $n$ gravitons and $m$ gauge bosons. Letting $V^{(g)}_{\epsilon}(z) \, e^{ip \cdot X}, V^{(\gamma)}_{\epsilon}(z) \, e^{ip \cdot X}$ denote the corresponding vertex operators\footnote{Here and in the following we work in the $(0)$ (heterotic) or $(0,0)$ (RNS-RNS) ghost number pictures. In practice, this means that we have performed the integration over the odd (Grassmann-valued) moduli and work with integrated vertex operators.}, the relevant amplitude (stripped of its momentum-conserving Dirac distribution) takes the form
\begin{eqaed}\label{eq:n+m-point amplitude}
    \mathcal{A}_{n,m} = \int_{\cF} \frac{\td^2\tau}{\tau_2^2} \left\langle \prod_{a=1}^n \int_{T^2_\tau}\td^2z_a V^{(g)}_{\epsilon_a}(z_a) \prod_{b=1}^m \int_{T^2_\tau}\td^2z_b V^{(\gamma)}_{\epsilon_b}(z_b)  \right\rangle_\text{ext} \cdot \mathcal{Z}_\text{int}(\tau;t) \, ,
\end{eqaed}
where the sum over sectors $\sum_s r(s)$ is implicit in the dot-product notation, since it will not affect our analysis. We also denote $z_{ij} \equiv z_i - z_j$ the relative position of insertions and the bosonic propagator on the torus by $G(z)$, which appears in the universal Koba-Nielsen factor contained in the \ac{CFT} correlator of \cref{eq:n+m-point amplitude},
\begin{eqaed}\label{eq:koba-nielsen_factor}
    \text{KN} \equiv \prod_{i < j} e^{i p_i \cdot p_j G(z_{ij})} \, .
\end{eqaed}
In order to compare the one-loop contribution to the tree-level contribution, we have not included the usual factor of $g_{s,d}$ in the normalization of vertex operators, which implements unitarity. \\

\noindent To proceed with the low-energy limit, one must exercise care due to the possibility of various \ac{IR} divergences. The desired contact term is the one with the most powers of momenta, namely $p^{2n+m}$, which is accompanied by the integral of a suitable modular function. This term may be isolated according to its behavior with respect to $\tau$, as we shall see shortly, or by recursively subtracting the lower-order terms (in particular, if any, poles) arising from lower-point amplitudes. In fact, our strategy allows us to obtain the correct scalings for all \ac{IR} terms in \cref{eq:n+m-point amplitude}, including poles. To see this, we classify the possible structures that can appear in the correlator of vertex operators, either in RNS-RNS or heterotic cases. Here and in the following we largely follow the notation of \cite{Berg:2016wux, Berg:2016fui}. In the relevant ghost number picture, graviton vertex operators contain terms such as $\partial X$ and $(p \cdot \psi)\psi$ for left-movers and right-movers, where $X$ and $\psi$ denote worldsheet scalars and spinors in the external sectors. Similarly, vertex operators for gauge bosons carry factors of Kac-Moody currents $J$ and $\bar{J}$, which must Wick-contract among themselves. The left-moving and right-moving sectors factorize up to contractions
\begin{eqaed}\label{eq:dXdbarX}
    \langle \partial X(0) \overline{\partial}X(z)\rangle = \frac{\pi}{\tau_2} + \delta(z) \, .
\end{eqaed}
As we have seen, and discussed in \cite{Basile:2024lcz} (cf. \cref{app:exp_suppression} for more details), the leading and subleading scalings in the large-$t$ limit is controlled by the large-$\tau_2$ behavior of the external-sector correlator, i.e. its \ac{IR} behavior at the cusp of moduli space. Hence, since the volume of the torus scales as $\tau_2$, contractions of the above type do not change this behavior, and effectively leave one less insertion in the Wick contractions. As such, we can ignore these terms and look at left-moving and right-moving contractions separately. Contractions of $X$ and $\psi$ fields lead to derivatives of propagators $\partial^2 G$ and the Szeg\"{o} kernel $S_s$ \cite{Berg:2016fui}. The latter decays exponentially at the cusp, and thus we can focus on terms involving $\partial^2G$. Finally, contractions of currents produce combinations of structure constants. These comprise the Wick contractions that do not affect the Koba-Nielsen part of vertex operators; then, each of the $\partial X$ insertions can contract with an exponential to yield combinations of terms of the form $p \cdot \partial G$. All in all, we see that the maximal power of momenta that can appear in the correlator is $p^{2n+m}$, since there are $n$ left-moving and right-moving such terms from graviton insertions, and $m$ terms from gauge-boson insertions (either in the left-moving or right-moving sector, depending on conventions). \\

\noindent With these building blocks at hand, we can begin our analysis of the \ac{IR} behavior of the integrand. The general term takes the schematic form
\begin{eqaed}\label{eq:schematic_term}
    (\partial^2 G)^{a_\text{L}} (p \cdot \partial G)^{b_\text{L}} (\overline{\partial^2 G})^{a_\text{R}} (p \cdot \overline{\partial G})^{b_\text{R}} \times \text{KN} \, ,
\end{eqaed}
which brings a power of momenta $p^{b_\text{L}+b_\text{R}}$. Further expanding the Koba-Nielsen factor produces terms of the form $(p \cdot p \cdot G)^k$, leading to an additional power of $p^{2k}$. The structure of the amplitude, specifically the number of graviton and gauge-boson legs, fixes
\begin{eqaed}\label{eq:momentum_powers}
    & 2a_\text{L}+b_\text{L} =  n+m \, , \\
    & 2a_\text{R}+b_\text{R} = n \, .
\end{eqaed} \\

\begin{figure}[!t]
    \centering  \includegraphics[width=0.82\linewidth]{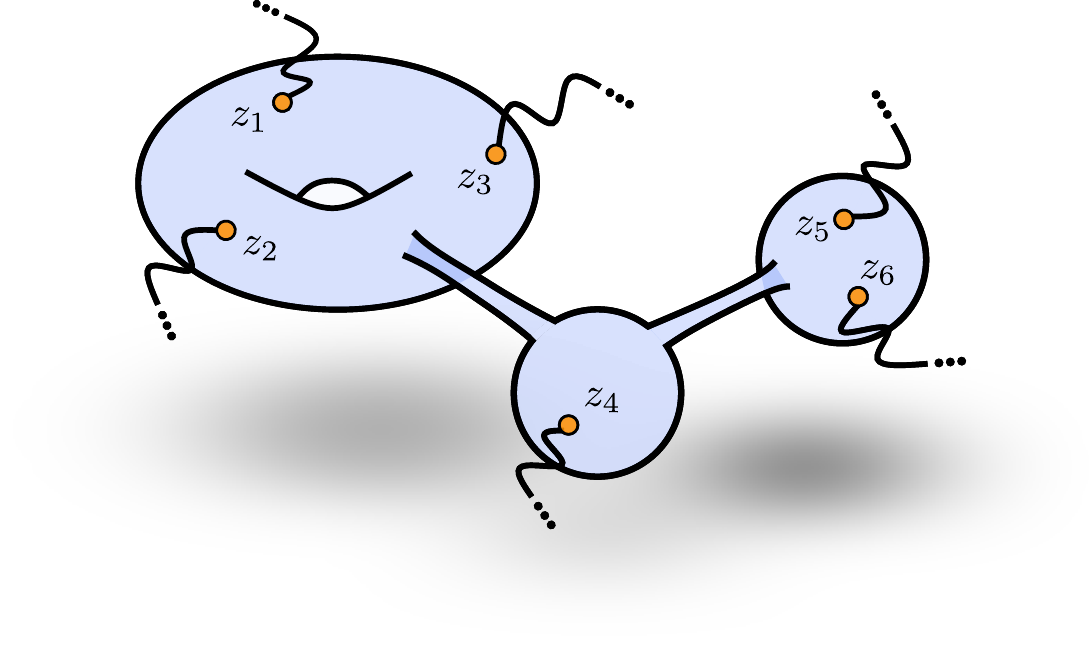}
    \caption{A complete factorization channel for one-loop closed-string amplitudes, which can be built via the sewing construction of \cite{Sonoda:1988mf, Sonoda:1988fq}. These configurations in worldsheet moduli space dominate the low-energy limit of the amplitude in the presence of kinematic poles. These are produced by massless states propagating in each thin long tube of the worldsheet. In particular, the depicted half-ladder arrangement leaves the maximum possible number of legs on the torus at fixed pole order.}
    \label{fig:factorization}
\end{figure}

\paragraph{Kinematic poles.} When expanding the integral for small momenta $p_i \to 0$, the amplitude may develop massless poles. These arise from exchanges of massless particles, and arrange into the simple poles of a maximal factorization channel (namely, a channel featuring the maximum number of poles), as depicted in \cref{fig:factorization}. A conformally equivalent picture is that these singularities are associated with collisions of vertex-operator insertion points. In addition, the amplitude itself may diverge in the presence of tadpoles if the one-loop vacuum energy does not vanish; they are universal, reflecting the fact that the classical vacuum we started from does not persist beyond tree level, and can be subtracted as explained above\footnote{Similarly, \ac{IR} divergences in four dimensions causes by undressed Fock states do not modify the moduli dependence we are interested in.}. In the presence of massless poles, the replacement $\text{KN} \to 1$ is invalid. Instead, at the level of the integrand, these poles arise from collisions of insertion points\footnote{Here we only treat two-point collisions explicitly. Multiple collisions can be in principle treated as limits of successive collisions, leading to a complete factorization pattern as depicted in \cref{fig:factorization}.}, where the worldsheet function listed above develop singularities of the form $\abs{z}^{-w}$ (up to angular dependence in $z/\overline{z}$). As a result, including a contribution $\abs{z}^s$ from the Koba-Nielsen factor, the local integral around a collision takes the form
\begin{eqaed}\label{eq:collision_integral}
    \int\td^2z \, \abs{z}^{s-w} \sim \frac{\pi}{s-w+2} \, ,
\end{eqaed}
where the factor of $\pi$ arises from quotienting the complex plane to obtain the fundamental cell of the torus\footnote{In more detail, taking into account that $\td^2z = 2\td x \td y$ for $z=x+iy$, the factor of $1/4$ relative to the result in the complex plane arises either by covering the origin with four fundamental cells, or by counting the contribution of a disc around the corner of the fundamental cell modulo lattice translations.}. This is a tachyonic pole for $w>2$, which means that these terms must cancel in the final result, since we assumed that the physical spectrum is tachyon-free. For $w=2$, \cref{eq:collision_integral} yields a massless pole. As depicted in \cref{fig:factorization}, the maximal poles arise from the maximal number of such contributions, each of which is a simple pole as befits tree-level and \ac{IR} locality.

\paragraph{Behavior at the cusp.} As shown in \cite{Berg:2016wux, Berg:2016fui}, for generic values of their arguments (i.e. away from collisions) the large-$\tau_2$ behavior of each of the building blocks in \cref{eq:schematic_term} is $G \sim \tau_2$, while each derivative removes one power of $\tau_2$. As a result, away from collisions, the Koba-Nielsen can be expanded in momenta, producing terms of the form $(p \cdot p \cdot G)^k$, and the corresponding worldsheet integrand scales as $\tau_2^{k-a_\text{L}-a_\text{R}}$. Integrating over positions then gives an additional factor of $\tau_2$ per insertion point, while the total external partition function gives a factor $\tau_2^{1 - \frac{d+c}{2}}$ coming from the path integral over zero-modes and the normalization of the reduced internal partition function $\mathcal{Z}_\text{int}^{(s)} \sim \tau_2^{\frac{c}{2}}$. In order to express these exponents in terms of more physically meaningful quantities, let $p^q$ denote the power of momentum in the generic Wick contraction (including any expansion of the Koba-Nielsen factor). Then $q = b_\text{L} + b_\text{R}+2k$ and the total power of $\tau_2$ in the worldsheet integrand at the cusp is $\tau_2^{\frac{q-(2n+m)}{2}}$, where we used \cref{eq:momentum_powers}. The integral over insertions further yields a factor of $\tau_2$ for each insertion point excluding collisions. Hence, for poles of total order $w$, the behavior of the worldsheet integral (which here we denote $B(\tau)$) at the cusp, excluding the internal partition function $\mathcal{Z}_\text{int}$, is
\begin{eqaed}\label{eq:behavior_cusp}
 \tau_2^{\frac{c'}{2}} \, , \quad c' \equiv q - (2n+m) + 2(n+m-w) + 2 - (d+c) \, .
\end{eqaed}

\paragraph{Scalings at infinite distance.} Applying the method of \cite{Basile:2024lcz}, integrating the function $B(\tau)$ against the internal partition function $\mathcal{Z}_\text{int} \equiv t^{\frac{c}{2}}(A + F(\tau;t))$ produces a function of the modulus $t$ which satisfies the asymptotic differential equation
\begin{eqaed}\label{eq:general_ode_correction}
    (\mathcal{D}_0 - w_{c'}) \int_\cF \frac{\td^2\tau}{\tau_2^2} \, B(\tau) \cdot F(\tau;t) = 0 \, ,
\end{eqaed}
leading to the scalings $t^{\frac{c+c'-2}{2}}$ and $t^{\frac{c-c'}{2}}$ for the correction to the volume term encoded by $F$. Logarithmic thresholds in $D=d+c$ and $d$ dimensions are also included when $c'=2$ or $c+c'=2$ respectively, since the \ac{IR} regularization of the integral involves the ``anomalous'' Eisenstein series $\hat{E}_1$. For the highest contact term, where $w=0$ and $q = 2n+m$ in \cref{eq:behavior_cusp}, the first of the two terms in the general solution to \cref{eq:general_ode_correction} turns out to precisely the \ac{KK}-like term of \cref{eq:wilson_coeff_EFT_expectation}, while the latter is either subleading or unphysical, since it is larger than the volume scaling for $c' < 0$ and thus cannot appear. Indeed, this can be seen by applying the Cauchy-Schwarz inequality to the (regularized) integral in \cref{eq:general_ode_correction} seen as an inner product on $L^2(\cF, \td^2\tau/\tau_2^2)$, since $\int_\cF F = O(t^{-1})$ as obtained in \cite{Aoufia:2024awo} from the analysis of $R^4$ terms. In fact the analysis of \cite{Aoufia:2024awo} is a special case of \cref{eq:general_ode_correction}, where $n=4, m=w=0$ and thus $c' =10-d-c$. Similarly, the analysis of \cite{Basile:2024lcz} is a special case where $n=m=w=0$, so that $c'=2-d-c$. For a general contact term $R^n F^m$, we have $w=0$ and $c' = 2(n+m) + 2 - (d+c) $, which leads precisely to the expected \ac{KK}-like term in \cref{eq:wilson_coeff_EFT_expectation}. \\

\noindent As for pole terms, \cref{eq:behavior_cusp} shows that for a given total power of momenta $q$ the growth in $\tau_2$ decreased by a power of $-w$ for a pole of total order $w$. This is reflected in the \ac{KK}-like term of the solution to \cref{eq:general_ode_correction}, which should match the \ac{KK}-like scaling of the most dominant lower-order contact term compatible with the power of momenta and order of the pole. This would be a term of the form $R^{n'}F^{m'}$ with $n' \leq n$, $m' \leq m$, fixed $2n'+m' = q$ and maximal $n'+m'$, since the \ac{KK}-like term in the Wilson coefficient scales with the number of legs (which is proportional to the scaling dimension of the operator). This is a term of the form $R^{n-w}F^m$. For this term $q=2n+m-2w$, and thus the \ac{KK}-like term scales as if $w$ legs were removed from the operator; this is indeed what happens. To see this, up to special cancellations, a pole of total order $w$ is dominated by a factorization channel where the torus retains the highest power of momentum and number of legs, since the \ac{KK}-like term of \cref{eq:wilson_coeff_EFT_expectation} grows faster with increasing number of legs. This means that each external insertion on a sphere must take away one external leg and power of $R$ or $F$ from the torus insertion, ending up in a half-ladder factorization channel (as depicted in \cref{fig:factorization}). For a pole of order $w$ there are $w$ spheres with two external legs each (consistently with the sewing construction of \cite{Sonoda:1988mf, Sonoda:1988fq}), so that the total number of legs on the torus (including the one along the spine of the half-ladder) is
\begin{eqaed}\label{eq:legs_retained_torus}
    (n + m)_\text{total external} - (w+1)_\text{spheres external} + (1)_\text{spine} = n + m - w
\end{eqaed}
as predicted. Therefore, \cref{eq:wilson_coeff_EFT_expectation} holds for the \ac{IR} expansion of \cref{eq:n+m-point amplitude} including pole terms, with no need of subtracting them beforehand.

\section{UV/IR relations and the swampland}\label{sec:UV/IR_relations_and_swampland}

The upshot of the analysis in the preceding section is that Wilson coefficients of curvature operators $R^n F^m$ scale according to \cref{eq:wilson_coeff_EFT_expectation}, which in Planck units translates into \cref{eq:wilson_coeff_EFT_expectation_planck}. From a purely \ac{EFT} standpoint, it is \emph{a priori} non-trivial that no (\ac{UV}-sensitive) terms in between those discussed in \cite{Calderon-Infante:2025ldq} would enter and (possibly) dominate the asymptotics. This reflects both the \ac{ESC} and the \ac{UV}/\ac{IR} mixing present in string theory. For instance, the vacuum energy (density) of a $d$-dimensional \ac{EFT} in a decompactification limit to $D=d+c$ dimensions can take the schematic form
\begin{eqaed}\label{eq:EFT_V_generic}
    V_\text{EFT} \sim m_\text{KK}^d \, \sum_{k=0}^{D} a_{k} \left(\frac{M_{{\text{Pl}};D}}{m_\text{KK}}\right)^k 
\end{eqaed}
up to logarithmic terms; all but the last one vanish in the decompactification limit, leaving a generically Planckian higher-dimensional vacuum energy. The vanishing of additional (possibly dominant, or otherwise \ac{UV}-sensitive) terms is already suggestive of \ac{UV}/\ac{IR} effects. In fact, the analysis presented in the preceding section further reinforces this picture: the universal scalings of \cref{eq:wilson_coeff_EFT_expectation_planck} and \cref{eq:wilson_coeff_EFT_expectation} turn out to apply to \ac{IR} data as well, namely gauge couplings and the vacuum energy \cite{Basile:2024lcz}. While our technical results were derived for closed strings at one-loop order, we can argue that some parametric inequalities persist with the addition of higher-loop and/or open-string effects.

\subsection{Estimating higher loops}\label{sec:higher_loops}

In order to estimate the effects of higher-loop contributions, we can leverage our one-loop results, which extend those of \cite{Aoufia:2024awo, Basile:2024lcz}, and the findings of \cite{Ooguri:2024ofs}, which together establish that all species limits in the worldsheet conformal manifold are decompactifications. One can then expect that higher-loop contributions rearrange into the perturbative expansion of the higher-dimensional theory, which is weighted by the higher-dimensional string coupling $g_{s,D}^2 = g_{s,d}^2 \mathcal{V}\,$\footnote{Note that, when the $D$-dimensional background arises from a ten-dimensional compactification, the higher-dimensional string coupling $g_{s,D}$ is parametrically the same as the 10d coupling $g_{s,10}$, since they differ by $O(1)$ volume factors.}. Indeed, this follows by estimating the large-volume scaling of integrals of genus-$h$ Narain partition functions over the moduli space $\mathcal{M}_h$ of genus-$h$ Riemann surfaces\footnote{Strictly speaking, for sufficiently high loop order the integration over odd moduli cannot be ignored or packaged into picture-changing operators (see however \cite{Wang:2022aem} for recent developments). This is due to the fact that the supermoduli space of super-Riemann surfaces fails to be holomorphically split \cite{Witten:2012ga, Witten:2012bh, Donagi:2013dua}. Although we expect the leading volume-like scaling to be unaffected, we leave the investigation of these interesting aspects (including the behavior of the measure close to degenerations \cite{Witten:2013tpa}) to future work.}. This is because the methods we used to derive the asymptotic differential equation in \cite{Aoufia:2024awo}, based on differentiating the internal partition function with respect to $\tau$ and $t$, can fail when the worldsheet conformal structure $\tau$ is described by non-commuting matrices $\Omega \in \mathcal{H}_h$ in the Siegel upper-half plane. Nevertheless, by using the explicit form of the higher-genus Narain lattice sums $\cZ_{c,c}^{(h)} = (\det \mathrm{Im} \Omega)^{c/2} \, \Gamma_{c,c}^{(h)}$ (see, e.g., \cite{Maloney:2020nni}) we can make use of the partial differential equation
\begin{eqaed}\label{eq:genus-h_PDE}
    \left(\Delta_\text{Narain} + w_{c,h} - \Delta_\Omega \right) \cZ_{c,c}^{(h)} = 0 \, ,
\end{eqaed}
where $w_{c,h} \equiv \frac{hc}{2}\left(\frac{1+h}{2}-\frac{c}{2}\right)$ reduces to $w_c$ for $h=1$. Importantly, the Laplacian $\Delta_\text{Narain}$ on Narain moduli space is the same for all $h$; therefore, as in \cite{Aoufia:2024awo} and \cref{sec:higher-derivative_corrections_species_limits}, the integral of $\Gamma_{c,c}^{(h)}$ (eventually regularized by subtracting the $\text{Sp}(2h,\mathbb{Z})$ Eisenstein series \cite{Maloney:2020nni}) satisfies a differential equation analogous to the ones we have been discussing. \\

\noindent Let us consider the case $c=1$ for simplicity; then the identification of the volume direction $t = \alpha'r^2$ is trivial, and $\Delta_\text{Narain} = - \frac{1}{4} (r \partial_r)^2 = - (t \partial_t)^2$. This leads to
\begin{eqaed}\label{eq:genus-h_c=1}
    (t \partial_t)^2 \int_{\mathcal{M}_h} \td\mu_h(\Omega) \, \cZ_{1,1}^{(h)} = \frac{h^2}{4} \, \int_{\mathcal{M}_h} \td\mu_h(\Omega) \, \cZ_{1,1}^{(h)} \, ,
\end{eqaed}
whose general solution is a linear combination of $t^{\frac{h}{2}}$ and $t^{-\frac{h}{2}}$. This shows that, indeed, the generic leading scaling of the relative $h$-loop contribution to any Wilson coefficient, according to the considerations in \cref{sec:higher-derivative_corrections_species_limits}, is $(g_{s,d}^2 \mathcal{V})^{h} = g_{s,D}^{2h}$. For more general Narain sums with $c>1$ it is more complicated to extract the overall volume dependence. However, from the explicit expression (here $\Omega_{ij} \equiv x_{ij} + i y_{ij})$ \cite{Maloney:2020nni}
\begin{eqaed}\label{eq:narain_sum_genus-h}
    \Gamma_{c,c}^{(h)} = \sum_{\mathbf{n}, \mathbf{w} \in \mathbb{Z}^{h \times c}} e^{2\pi i x_{ij} n^i_p w^{jp} - \pi y_{ij} (G^{pq} n^i_p n^j_q + G_{pq} w^{ip}w^{jq})}
\end{eqaed}
we can perform a Poisson resummation to extract the large-volume behavior. In \cref{eq:narain_sum_genus-h} we neglected the $B$-field, since for our purposes we only need the metric $G$ on the $c$-torus. Neglecting the winding modes $w$, we see that Poisson resummation leads to a prefactor of $(\det G)^{\frac{h}{2}}$ which indeed corresponds to the expected dominant contribution $\mathcal{V}^h$. Then, the integration over $\mathcal{M}_h$ gives a factor of Weil-Petersson volume computable via Mirzakhani recursion \cite{Mirzakhani:2006eta, Mirzakhani:2006fta}. This grows factorially as $(2h)!$ (accompanied by a power law) \cite{Mirzakhani:2010pla}, as also found in \cite{Gross:1988ib, Baker:1988tw, Gross:1988tx} and consistently with the expected $O(e^{-1/g_{s,d}})$ non-perturbative behavior via resurgence. \\

\noindent All in all, if the perturbative series for our Wilson coefficients of interest resums, we argued that the leading large-volume scalings for $g_{s,D} \lesssim 1$ take the form
\begin{eqaed}\label{eq:resummed_wilson_coefficients}
    \sum_{h=0}^\infty a^{(h)}_{n,m}(t) \, g_{s,d}^{2h-2} \overset{t \to \infty}{\sim} \mathcal{V} \, \frac{f(g_{s,D})}{g_{s,D}^2} = \frac{1}{g_{s,d}^2} \, f(g_{s,D}) \, ,
\end{eqaed}
where we factored out the tree-level term so that (generically) $f(g_{s,D}) = O(1)$. Thus, at least the volume-like behavior is expected to persist even for $g_{s,D} = O(1)$. In order to study the strong-coupling behavior $g_{s,D} \gg 1$, one must likely resort to dualities (or input a swampland principle which accounts for them, see e.g. \cite{Bedroya:2023tch, Delgado:2024skw}), thus reproducing a (S-)dual weakly coupled description or a further decompactification to M-theory in $D+1=d+c+1$ dimensions. We leave these developments to future work.

\subsection{Parametric inequalities and phenomenology}\label{sec:parametric_inequalities}

The above considerations lead us to universal volume-like scalings, accompanied by the \ac{KK}-like terms of \cref{eq:wilson_coeff_EFT_expectation_planck}. On the one hand, the higher-curvature terms identify the species scale with the string scale, $\Lambda_\text{sp} = M_s$, so that
\begin{eqaed}\label{eq:string_scale_general}
    \frac{1}{g_{s,d}^2} = \frac{\mathcal{V}}{g_{s,D}^2} = \left(\frac{\Lambda_\text{sp}}{M_{{\text{Pl}};d}}\right)^{2-d} = N_\text{sp}
\end{eqaed}
is the number of species. Note that $\mathcal{V} \gtrsim 1$, since $\Delta_\text{gap} \lesssim 1$ as found in \cite{Hellerman:2009bu, Hellerman:2010qd}. On the other hand, the \ac{IR} gauge couplings $g$ and vacuum energy $V$ satisfy
\begin{eqaed}\label{eq:gauge_coupling_V}
    & V = a \, g_{s,D}^2 \, \Lambda_\text{sp}^2 \, M_{{\text{Pl}};d}^{d-2} + b \, m^d + \dots \, , \\
    &  g^2 = c \, \Lambda_\text{sp}^2 \, M_{{\text{Pl}};d}^{d-6} + \dots \, ,
\end{eqaed}
where the dots include subleading (possibly logarithmic) contributions and the neglected open-string terms (if any), and $a,b,c$ are (generically non-vanishing) constants. The first term of $V$ includes an additional factor of $g_{s,d}^2$ because the tree-level contribution vanishes\footnote{This is because we are evaluating the vacuum energy around flat spacetime. More generally, tree-level zero-point \cite{Eberhardt:2020bgq, Eberhardt:2021jvj, Eberhardt:2021ynh, Ahmadain:2022tew, Ahmadain:2022eso, Ahmadain:2022eso} and two-point \cite{Erbin:2019uiz, Giribet:2023gub} amplitudes may be non-trivial. At any rate, the results in this paper are not sensitive to these subtleties, since they concern the (universal asymptotics of the) prefactors.}. Still, this term is parametrically too large for a \emph{bona fide} \ac{EFT} description, since in the limits we consider $g_{s,D}$ does not vanish sufficiently quickly. Thus, to discuss cases when it cancels\footnote{See \cite{Basile:2024lcz} for a discussion of the phenomenological implications of $a=0$ in \cref{eq:gauge_coupling_V} at various loop orders.} we explicitly kept the \ac{KK}-like (or Casimir-like) term in $V$, whose universal presence was found in \cite{Basile:2024lcz}; this term is present also for weakly coupled string limits, with the mass gap $m = M_s$ rather than $m = m_\text{KK}$ and the two terms in \cref{eq:gauge_coupling_V} scale in the same fashion. As for the gauge coupling, while it is possible that $c=0$ \cite{Kiritsis:1998en}, the generic case $c \neq 0$ does not necessarily endanger the validity of the \ac{EFT}. \\

\noindent In order to account for the neglected terms in \cref{eq:gauge_coupling_V}, we argue for parametric inequalities for $V$ and $g$ as follows. The presence of the Casimir-like term $m^d$ in the vacuum energy is rather robust, since $b=0$ requires unbroken either spacetime supersymmetry, super-no-scale settings \cite{Angelantonj:2003hr, Kounnas:2016gmz, Florakis:2021bws, Florakis:2022avh, Dudas:2025yqm}\footnote{It is worth noting that, although non-generic, super-no-scale models offer an intriguing scenario to realize numerical scale separation \cite{Abel:2024vov}. Other constructions of this type include those of \cite{Demirtas:2021nlu, Demirtas:2021ote} (see, however, \cite{Lust:2022lfc, Bena:2024are, Cribiori:2026btb}).}, or other constructions that remove the whole one-loop contribution \cite{Kachru:1998hd, Larotonda:2026hxy}. Moreover, even if $b=0$ at one-loop order, an analogous contribution is expected to arise at higher loops on dimensional grounds. Therefore, unless the neglected (open-string) contributions are finely tuned to cancel both terms in \cref{eq:gauge_coupling_V}, one obtains $V \gtrsim m^d$, where in general $m = m_\text{KK} \text{ or } M_s$ is the cutoff of the $d$-dimensional \ac{EFT}. As for gauge couplings, we do not expect additional contributions to the inverse gauge couplings $g^{-2}$ to overwhelm the maximal volume-like scaling. For instance, the contribution to the (bulk) gauge couplings arising from D-branes brings along the volumes of wrapped internal cycles; in order to overwhelm the total volume, the orthogonal cycles would need to be much smaller than the string length, likely violating the \ac{CFT} bootstrap bounds on the spectral gap \cite{Hellerman:2009bu, Hellerman:2010qd}. This argument is in the spirit of \cite{Collins:2022nux} and, in the context of moduli stabilization, has bearing on the issue of parametric scale separation (see \cite{Coudarchet:2023mfs} for a review, and \cite{Bedroya:2025ltj, Bedroya:2025fie, Apers:2026lgi, Cribiori:2026btb} for recent developments). We expect analogous arguments to be valid other kinds of contributions we neglected, such as non-perturbative effects and fluxes, but we leave a more systematic investigation of these aspects to future work. \\

\noindent The upshot of the above considerations is that the unknown terms in \cref{eq:gauge_coupling_V} can be accounted for by the parametric inequalities
\begin{eqaed}\label{eq:parametric_inequalities_IR}
    & V \gtrsim m^d \, , \\
    & g^2 \gtrsim \Lambda_\text{sp}^2 M_{{\text{Pl}};d}^{d-6} \, .
\end{eqaed}
The latter inequality implies that the species scale be bounded by the gauge coupling(s) in Planck units, which takes the form of a (parametric) magnetic weak-gravity bound \cite{Arkani-Hamed:2006emk}, which follows from the completeness principle \cite{Polchinski:2003bq} applied to its electric counterpart (by now almost fully established at the worldsheet level \cite{Heidenreich:2024dmr}). The former inequality is the essential ingredient in the derivation of the dark dimension scenario \cite{Montero:2022prj, Anchordoqui:2025nmb}, and is therefore important to place it on firmer and more general grounds as we have done in this paper. This includes the extended analysis of \cref{app:additive_towers_anisotropic_limits}, which shows that \cref{eq:parametric_inequalities_IR} is robust and the Casimir-like term $m^d$ is generically present. When multiple, subleading light towers are present, their (otherwise generically dominant) contributions are incompatible with observational constraints on (the effective number of) mesoscopic extra dimensions \cite{Montero:2022prj, Anchordoqui:2025nmb}; nevertheless, were they present, \cref{eq:parametric_inequalities_IR} would still hold. While it is still unclear how to reliably decouple the supersymmetry-breaking and Kaluza-Klein scales (see, however, \cite{Dudas:2025yqm} for a one-loop realization), our point is that \cref{eq:parametric_inequalities_IR}, combined with observational data, favors this scenario regardless of what mechanism achieves this decoupling. Moreover, the presence of unstabilized fields may be compatible with dynamical dark energy according to the model of \cite{Bedroya:2025fwh}. According to \cref{eq:gauge_coupling_V}, the dark dimension scenario would lead to a very weak bound on the (electromagnetic) gauge coupling; indeed, the one-loop estimate of the threshold would set it around $10^{-20}$, far from a realistic value. However, in order to avoid detection of \ac{KK} replicas of the standard model, the gauge sector ought to be localized\footnote{The localization of the standard-model brane has interesting phenomenological consequences for the dark sector \cite{Gonzalo:2022jac, Anchordoqui:2022txe, Anchordoqui:2024dxu} and, possibly, grand unification \cite{Heckman:2024trz}.}, and thus one expects that the relevant gauge couplings be set by $g_{s,D}$ rather than $g_{s,d}$. Alternatively, it could be set by localized Neveu-Schwarz branes in a little-string scenario along the lines of \cite{Antoniadis:2001sw} (see also \cite{Lust:2008qc, Anchordoqui:2009mm, Basile:2024lcz}). \\

\noindent The robust nature of the inequality $V \gtrsim m^d$ can be ascribed to holographic bounds. Indeed, it can be interpreted as the \ac{CKN} bound \cite{Cohen:1998zx} applied to the \ac{EFT} with the Hubble scale $M_{{\text{Pl}};d}^{d-2}H^2 = V$ as \ac{IR} scale and $m$ as \ac{UV} scale. Moreover, using that $\Lambda_\text{sp} \lesssim m^{\frac{1}{d-1}}$ in $d$-dimensional Planck units\footnote{This follows from the fact that for decompactifications $\Lambda_\text{sp} = m^{\frac{c}{d+c-2}} \lesssim m^{\frac{1}{d-1}}$ in $d$-dimensional Planck units, and weakly coupled string limits are accounted for by formally sending $c$ to infinity. The inequality still holds when the string coupling $g_{s,D}$ is included in these expressions, as well as for anisotropic multiple decompactifications (cf. \cref{app:additive_towers_anisotropic_limits}). Amusingly, maximizing the denominator in the exponent using $d+c \leq 11$ leads to the weaker bound $\Lambda_\text{sp} \lesssim \abs{V}^{\frac{1}{36}}$ which is roughly the expected scale where putative grand-unification scenarios take place.}, one finds
\begin{eqaed}\label{eq:species_scale_holographic_bound}
    \Lambda_\text{sp} \lesssim \abs{V}^{\frac{1}{d(d-1)}}
\end{eqaed}
in $d$-dimensional Planck units, which happens to be the \ac{CEB} bound \cite{Bousso:1999xy} applied to the \ac{EFT} with \ac{IR} scale $V^{\frac{1}{d}}$ and \ac{UV} scale $\Lambda_\text{sp}$. These holographic bounds have received recent interest in the context of cosmology \cite{Andriot:2025cyi, Cribiori:2025oek}, and are both saturated by the dark dimension scenario. Here we derived them from our worldsheet analysis, extending the results of \cite{Aoufia:2024awo, Basile:2024lcz} into a more complete picture of how swampland constraints arise from the \ac{UV}/\ac{IR} mixing inherent in string theory.

\section{Conclusions}\label{sec:conclusions}

The analysis in this paper significantly extended the scope of \cite{Aoufia:2024awo, Basile:2024lcz}, deriving generic scaling relations for infinitely many higher-derivative operators of gauge-gravitational effective actions. The generality of our approach hinges on two main factors. Firstly, the worldsheet formulation allows us to assign gravitational \acp{EFT} in spacetime to internal worldsheet \acp{CFT}, and thus by studying universal properties of the latter we can glean insights about the former in any \ac{IR} phase\footnote{At least any \ac{IR} phase accessible from the worldsheet. However, as we have argued in \cref{sec:UV/IR_relations_and_swampland}, some of our results are expected to apply to $O(1)$ string coupling, and perhaps to strong coupling via dualities---although non-supersymmetric settings are more elusive in this respect.}. Secondly, we only look at parametrics and generic terms; the numerical prefactors are $O(1)$ in any available limit, except when they (non-generically) vanish. There are several lessons we can draw from our findings, as well as implications to discuss in relation to future work. \\

\noindent \paragraph{Infinite-distance limits versus species limits.} Given the parametric nature of our analysis, its results trivialize away from species limits, where they reduce to the tautology that, absent any limit, all quantities are simply $O(1)$. The non-trivial scalings arising in species limits, however, establish the \ac{ESC} of \cite{Lee:2019wij} at the worldsheet level. Once again, we remark that this is a non-trivial statement, since non-geometric vacua exist; here we extended the results of \cite{Ooguri:2024ofs, Aoufia:2024awo, Basile:2024lcz} to the effect that any limit in the worldsheet conformal manifold are decompactifications. To connect to the swampland distance conjecture of \cite{Ooguri:2006in}, the work of \cite{Ooguri:2024ofs} translated to the worldsheet setting implies that species limits lie at infinite distance, since they require a vanishing spectral gap \cite{Aoufia:2024awo} and bring along an infinite tower of light species. The considerations of \cite{Stout:2022phm} provide evidence for the converse proposition, but a general proof is still lacking and provides a clear avenue for future research. \\

\noindent \paragraph{The geometry of field spaces.} Related to the preceding point, our approach does not directly probe the metric of scalar field space, encoded in the two-derivative terms of the scalar effective action. Actually, this is the only missing sector of the (bosonic\footnote{The fermionic sector, while interesting and necessary to remove physical tachyons \cite{Angelantonj:2023egh, Leone:2023qfd, Leone:2024xae} (see also \cite{Leone:2025mwo} for a recent review), is less relevant for the (semiclassical) gauge-gravitational physics taking place at very long distance scales. The fermionic sector is also naturally more complicated to deal with at higher-loop order, due to the integration over odd moduli of super-Riemann surfaces \cite{Witten:2012ga, Donagi:2013dua, Witten:2012bh} as well as the construction of an appropriate measure in the presence of Ramond punctures \cite{Donagi:2024bbe}.}, local part of the) \ac{IR} data; from the worldsheet perspective, this is given by the Zamolodchikov metric \cite{Zamolodchikov:1986gt} on the conformal manifold, whose detailed calculation depends on the internal sector \cite{Cvetic:1989ii, Friedan:2012hi}. Unsurprisingly, a sufficient amount of unbroken spacetime supersymmetry simplifies matters significantly, allowing to express the corresponding K\"{a}hler potentials in terms of suitable worldsheet partition functions on the sphere \cite{Gerchkovitz:2014gta}, whose calculation can be amenable to localization. Beyond specific examples, for our purposes of generality we can use the results of \cite{Ooguri:2024ofs} to infer that in normal coordinates $(t, \varphi)$ the line element should take the form $\td s^2 = \td t^2/t^2 + f(t) \, g_{ij}(\varphi) \, \td\phi^i \td\phi^j$ with ${f(t) \sim t^{-\alpha} \to 0}$ at large $t$. This is because the \ac{CFT} spectral gap vanishes exponentially in the Zamolodchikov distance\footnote{Note that, from a bottom-up perspective, this universal scaling in terms of a parameter setting a physical scale (such as $t$) was argued for on information-theoretic grounds \cite{Stout:2022phm}.}. The dichotomy of infinite-distance limits dictated by the \ac{ESC}, as confirmed herein, tells us that the first term is just the kinetic term for the radion, which is accompanied by a dilaton dependence which is fixed by string perturbation theory. It would be interesting to assess our expectation for the a-priori model-dependent function $f(t)$, and in particular whether something can be learned without any further input on the conformal manifold (except that its dimension be larger than one); perhaps universal properties of conformal blocks \cite{Perlmutter:2015iya} for light states could be useful to this end. Ultimately, a deeper understanding of the general structure of the field-space geometry of stringy \acp{EFT} is likely to shed light on the expected equivalence between infinite-distance limits and species limits, as outlined in the preceding point; specifically, the latter imply the former \cite{Ooguri:2024ofs} and bring along the expected light towers of species \cite{Stout:2022phm, Aoufia:2024awo}, but the converse statement is yet to be proven in general, as we discussed.

\noindent \paragraph{\ac{UV}/\ac{IR} mixing, holography and the swampland.} Beyond the interest for the classification of species limits in string theory, the parametric inequalities discussed in \cref{sec:UV/IR_relations_and_swampland} suggest that the universal low-energy properties of string theory indeed take the form of holographic bounds and other swampland inequalities, as expected from bottom-up considerations \cite{McNamara:2020uza, McNamaraThesis, Basile:2023blg, Basile:2024dqq, Bedroya:2024ubj, Herraez:2024kux, Herraez:2025clp}. From the worldsheet perspective, \ac{UV}/\ac{IR} mixing is already apparent from modular invariance, which relates the light and heavy sectors of the spectrum and is ultimately responsible for the emergence of light towers (in the form of mesoscopic extra dimensions or weakly coupled string modes) and the generic scalings we obtained. In addition to its implications for the (magnetic) weak-gravity conjecture and the dark dimension proposal, \ac{UV}/\ac{IR} bounds connecting high-energy and low-energy quantities can have phenomenological implications for scale separation \cite{Andriot:2025cyi} and inflation \cite{Cribiori:2025oek} in the context of cosmology. It is thus important to further elucidate how \ac{UV}/\ac{IR} mixing features in string theory, and to which extent, in order to facilitate connections to observations. \\

\noindent \paragraph{Strong coupling and dualities.} In \cref{sec:UV/IR_relations_and_swampland} we argued that our results extend to higher loops, and---upon resumming the perturbative series, which likely involves resurgence ingredients---take the same qualitative form, insofar as we steer clear of the strongly coupled regime. It would be very interesting to extend our investigations to this case. In controlled (supersymmetric) settings, the strong-coupling behavior of Wilson coefficients is governed by a weakly coupled S-dual frame or a decompactification to strongly coupled M-theory in one dimension higher \cite{Green:1997di, Green:1997tv, Green:1999pv, Obers:1998fb, Obers:1999um, Green:2008uj, Castellano:2023aum, vandeHeisteeg:2023dlw, Calderon-Infante:2025ldq, Castellano:2024bna, Castellano:2025ljk, Castellano:2025yur, Aoufia:2025ppe}. In order to investigate this regime in more generality and with less control, we can approach the problem both from the top down (by assuming that string dualities persist in general, potentially non-supersymmetric settings) and from the bottom up (attempting to bootstrap dualities from swampland principles along the lines of \cite{Bedroya:2023tch}). These conceptual hurdles are accompanied by the mathematical difficulties in analyzing higher-loop contributions quantitatively; namely, the holomorphic non-splittability of the supermoduli space of super-Riemann surfaces \cite{Witten:2012ga, Donagi:2013dua, Witten:2012bh} seems to require a dedicated treatment and adaptation of our methods. We look forward to addressing these issues in the future. \\

\noindent \paragraph{Topological strings and BPS black-hole entropy.} A natural arena to apply the general worldsheet framework developed in this paper is the topological string on Calabi-Yau manifolds, since its simpler structure relative to the physical string is much better understood \cite{Cecotti:1991me, Bershadsky:1993ta, Bershadsky:1993cx, Ooguri:1995cp, Aganagic:2002qg, Grimm:2007tm, Angelantonj:2019qfw, Angelantonj:2022dsx, Iwaki:2023cek, Gu:2023mgf, Marino:2024tbx, Hattab:2024chf, Hattab:2024ewk, Hattab:2024yol}. Moreover, the intimate connection between topological string amplitudes and certain F-terms in the effective action of type II Calabi-Yau reductions \cite{Antoniadis:1993ze, Antoniadis:2007ta} led to a wealth of developments on \ac{BPS} black-hole microphysics \cite{Ooguri:2004zv, Dabholkar:2004yr, Castellano:2025ljk, Castellano:2025yur} and their connections to the species scale \cite{vandeHeisteeg:2022btw, Cribiori:2022nke, Castellano:2023stg, Castellano:2023jjt, Basile:2025bql}. The machinery we have developed could be applied to understand the universal features of topological string amplitudes---perhaps to all orders---and the resulting implications for minimal black holes \cite{Cribiori:2022nke, Cribiori:2023ffn}. This direction for future research is also related to the preceding point, since the M-theory limit of Calabi-Yau compactifications can be efficiently explored from the topological string side. \\

\noindent All in all, the program initiated in \cite{Aoufia:2024awo, Basile:2024lcz} and significantly developed in this paper has the potential to shed light on a more global picture of the string landscape and the universal---or at least generic---low-energy properties of quantum gravity. With a great deal of optimism, this approach, in spirit orthogonal to precision searches of realistic models of particle physics and cosmology, may lead to stringy signatures for experiments accessible with foreseeable technology. Less ambitiously, this work highlights the deep connection between bottom-up swampland principles, motivated by black-hole physics and holography, and top-down constructions. These complementary aspects are tied together by the underlying consistency conditions of string theory, whose remarkable constraining power and elusive underpinnings are evidently yet to be completely grasped and harnessed.

\subsection*{Acknowledgements}
We would like to thank Muldrow Etheredge, Bernardo Fraiman, Michael Haack and Yonatan Zimmerman for helpful discussions.We thank Melissa the cat to fit the following grant number in the margin.
The work of CA is supported through the grants CEX2020-001007-S, PID2021-123017NB-I00 and PID2024-156043NB-I00, funded by MCIN/AEI/10.13039/501100011033, and ERDF, EU. This work was supported by a short term scientific mission grant from the COST action CA22113 THEORY-CHALLENGES.
The work of IB is supported by the Origins Excellence Cluster and the German-Israel-Project (DIP) on Holography and the Swampland.
The work of GL is funded by the
European Union - NextGenerationEU/PNRR mission 4.1; CUP: C93C24004950006.
The work of ML is supported by the Ayuda RYC2023-043268-I funded by MICIU/AEI/10.13039/501100011033 and FSE+, as well as through the grants PID2024-156043N B-I00, PID2021-123017NB-I00 and CEX2020-001007-S, funded by MCIN/AEI/10.13039/50110 0011033, and ERDF, EU. CA would like to thank the II. Institute for Theoretical Physics at the University of Hamburg and the DESY
Theory Group for hospitality during late stages of this work.
IB and ML acknowledge the organizers of the Corfu2025 conference on Quantum Gravity and Strings, where discussions on this project were initiated. 

\appendix

\section{Eisenstein series and regularization}\label{app:eisenstein_series_and_regularization}

In this appendix we discuss a method of \ac{IR} regularization of naively divergent modular integrals by means of the Rankin-Selberg-Zagier method \cite{Rankin_1939, selberg1940bemerkungen,zagier1981rankin}, mainly following the notation of \cite{Benjamin:2016fhe,Angelantonj:2010ic,Angelantonj:2011br}. In string theory, such integrals arise whenever loop amplitudes---which (up to tadpole subtraction) are finite order by order---are split into an analytic and non-analytic piece in order to extract (field-dependent) \ac{EFT} Wilson coefficients (see \cref{footnote:divergencies} and e.g. \cite{Green:2010wi} for a discussion of this point in maximal supergravity). In the main text, we encountered these integrals when computing e.g. four-dimensional gauge couplings in \cref{sec:gauge_couplings_species_limits} or the torus partition functions of Narain lattice \acp{CFT}. Historically, regularization of integrals over the ${\text{SL}}(2,\bZ$) fundamental domain $\cF$ has been performed by cutting away the cusp at a fixed but large $\tau_2 =L$. While practical, this procedure is not modular invariant and depends on the choice of a cutoff. The regularization introduced by Rankin and Selberg, and later generalized by Zagier, instead amounts to defining a new square-integrable function by subtracting off a suitable modular invariant combination from the integrand. Although it might seem arbitrary, we will see that this subtraction is natural and dovetails beautifully with the spectral theory of ${\text{SL}}(2,\bZ)$-invariant functions on the upper-half plane. In the following, as in the main text, we will take the volume of the fundamental domain to be ${\mathrm{Vol}}(\cF) = 3/\pi$.

\paragraph{Functions of rapid decay.} We start by introducing modular integrals of functions $f(\tau)$ of \emph{rapid decay}, meaning that they decay faster than any polynomial at the cusp $\tau_2 \to \infty$. The \acfi{RS} transform of this function is defined as the integral 
\begin{eqaed}\label{eq:rsz_integral}
	\cR_s[f] \equiv \int_{\cF} \td \mu \, f(\tau) E_s(\tau) \, ,
\end{eqaed}
with $E_s(\tau), \ s \in \bC$ the real-analytic Eisenstein series defined as the Poincaré sum
\begin{eqaed}\label{eq:Es_def}
	E_s(\tau) = \sum_{\gamma \in \Gamma_\infty \backslash P{\text{SL}}(2,\bZ)} \mathrm{Im}(\gamma \tau)^s = \tau_2^s + \sum_{(c\ge 1,d)=1}\frac{\tau_2^s}{|c\tau+d|^s} \, .
\end{eqaed}
In the above sum, $\Gamma_{\infty}$ is the subgroup of upper-triangular matrices fixing $\tau_2$. This definition converges as long as $\mathrm{Re}\, s >1$, however it admits a meromorphic continuation over the entire plane whose Fourier decomposition with respect to $\tau_1$ reads
\begin{eqaed}\label{eq:Es_fourier}
 E_s(\tau) = \tau_2^s + \frac{c_{1-s}}{c_s}\tau_2^{1-s} + \dots \, ,
\end{eqaed}
with $c_s \equiv \pi^{-s}\Gamma(s)\zeta(2s)$ and where the dots denote terms expressed as a series over $\cos (2\pi n \tau_1), \ n\ne 0$ and with coefficients exponentially suppressed in $\tau_2$. The decomposition above can also be identified with the large-$\tau_2$ expansion of the Eisenstein series. The first term of this expansion is commonly denoted as \emph{constant term}, referring to its Fourier expansion. The Eisentein series are eigenfunctions of the ${\text{SL}}(2,\bZ)$-Laplacian $\Delta_\tau \equiv -4\tau_2^2 \partial_\tau \partial_{\overline \tau} = - \tau_2^2(\partial_{\tau_1}^2+\partial_{\tau_2}^2)$ with eigenvalue
\begin{eqaed}\label{eq:eigen_e}
\Delta_\tau E_s(\tau) = s(1-s) E_s \, .
\end{eqaed}
The meromophic continuation of the Eisenstein series has a simple pole at $s=1$. The regular part of $E_1$ is extracted from the Kronecker-limit formula, which subtracts its $\tau$-independent residue as
\begin{eqaed}\label{eq:E1_def}
\hat{E}_1(\tau) \equiv \lim_{s\to1} \left( E_s - \frac{3}{\pi(s-1)}\right) = \tau_2 - \frac{3}{\pi}\log \tau_2 + \text{const.} + \dots \, ,
\end{eqaed}
where again the dots denote non-zero Fourier modes. The regularized $\hat{E}_1$ is not an eigenfunction of the Laplacian; instead, it obeys the ``anomalous'' equation
\begin{eqaed}\label{eq:eigen_e1}
\Delta_\tau \hat{E}_1 = -\frac{3}{\pi} \, .
\end{eqaed}
In the main text, we also introduced the \emph{quasi-modular, holomorphic} Eisenstein function of weight two $\mathsf{E}_2$, defined as the (regularized) modular form \cite{DHoker:2022dxx}
\begin{eqaed}\label{eq:E_2-def}
    \mathsf{E}_2 = \lim_{N\to \infty}\lim_{M \to \infty}\frac{1}{2\zeta(2)} \sum_{n=-N}^N \sum_{\substack{m=-M\\ (m,n)\ne 0}}^M \frac{1}{ |c\tau + d|^2} \, ,
\end{eqaed}
which can be seen as defining the holomorphic function $\tau_2^{-2} E_2(\tau)$ in terms of the non-holomorphic counterpart. Under modular transformations, $\mathsf{E}_2(\tau)$ acquires an inhomogeneous piece, namely
\begin{eqaed}\label{eq:E_2-transformation}
    \mathsf{E}_2\left( \frac{a\tau + b}{c\tau+d}\right) = (c\tau+d)^2 \, \mathsf{E}_2(\tau) + \frac{12}{2\pi i} \, c(c\tau + d) \, ,
\end{eqaed}
which allows us to use this function as a modular connection as in \cref{sec:gauge_couplings_higher_derivative_corrections}. We close the discussion on the Eisenstein series by mentioning that one can choose another regularization of the series above, this time preserving modular invariance instead of holomorphicity. We call this function $\hat{\mathsf{E}}_2$, and it is related to $\mathsf{E}_2$ by \cite{Kiritsis:1997hj,DHoker:2022dxx}
\begin{eqaed}\label{eq:e2hat_vs_e2}
    \hat{\mathsf{E}}_2 = \mathsf{E}_2 - \frac{3}{\pi \tau_2} \, .
\end{eqaed}
We see that depending the regularization, we have a choice as whether to preserve modular invariance of holomorphicity. This is an example of an anomaly in the context of modular forms.
\\

\noindent Let us now come back to the regularization of modular integrals of rapid decay functions. Since the Eisenstein series are expressed as a Poincaré sum, we can use the \emph{unfolding trick} to express the \ac{RS} transform of $f$ as an integral over the strip ${S}=\{ \tau_2 \geq 0, -\frac12 \leq \tau_1 \leq \frac12\}$ according to
\begin{eqaed}\label{eq:RS_f_unfolding}
	\cR_s[f] = \int_S \td \mu \, \tau_2^s \, f(\tau) = \int_0^{\infty} \td\tau_2\, \tau_2^{2-s}f_0(\tau_2) \, ,
\end{eqaed}
where $f_0$ denotes the zero-mode of the Fourier decomposition of $f$ with respect to $\tau_1$. We see that the result is a Mellin transform of $f_0$. Notice that the meromorphic continuation of the Eisenstein series described above endows the \ac{RS} transform with a similar continuation. In particular, we see that the modular integral of $f$ can now be expressed by extracting the residue of the simple pole at $s=1$, namely
\begin{eqaed}\label{eq:residue_integral}
\Res_{s=1} \cR_s[f] = \frac{3}{\pi}\int_{\cF} \td \mu\, f(\tau) \, .
\end{eqaed}
\paragraph{Functions of slow growth.} Unfortunately, the modular functions encountered in string theory are rarely of rapid decay. The generalization of the above procedure to the case of functions of \emph{slow growth} is due to Zagier. We define a slow-growth function at the cusp according to the asymptotic behavior
\begin{eqaed}\label{eq:slow-growth_def}
f(\tau) \overset{\tau_2 \to \infty}{\sim} \sum_{i=1}^m \frac{c_i}{n_i!} \, \tau_2^{\alpha_i} \log^{n_i}(\tau_2) + \text{sub-polynomial} \, ,
\end{eqaed}
where $\alpha_i,\ c_i \in \bC$, $n_i \in \mathbb{N}$. Denoting by $\phi(\tau_2)$ the term making the function non-integrable, we simply define the \ac{RS} transform to omit the problematic contribution as
\begin{eqaed}\label{eq:RS_transform_subtracted}
\cR_s[f] \equiv \int_0^{\infty} \td \tau_2 \, \tau_2^{2-s} \, (f_0(\tau_2) - \phi(\tau_2)) \ .
\end{eqaed}
This modification can be shown to be equivalent to the regularized integral \cite{Angelantonj:2011br}
\begin{eqaed}\label{eq:renormalized_integral}
\mathrm{RN}\int_{\cF} \td \mu \, f(\tau) E_s(\tau) &\equiv \int_{\cF_L} \td \mu \, f(\tau) E_s(\tau) - \int_{\cF-\cF_L} \td \mu \, (f(\tau) E_s(\tau)-\phi(\tau_2) \phi_s(\tau_2))\\&-h_s^L- \frac{c_{s-1}}{c_s} \, h_{1-s}^L = \cR_s[f] \, ,
\end{eqaed}
where we defined $\cF_L$ as the fundamental domain cut off at $\tau_2 \le L$, $\phi_s$ the constant term of $E_s$ and $h_s^L$ the cut-off Mellin transform
\begin{eqaed}
h_s^L \equiv \int_0^L \td \tau_2 \, \tau_2^{s-2} \phi_s \ .
\end{eqaed}
One can verify that the right-hand side of the above equation is independent of $L$, invariant under $s\to 1-s$, and has simple poles at $s=0,1$ and other poles of degree $n_i+1$ at $s=\alpha_i, 1-\alpha_i$. With these definitions, we are now able to give a useful characterization of the regularized integral. Firstly, notice that it correctly reduces to the usual \ac{RS} transform if $f$ is of rapid decay. Secondly, since by direct computation 
\begin{eqaed}\label{eq:RS_transform_1}
\cR_s[1] = \cR_s[E_r(\tau)] = 0 \, ,
\end{eqaed}
we can subtract a suitable combination of Eisenstein series
from $f$, making it integrable without changing its renormalized integral. Specifically, we define
\begin{eqaed}\label{eq:tildef}
\tilde f(\tau) \equiv f(\tau) -\sum_{i| \alpha_i>\frac12} c_i \, \partial_s^{n_i} E_s(\tau)|_{s=\alpha_i} \, ,
\end{eqaed}
so that $\tilde f$ is square-integrable over $\cF$. The combination of Eisenstein series we chose can be seen as the modular completion of the diverging terms at the cusp. Integrals involving $\tilde f$ can now be computed using the standard \ac{RS} transform, and in particular we can compute its mode expansion using the spectral decomposition of ${\text{SL}}(2,\bZ)$ functions according to \cite{Benjamin:2016fhe}
\begin{eqaed}\label{eq:spectral_expansion_f}
\tilde f(\tau) = \tilde f_0 + \sum_{n=1}^{\infty} \frac{(f,\nu_n)}{(\nu_n, \nu_n)}\nu_n(\tau) + \frac{1}{4\pi i}\int_{\Re \, s = \frac{1}{2}} \td s \, \cR_{1-s}[f]E_s(\tau) \, ,
\end{eqaed}
where the integral in $s$ is performed over the line $s \in 1/2 + i\bR$. The functions $\nu_n(\tau)$ are \emph{Maass cusp forms}, which decay exponentially at the cusp and are characterized by the eigenvalue equation 
\begin{eqaed}\label{eq:maass_eigenvalue}
\Delta_\tau \nu_n = \left( \frac14 + \lambda^2_n \right) \nu_n \ , \quad \lambda_n \ge 0 \, .
\end{eqaed}
The interested reader can find more details about the harmonic analysis of automorphic functions in \cite{Benjamin:2016fhe,DHoker:2022dxx}. All in all, we see that \cref{eq:spectral_expansion_f} provides a spectral decomposition of $f$ by means of \cref{eq:tildef}. In practice, in the main text we resort to \cref{eq:tildef} as a modular-invariant \ac{IR} regularization of the integrals at stake, including the ``anomalous case'' which crucially produces logarithmic thresholds thanks to \cref{eq:eigen_e1}.

\section{Boundary terms and suppression of heavy sectors}\label{app:exp_suppression}

In this appendix we carefully treat boundary terms in the modular integrals involving Laplacian operators and derive the exponential suppression employed in \cref{sec:gauge_couplings_species_limits} and \cref{sec:gauge_couplings_higher_derivative_corrections} to derive the main technical tool of the paper, namely the asymptotic differential equations for modular integrals. When the internal partition function is integrated against an external sector, the vacuum contribution from the latter dominates in the large-$t$ limit. In order to show this, we proceed as follows. For convenience, in this appendix we will simplify the notation by dropping subscripts and other indices referring to sums over (spin-structure) sectors, and denote the modular integral by $\int_\cF \td\mu$ as above, the reduced internal partition function ($\cZ_\text{int}$ or $\cZ_{\text{c}}$ in the main text) by $\cZ(\tau;t)$ and the relevant external sectors by $B(\tau)$, often dropping their arguments for visual clarity. We denote by $c$ (resp. $c'$) the effective central charge which controls the large-$\tau_2$ scaling of $\cZ \sim \tau_2^{c/2}$ (resp. $B \sim \tau_2^{c'/2}$). Finally, the relevant integrals are regularized by subtracting an appropriate Eisenstein series. \\

\noindent \paragraph{Details on asymptotic differential equations and their general solutions.} To begin with, let us recall the procedure used in the main text in this lighter notation. Isolating the vacuum contribution $B_\text{vac}$ in the regularized integral $I(t) \equiv \int_\cF \td\mu \, \widetilde{B \cZ} = \int_\cF \td\mu (B \cZ - E_{(c+c')/2} ) $ allows us to obtain a closed-form asymptotic differential equation for $I(t)$ by acting with $\mathcal{D}_c$ according to
\begin{eqaed}\label{eq:ODE_steps}
    \mathcal{D}_c I(t) = \int_\cF \td\mu \, B \cD_c \cZ \sim \int_\cF \td\mu \, B(\Delta_\tau - w_c) \cZ \, .
\end{eqaed}
To proceed, we would like to integrate the Laplacian by parts, but to rigorously show that the regularization takes case of \ac{IR} divergences we need to treat them carefully. To this end, we observe that when $c+c'\geq 2$ (so that the Eisenstein regularization is needed for convergence)
\begin{eqaed}\label{eq:laplacian_boundary}
    0 & = \int_\cF \td\mu \, \Delta_\tau \left(B \cZ - E_{\frac{c+c'}{2}}\right) \\
    & = \int_\cF \td\mu \left(\Delta_\tau B \cZ + B \Delta_\tau \cZ - 2 \nabla B \cdot \nabla \cZ - w_{c+c'} E_{\frac{c+c'}{2}}\right) + \, \delta_{c+c',2} \, ,
\end{eqaed}
where the minus sign in front of the inner product of gradients arises because $\Delta_\tau$ is defined with a minus sign \cite{Benjamin:2021ygh} and the final term accounts for the ``anomalous'' Eisenstein series $\hat{E}_1$, which we implicitly use when necessary and satisfies $\Delta_\tau \hat{E}_1 = - 3/\pi$. The gradient term can then be evaluated as follows: cutting off the fundamental domain $\cF$ to the subdomain $\cF_L$ with $\tau_2 \leq L$ (with $L$ to be sent to infinity), the boundary term simplifies to the integral over the segment $\{\tau_2=L\}$ of the normal derivative, yielding
\begin{eqaed}\label{eq:boundary_term}
    \int_{-\frac{1}{2}}^{\frac{1}{2}}\td \tau_1 \,\left(B \partial_{\tau_2} \cZ \right)_{\tau_2=L} \overset{L \to \infty}{\sim} \frac{c}{2} \, L^{\frac{c+c'-2}{2}} \, .
\end{eqaed}
As we have obtained in \cref{eq:behavior_cusp}, we have $c+c'-2 = N - d$ when $I(t)$ computes the one-loop contribution to a (string-frame) dimension-$N$ Wilson coefficient of a $d$-dimensional \ac{EFT}. Hence, insofar as we consider operators with $N \leq d$, the limit $L \to \infty$ of this boundary term is finite and yields $(c/2) \, \delta_{c+c',2}$. When $c+c'>2$ this procedure cannot applied; similarly, splitting $\cZ = \widetilde{\cZ} + E_\frac{c}{2}$ does not remove the divergence in \cref{eq:boundary_term} for $N$ sufficiently large. However, we can isolate the divergence by adding and subtracting suitable combinations of $w_c$ and $w_{c'}$, since the divergence in \cref{eq:laplacian_boundary} is governed by the eigenvalues $\Delta_\tau \tau_2^{a/2} = w_a \, \tau_2^{a/2}$ arising from the vacuum contributions. Therefore, writing $\Delta_\tau B = (\Delta_\tau - w_{c'})B+w_{c'}B$ and analogously for $\cZ$, and writing $B \Delta_\tau \cZ - \nabla B \cdot \nabla \cZ = - \nabla \cdot (B \nabla Z)$, we obtain
\begin{eqaed}\label{eq:integration_by_parts}
    - \, \delta_{c+c',2} = & \int_\cF \td\mu \, \Big( (\Delta_\tau-w_{c'})B \cZ - B(\Delta_\tau - w_c) \cZ + (w_{c'}-w_c)B \cZ \\
    & - w_{c+c'} E_{\frac{c+c'}{2}} - 2 \nabla \cdot (B \nabla \cZ)\Big) \, .
\end{eqaed}
Then, using that
\begin{eqaed}\label{eq:w_split}
    w_{c+c'} = w_{c'} - w_c - \frac{c(c+c'-2)}{2} \, ,
\end{eqaed}
we can recombine the $B\cZ$ and Eisenstein terms according to
\begin{eqaed}\label{eq:integration_by_parts_2}
    - \, \delta_{c+c',2} = & \int_\cF \td\mu \, \Big( (\Delta_\tau-w_{c'})B \cZ - B(\Delta_\tau - w_c) \cZ + (w_{c'}-w_c) \left(B \cZ - E_{\frac{c+c'}{2}} \right) \\
    & + \frac{c(c+c'-2)}{2} \,  E_{\frac{c+c'}{2}} - 2 \nabla \cdot (B \nabla \cZ)\Big) \, .
\end{eqaed}
The upshot is that we can effectively integrate the Laplacian terms by parts at the price of adding the left-hand side (when present) and the integral in the second line. Despite appearances, the latter does not actually depend on the modulus $t$, since $\cZ - \tau_2^{c/2}$ is exponentially suppressed at large $\tau_2$. Hence, the boundary term is independent of $t$, and the potential divergence in \cref{eq:boundary_term} is cancelled by the Eisenstein series. When $c+c'=2$ there is no divergence, and the second line of \cref{eq:integration_by_parts_2} evaluates to $-c$ due to \cref{eq:boundary_term}. \\

\noindent Applying these results to \cref{eq:ODE_steps}, we obtain
\begin{eqaed}\label{eq:ODE_integrated_by_parts}
    \left(\cD_c + w_c - w_{c'}\right) I(t) \sim \int_\cF \td\mu \, (\Delta_\tau - w_{c'})B \cZ + (1-c)\delta_{c+c',2} + \text{const.} \times \left(1-\delta_{c+c',2}\right) \, .
\end{eqaed}
As discussed above, written this way the ($B$-dependent) constant term vanishes when $c+c'=2$, which is crucial to obtain the correct logarithmic thresholds, especially when $c=1$ when a non-vanishing additive constant would have produced a $(\log t)^2$ term in the general solution to \cref{eq:ODE_integrated_by_parts}. In the first term on the right-hand side, $\cZ$ is integrated against the exponentially suppressed function $(\Delta_\tau - w_{c'})B$, which justifies substituting its harmonic decomposition as in \cite{Aoufia:2024awo}, which results in the familiar leading volume-like scaling $A \, t^{c/2}$. This can be also derived by applying the same method as above to integrals of the form $\int_\cF \td\mu \, C \cZ$, with $C$ exponentially suppressed: acting with $\cD_c+w_c$, one can now freely integrate by parts and use the exponential suppression of the right-hand side which we will derive in detail below. Moreover, as we will show in the following, the remaining terms coming from the integral vanish faster than any power for large $t$. \\

\noindent As a result, on the right-hand side we can keep only the volume-like and constant terms. Hence, in terms of integration constants $b_i$ and constants determined by the inhomogeneous terms, up to additive constants that do not affect the $t$-dependence of the Wilson coefficients the general asymptotic solution takes the form
\begin{eqaed}\label{eq:It_general_solution_system}
    I(t) \sim \begin{cases}
    \frac{4A}{c'(c'-2)} \, t^{\frac{c}{2}} + b_1 \, t^{\frac{c+c'-2}{2}} + b_2 \, t^{\frac{c-c'}{2}} \, , \quad c' \neq 2 \wedge c+c' \neq 2 \, , \\
    - \, A \, t^{\frac{c}{2}} \, \log t + b_1 \, t^{\frac{c}{2}} + b_2 \, t^{\frac{c-2}{2}} \, , \quad c' = 2 \, , \\
    \frac{4A}{c(c-2)} \, t^{\frac{c}{2}} - \, \log t + b_2 \, t^{c-1} \, , \quad c+c' = 2 \, .
    \end{cases}
\end{eqaed}
For $c'>0$ the $b_2$ terms are subleading, while for $c'<0$ they are superleading but unphysical. Physically, this follows from the fact that they lead to a divergent higher-dimensional effective action in the decompactification limit. Mathematically, one can apply the Cauchy-Schwarz inequality to $\int_\cF \td\mu \, B \widetilde{\cZ}$ together with the scalings derived in \cite{Aoufia:2024awo}, as discussed in \cref{sec:gauge_couplings_higher_derivative_corrections}. More precisely, the above solutions are valid when $c' \neq 0$. When $c'=0$ the integral behaves as the simplest one $\int_\cF \td\mu \, \cZ$ studied in \cite{Aoufia:2024awo} in the context of quartic Riemann operators. In this case, the unphysical $b_2$ terms would become logarithmic. Finally, as mentioned above \cref{eq:laplacian_boundary}, the last term in \cref{eq:ODE_integrated_by_parts} arises from the Eisenstein regularization, and is thus absent when $c+c' < 2$ or, equivalently, $N<d$. Actually, this is also true for $c+c' = 2$, since the coefficient in front of the Eisenstein series vanishes and the (finite) boundary contribution in this case is accounted for in the second-to-last term in \cref{eq:ODE_integrated_by_parts}. Moreover, the additive constant vanishes for $B=1$, and more generally for $c'=0, c=1$, while it is subleading in the solution for $c'=0, c>1$. Thus we reproduce the results of \cite{Aoufia:2024awo, Basile:2024lcz} as special cases. In general, as discussed in the main text, the above conclusions hold sector by sector, and the full result is given by a linear combination over the sectors. \\

\noindent \paragraph{Exponential suppression of heavy sectors.} The above derivation hinges on the fact that integrals of the form $\int_\cF \, C \cZ$, with $C = B - B_\text{vac}$ exponentially suppressed for $\tau_2 \to \infty$, yield only the volume-like scaling $t^{\frac{c}{2}}$. In the notation of \cite{Basile:2024lcz} and \cref{eq:general_ode_correction}, the correction $F$ in $\cZ \equiv t^{\frac{c}{2}}(A + F(\tau;t))$ is such that
\begin{eqaed}\label{eq:integrals_tends_to_zero}
    \int_\cF \td\mu \, C F \; \overset{t \to \infty}{\longrightarrow} \; 0 
\end{eqaed}
faster than any power. We will now show this property, and in addition argue that the suppression is controlled by the Bessel-like exponential behavior $e^{-\sqrt{t}}$. \\

\noindent To begin with, since the reduced partition function contains contributions from bosonic states, we will assume that $F \geq 0$. Moreover, since the light spectrum dominates the partition function for large $t$, we have $F(\tau;t) \sim F(\tau_2/t)$, and from the asymptotic differential equation of \cite{Aoufia:2024awo} we have
\begin{eqaed}\label{eq:integral_F}
    \int_\cF \td\mu \, F = O(t^{-1}) \, .
\end{eqaed}
Performing the change of variables $\tau_2 = tu$, this implies that the integral of $F$ over the strip $S=\{\tau_2 \geq 0\}$ is finite, since the fundamental domain $\cF$ reduces to the strip as $t \to \infty$ and the measure $\td\mu = t^{-1} \td\tau_1 \td u/u^2$. Since $C$ is exponentially suppressed it is bounded by a constant, and therefore the integral in \cref{eq:integrals_tends_to_zero} does in fact vanish as $t \to \infty$. Performing the same change of variables, one finds that the integral over the strip
\begin{eqaed}\label{eq:strip_integral_CF}
    \int_S \frac{\td\tau_1 \td u}{u^2} \, C(\tau_1,tu) F(u) = O(1) \, .
\end{eqaed}
We then split the strip at some $O(1)$ value of $\tau_2$, say $\tau_2 = 1$. The contribution from the \ac{IR} region $\tau_2 > 1$ is exponentially suppressed in $t$, since $F$ is integrable and $C$ is suppressed as $e^{-tu}$. In the remaining contribution, the behavior of the integrand around $\tau_2 = 0$ is controlled by a power of $u$ from $C$ times $F$, which implies that $F$ must vanish at the origin faster than any power\footnote{This also follows applying the dominated convergence theorem to the strip integral in \cref{eq:integrals_tends_to_zero}, since the integrand vanishes pointwise as $t \to \infty$.}. \\

\noindent We now argue that the limit in \cref{eq:integrals_tends_to_zero} is approached with a Bessel-like exponential decay. Since $F$ is part of a partition function, it is natural to infer that this small-argument behavior is given by an exponential of the form $F(u) \sim e^{-1/u}$. Indeed, modular invariance of the canonical partition function $Z_\text{can} = \cZ(\tau_1=0)$ implies the presence of terms of this type, and this behavior is precisely realized in examples such as $\cZ = \Gamma_{1,1}$ as can be ascertained by Poisson resummation. As a result the integral in \cref{eq:strip_integral_CF} behaves as
\begin{eqaed}\label{eq:saddle_bessel}
    \int_0^\infty \td u \, e^{-tu - \frac{1}{u}} = \frac{2}{\sqrt{t}} \, K_1\left(2\sqrt{t}\right) \sim e^{-2\sqrt{t}} \, ,
\end{eqaed}
The Bessel-like exponential decay $e^{-\text{const.} \times \sqrt{t}}$ is in fact universal and independent of power-like terms in the integrand, since it is controlled by the saddle-point asymptotics exposed by rescaling $u \to \sqrt{t} u$.

\section{General spectra: subleading towers and anisotropic limits}\label{app:additive_towers_anisotropic_limits}
In this appendix we detail how derivations of asymptotic differential equations in the main text, as well as the preceding appendix, are affected by the presence of subleading and/or multiple light tower of species in the internal \ac{CFT}. The former, dubbed ``additive'' species in \cite{Castellano:2021mmx, Castellano:2022bvr}, ought not affect the leading scalings, since the effective number of species that participate in the species limit is dominated by the leading tower. The latter, dubbed ``multiplicative'' species in \cite{Castellano:2021mmx, Castellano:2022bvr}, describes anisotropic multiple decompactifications occurring at different rates. As such, the \ac{EFT} estimate for Wilson coefficients again takes the same form as in \cref{eq:wilson_coeff_EFT_expectation_planck}, with the proviso that the species scale be defined accordingly with the presence of multiple towers competing in the effective number of species. In the following we use the same notation as in \cref{app:exp_suppression}.

\subsection{Subleading light towers (additive species).}\label{sec:subleading_light_towers}
Subleading towers are sectors of the spectrum of the internal worldsheet \ac{CFT} whose conformal weights $\Delta(t) \sim \Delta_* t^{-a}$ as $t \to \infty$, with $0<a<1$. (More generally, only $t^{-1} \ll \Delta(t) \ll 1$ is needed.) Furthermore, these states do not come with unbounded prefactors in the sum, which is instead the hallmark of multiplicative species (see below). From the \ac{EFT} perspective, this means that the effective number of species is additive, and is dominated by the leading tower \cite{Castellano:2021mmx, Castellano:2022bvr}. Hence, we should expect that, whichever the leading term in \cref{eq:wilson_coeff_EFT_expectation}, it not be overcome by the contribution of subleading towers. A quick way to ascertain this is the argument in \cite{Aoufia:2024awo} based on the \ac{RS}-transform: since the modular integral of $\cZ$ scales according to the leading-tower contribution, subleading states cannot overcome this scaling. \\

\begin{figure}[!ht]
    \centering  \includegraphics[width=0.62\linewidth]{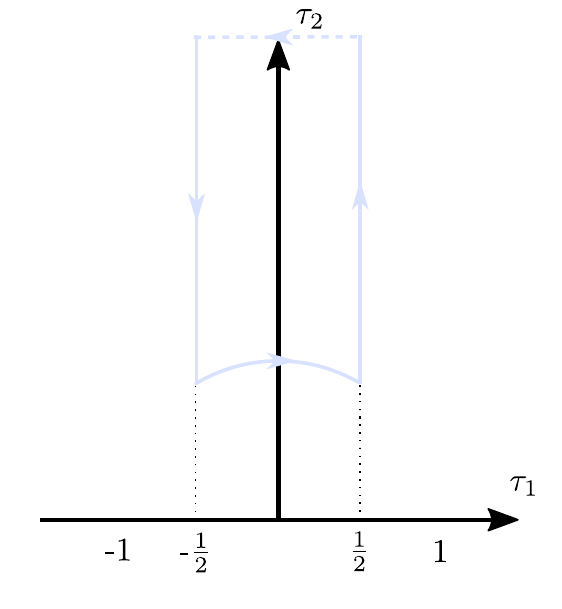}
    \caption{The boundary of the (cut-off) ${\text{SL}}(2,\bZ)$ fundamental domain $\cF$ depicted in \cref{fig:fundamental_domain}. The contributions from vertical lines in the integral of the subleading parts of the reduced partition function cancel by T-modular invariance (i.e. periodicity in $\tau_1$, whereas the contribution from the arc is \emph{a priori} non-vanishing, since the split in \cref{eq:sum_states_split} is not modular invariant.}
    \label{fig:fundamental_domain_boundary}
\end{figure}

\noindent \paragraph{Asymptotic analysis of subleading terms.} To derive the above result in some more detail using asymptotic differential equations, recall that applying the differential operator $\cD_c$ defined in \cref{eq:asymptotic_diff_eq} to the terms in the reduced partition function comprising subleading light states results in $O(\Delta)$ terms in the sum; parametrizing $\Delta \sim t^{-a}$ for concreteness (any suitable function of $t$ would do the job), each subleading tower brings terms bounded by $t^{-a} \abs{\cZ}$ in the asymptotic differential equation, which are negligible with respect to (say) $w_c \cZ$. In order to expose these subleading terms, we split the sum over states according to
\begin{eqaed}\label{eq:sum_states_split}
    \cZ = \cZ_\text{leading} + \cZ_\text{sub} + \cZ_\text{const} + \cZ_\text{heavy} \, ,
\end{eqaed}
and we neglect the heavy contributions which yield exponentially suppressed corrections. Focusing on the second and third terms, we can derive a differential equation similar to \cref{eq:asymptotic_diff_eq}, but care must be exercised since the split in \cref{eq:sum_states_split} is not compatible with modular invariance. To account for this, we proceed as follows. Firstly, since the subleading light spectrum falls off as $t^{-a}$, we replace the differential operator $\cD_c$ in \cref{eq:asymptotic_diff_eq} by
\begin{eqaed}\label{eq:new_Dc_a}
    \cD_{c,a} \equiv - \, a^{-2} \, t^2 \partial_t^2 - \left(a^{-2} + a^{-1}(1-c)\right) t \partial_t \, .
\end{eqaed}
This has the effect of replacing $t$ by $t^a$, and thus the analog of \cref{eq:asymptotic_diff_eq} holds for $\cZ_\text{light} + \cZ_\text{const}$ up to $O(1)$ terms coming from $(\Delta_\tau + w_c)\cZ_\text{const}$. Upon integrating over the fundamental domain, we cannot \emph{a priori} discard the boundary term altogether: the cusp is safe (the vacuum contribution is part of $\cZ_\text{leading})$, and the vertical lines cancel out due to termwise T-modular invariance, but the contribution from the arc remains (see \cref{fig:fundamental_domain_boundary}). However, since the arc integral of the total sum in \cref{eq:sum_states_split} does vanish, we can bound the arc integral of $\cZ_\text{subleading}$ by that of $\cZ_\text{leading}$ plus $O(1)$ terms from the (asymptotically) constant sector $\cZ_\text{const}$. In turn, the arc integral of $\cZ_\text{leading}$ is given by
\begin{eqaed}\label{eq:arc_integral_leading}
    \int_{\frac{\pi}{3}}^{\frac{2\pi}{3}} \frac{d \theta}{\sin^2 \theta} \, \cZ_\text{leading}\left(\frac{\sin \theta}{t}\right) = 2 \int_{\frac{\pi}{3}}^{\frac{\pi}{2}} \frac{d \theta}{\sin^2 \theta} \, \cZ_\text{leading}\left(\frac{\sin \theta}{t}\right) ,
\end{eqaed}
which scales at most like $t^{c/2}$ due to the Fourier expansion of \cref{eq:fourierexpcz}. In fact, one can see this more directly by bounding $\cZ_\text{leading}$ by the canonical partition function $Z_\text{can}$ according to
\begin{eqaed}\label{eq:bounding_arc_integral}
    & \int_{\frac{\pi}{3}}^{\frac{\pi}{2}} \frac{d \theta}{\sin^2 \theta} \, \cZ_\text{leading}\left(\frac{\sin \theta}{t}\right) = \int_{\frac{\sqrt{3}}{2}}^1 \frac{dy}{y^2\sqrt{1-y^2}} \, \cZ_\text{leading}\left(\frac{y}{t}\right) \\
    & \leq \frac{1}{t}\int_{\frac{\sqrt{3}}{2t}}^\frac{1}{t} \frac{dy}{y^2\sqrt{1-y^2t^2}} \, Z_\text{can}(y) \overset{t \to \infty}{\sim} \frac{1}{t}\int_{\frac{\sqrt{3}}{2t}}^\frac{1}{t} \frac{dy}{y^{2+\frac{c}{2}}\sqrt{1-y^2t^2}} \propto t^{\frac{c}{2}} \, ,
\end{eqaed}
where we used modular invariance of $Z_\text{can}$ to extract the asymptotics in the second line. \\

\noindent The upshot is that the modular integral $I_\text{subleading} \equiv \int_\cF \td \mu \left(\cZ - \cZ_\text{leading}\right)$ satisfies an asymptotic differential equation of the form
\begin{eqaed}
    \left(\mathcal{D}_{c,a} + w_c\right) I_\text{subleading} \sim f(t) \lesssim t^{\frac{c}{2}} \, ,
\end{eqaed}
and thus its contribution can at most rescale the volume-like term. Indeed, the homogeneous solution corresponding to the volume-like term is $t^{ac/2} = \cV^a \ll \cV$. The above procedure can be applied recursively for any number of subleading sectors as well. \\

\noindent As for the \ac{KK}-like terms, stemming from the richer structure of \cref{eq:It_general_solution_system}, repeating the derivation in \cref{app:exp_suppression} we see that they take the same form by solving \cref{eq:ODE_integrated_by_parts}, and any effect from the subleading spectrum on these terms is therefore subleading as expected from \ac{EFT} considerations.

\subsection{Anisotropic decompactifications (multiplicative species).}\label{sec:anisotropic_decompactifications}
Multiplicative species also comprise conformal weights $t^{-1} \ll \Delta(t) \ll 1$, but carry unbounded prefactors in the sum over states as opposed to the additive case \cite{Castellano:2021mmx, Castellano:2022bvr}. The prototypical example is an anisotropic multiple decompactification, where two factors of the internal manifold decompactify with string-units volumes $(\cV_1)^{2/c_1} \sim t$ and $(\cV_2)^{2/c_2} \sim t^a$, where the parameter $0<a<1$ quantifies the amount of anisotropy. As a result, the reduced partition functions multiply and the effective number of species is the product of the respective factors for each manifold, although the leading \ac{KK} tower pertains to the first factor. The simplest example is a double circle decompactification with string-units radii $R_1^2 = t \gg R_2^2 = t^{a} \gg 1$; in this limit, the light states have conformal weights $n_1^2/R_1^2 + n_2^2/R_2^2$ with $n_1,n_2$ integers. Hence, the lightest tower (driven by $R_1$) has $n_2=0$, while the subleading tower (driven by $R_2$) has a large degeneracy due to the leading tower. From the \ac{EFT} perspective of \cite{Calderon-Infante:2025ldq} this means that, up to properly defining the species scale to account for this change, whichever the leading term \cref{eq:wilson_coeff_EFT_expectation_planck} remains expected to hold. \\

\noindent \paragraph{Species counting.} To see this, we account for the multiplicative structure by expressing the reduced partition function as a product $\cZ(\tau;t) = \prod_i \cZ_i(\tau;t)$ over decompactifying sectors, each of which has a (non-chiral) central charge $c_i>0$ and a spectral gap $\Delta_{\text{gap},i}(t) \sim t^{-a_i}$ with $1=a_1 > a_2 > \dots > 0$ arranged in strictly decreasing order. Effectively, this means that several sectors decompactify with mass gaps $m_i/M_s = (m_1/M_s)^{a_i} = t^{-a_i/2} = (\cV_i)^{-1/c_i}$. Hence, the Planck scale in $D = d + \sum_i c_i$ is given by
\begin{eqaed}\label{eq:multiplicative_planck_scale}
    M_{{\text{Pl}};D}^{D-2} = M_{{\text{Pl}};d}^{d-2} \prod_i m_i^{c_i} \, ,
\end{eqaed}
and therefore the string scale, which is bounded by \cref{eq:multiplicative_planck_scale}, takes the form
\begin{eqaed}\label{eq:multiplicative_Ms}
    M_s^{d-2} = g_{s,D}^2 \, M_{{\text{Pl}};d}^{d-2} \prod_i \left(\frac{m_i}{M_s}\right)^{c_i} = g_{s,D}^2 \, M_{{\text{Pl}};d}^{d-2} \, t^{-\frac{c_\text{eff}}{2}} \, , \quad c_\text{eff} \equiv \sum_i a_i \, c_i \, .
\end{eqaed}
From \cref{eq:multiplicative_Ms}, we learn that (up to factors of $g_{s,D}$, which are irrelevant for our purposes) the number of species defined as $N_\text{sp} \equiv (M_{{\text{Pl}};d}/\Lambda_\text{sp})^{d-2}$ scale reads
\begin{eqaed}\label{eq:multiplicative_species_number}
    N_\text{sp} = t^{\frac{c_\text{eff}}{2}} = \prod_i \cV_i \, .
\end{eqaed}
However, the above steps raise a potential concern: as explained in \cite{Castellano:2021mmx, Castellano:2022bvr}, one ought to include one subleading light towers of species and iteratively check that the resulting species scale remain above the next subleading mass gap. Otherwise, the product in \cref{eq:multiplicative_species_number} must be truncated to only include the towers that effectively participate in the species limit. Fortunately, this concern turns out to be immaterial in our setting; since our anisotropy parameters $a_i$ are defined in terms of the ratios between mass gaps and the string scale (rather than the Planck scale, for which the expressions would be different unless $a_i = 1$), it can never be the case that a subleading light tower has mass gap above $M_s$. Effectively, in this fashion we are defining subleading light towers to be parametrically lighter than $M_s$ from the outset, which is the natural approach from the perspective of the worldsheet \ac{CFT}. Therefore, all of them must be included in \cref{eq:multiplicative_species_number}, and the resulting species scale (which we here denote as $\Lambda_\text{sp} = M_s$, albeit without the irrelevant factors of $g_{s,D}$) is given by
\begin{eqaed}\label{eq:multiplicative_species_scale}
    M_s = M_{{\text{Pl}};d} \left( \frac{m_1}{M_{{\text{Pl}};d}}\right)^{\frac{c_\text{eff}}{d+c_\text{eff}-2}} \, .
\end{eqaed}
The upshot is that the formula in \cref{eq:multiplicative_species_scale} for the species scale of the anisotropic decompactifications we are considering behaves as that of a decompactification from $d$ to $d+c_\text{eff}$ dimensions, although $D-d=c_\text{tot} \equiv \sum_i c_i$ dimensions actually decompactify. In particular, this ensures that \cref{eq:species_scale_holographic_bound} holds even in these cases. \\

\noindent \paragraph{Asymptotic analysis of multiplicative towers.} The product structure of the reduced partition function allows us to employ \cref{eq:fourierexpcz} to confirm that string-frame Wilson coefficients scale according to $a_{n,m}(t) \sim t^{c_\text{eff}/2}$, as expected from the above considerations. In contrast with \cref{sec:subleading_light_towers}, the unbounded degeneracy of subleading light towers---arising from the product structure---qualitatively modifies the asymptotic differential equation for $\cZ$; this is consistent with the fact that the leading term is not controlled by $t^{c_1/2}$ or $t^{c_\text{tot}/2}$, but rather $t^{c_\text{eff}/2}$. However, as in \cref{sec:subleading_light_towers}, we can still apply the arguments leading to \cref{eq:ODE_integrated_by_parts,eq:It_general_solution_system} to conclude that the leading \ac{KK}-like term takes the same form as in \cref{eq:wilson_coeff_EFT_expectation_planck,eq:wilson_coeff_EFT_expectation}. Indeed, letting $\cZ = t^{c_1/2} \left(A_1 + F_1(\tau;t)\right) \prod_{i \neq i} \cZ_i$, analogously as in \cref{app:exp_suppression} and \cite{Basile:2024lcz}, the \ac{KK}-like terms arise from integrals of the form
\begin{eqaed}\label{eq:multiple_factor_integral}
     I^{\text{KK}}_\text{anisotropic}(t) \overset{t \to \infty}{\sim} t^\frac{c_1}{2} \int_\cF \td\mu \, B_\text{vac} \left( \prod_{i \neq 1} \cZ_i \right) F_1 \, ,
\end{eqaed}
where $\cZ_i \sim \tau_2^{c_i/2} g_i(\tau_2/t^{a_i})$ for some functions $g_i$. Letting $a \equiv \max_{i \neq 1} a_i < 1$ and changing variables according to $\tau_2 = t^a u$, analogously as in \cref{app:exp_suppression}, as $t \to \infty$ one obtains the (strip) integral 
\begin{eqaed}\label{eq:multiple_change_variables}
    I^{\text{KK}}_\text{anisotropic}(t) \overset{t \to \infty}{\sim} t^{\frac{c_1 + a(c'+c_\text{tot}-c_1-2)}{2}} \int_S \frac{\td \tau_1 \td u}{u^2} \, B_\text{vac} \, \prod_{i \neq 1} \left( u^{\frac{c_i}{2}} \, g_i\left(t^{a-a_i}u\right) \right) F_1\left(\frac{u}{t^{1-a}}\right) .
\end{eqaed}
Since $a \leq a_i \ \forall i$, each $g_i$ is $O(1)$. Thus, we can recycle the results of \cref{app:exp_suppression} to conclude that the leading \ac{KK}-like term for a dimension-$N$ Wilson coefficient arising from \cref{eq:multiple_factor_integral} scales as $I^{\text{KK}}_\text{anisotropic}(t) \sim t^{p/2}$, with
\begin{eqaed}\label{eq:final_multiple_scaling}
    p & = c_1 + a(c'+c_\text{tot}-c_1-2) + (1-a)(c'+c_\text{tot}-c_1-2) \\
    & = c'+c_\text{tot}-2 = N-d \, .
\end{eqaed}
This is indeed the expected result from \ac{EFT}: to wit, the leading light spectrum carries no additional degeneracy, while it is the subleading light spectrum that contributes the unbounded prefactors in the sum over states according to e.g. $\cZ \sim (\prod_{i\neq 1} \cV_i) \cZ_1$.\\

\noindent As for subleading \ac{KK}-like terms, a portion of this prefactor does arise from (asymptotic) degeneracy: as exemplified by the double circle compactification discussed above, the $k$-th subleading tower carries an asymptotic degeneracy $\cV_{<k} \equiv \prod_{i<k} \cV_i$ and it effectively probes a spacetime of dimension $D_{<k} \equiv d+c_{<k} \equiv d+\sum_{i<k}c_i$. We now apply the above procedure to the \ac{KK}-like term associated to the $k$-th subleading tower, which is captured by the counterpart of \cref{eq:multiple_factor_integral} 
\begin{eqaed}\label{eq:mutiple_factor_integral_k}
    I^\text{KK}_{k}(t) \overset{t \to \infty}{\sim} t^\frac{c_{\leq k}^\text{eff}}{2} \int_\cF \td\mu \, B_\text{vac} \left( \prod_{i > k} \cZ_i \right) F_k \, ,
\end{eqaed}
where $c_{\leq k}^\text{eff} \equiv \sum_{i \leq k} a_i c_i \equiv c^\text{eff}_{<k} + a_k c_k$. Analogously as in the above derivation of \cref{eq:final_multiple_scaling}, letting $a \equiv \max_{i>k}a_i<a_k$ and changing variables according to $\tau_2=t^a u$, one ends up with the overall scaling $I^\text{KK}_{k}(t) \sim t^{p/2}$, with
\begin{eqaed}\label{final_final_multiple_scaling}
     p & = c_{\leq k}^\text{eff} + a(c'+c_\text{tot}-c_{\leq k}-2) + (a_k-a)(c'+c_\text{tot}-c_{\leq k}-2) \\
     & = c_{\leq k}^\text{eff} + a_k(c'+c_\text{tot}-c_{\leq k}-2) = c_{< k}^\text{eff} + a_k (N-D_{<k}) \, .
\end{eqaed}
Once again, this is the correct string-frame scaling expected from \ac{EFT}, since it arises from one-loop contributions of $\cV_{<k}$ \ac{KK} towers gapped at $m_k = M_s t^{-a_k/2}$ in $D_{<k}$ dimensions, implementing a decompactification of $c_k$ more dimensions. To see this more plainly, one can rearrange \cref{final_final_multiple_scaling} to yield
\begin{eqaed}\label{eq:KK_k_integral_EFT_match}
    I^\text{KK}_{k}(t) \sim \cV_{<k} \left(t^{a_k}\right)^{\frac{N-D_{<k}}{2}} = \cV_{<k} \left(\frac{M_s}{m_k}\right)^{\! N-D_{<k}} \, .
\end{eqaed}
Indeed, fully decompactifying $\cV_{<k}$ from $d$ to $D_{<k}$ dimensions, this string-units term correctly uplifts to the expected Planck-units term for a dimension-$N$ operator $\cO_N(R,F)$,
\begin{eqaed}\label{eq:Dk_uplift}
    S^{\text{KK}}_k = \int d^{D_{<k}}x \, \sqrt{-g} \, \frac{\cO_N(R,F)}{m_k^{N-D_{<k}}} \, .
\end{eqaed}
Much like, for sufficiently large $N$, the \ac{KK}-like term $S^{\text{KK}}_1$ arising from the lightest tower generically dominates the volume-like term, in a range of values of $N$ the expression in \cref{eq:Dk_uplift} will generically dominate $S^{\text{KK}}_1$. This happens whenever $N-d < p$, where $p$ is defined in \cref{final_final_multiple_scaling}. However, when this occurs the exponent $N-D_{<k}$ in \cref{eq:KK_k_integral_EFT_match} is negative, and thus the volume-like term dominates the asymptotics (unless its prefactor vanishes). To wit,
\begin{eqaed}\label{eq:range_N_negative}
    N-D_{<k} = N-d-c_{<k} < p - c_{<k} = c_{<k}^\text{eff} - c_{<k} + a_k(N-D_{<k}) \, ,
\end{eqaed}
and therefore
\begin{eqaed}\label{eq:range_N_negative_2}
    N-D_{<k} < \frac{c_{<k}^\text{eff} - c_{<k}}{1-a_k} < 0 \, ,
\end{eqaed}
so that (even including the degeneracy factor $\cV_{<k}$) the volume-like term $\cV = \prod_i \cV_i$ indeed dominates. \\

\noindent All in all, we find that the general formula of \cref{eq:wilson_coeff_EFT_expectation_planck,eq:wilson_coeff_EFT_expectation} still holds, in the sense that whichever term leads the asymptotic it will still provide the leading scaling in the more general settings we have examined in this appendix.

\section{Unphysical tachyons}\label{app:unphysical_tachyons}
In this appendix we discuss the possibility of \emph{unphysical tachyons} in the string spectrum, namely states with negative conformal weight $\Delta=h + \overline{h} < 0$ which do not satisfy level matching, i.e. $h \neq \overline{h}$. While these states do not cause divergences in the modular integrals at stake, they require a special treatment, since their contribution to the integrand is exponentially large at the cusp. In particular, the Rankin-Selberg-Zagier theorems \cite{Rankin_1939, selberg1940bemerkungen,zagier1981rankin} do not apply. In this appendix we thus extend the treatment given in \cref{app:exp_suppression} to include their contributions, and show that their contributions align with the ones given in \cref{eq:wilson_coeff_EFT_expectation_planck,eq:wilson_coeff_EFT_expectation} as expected from \ac{EFT} considerations \cite{Calderon-Infante:2025ldq}. \\

\noindent \paragraph{Treating the \ac{IR} region.} To begin with, let us recall why unphysical tachyons do not cause divergences in modular integrals \cite{Angelantonj:2010ic, Angelantonj:2011br}. This is particularly relevant for heterotic strings, for which these states are always present as ``protogravitons'' \cite{Dienes:1990ij, Abel:2015oxa}. Such states are encoded by terms in the partition function which give integrals of the form
\begin{eqaed}\label{eq:unphysical_tachyon_structure}
    \int_\cF \td\mu \, \frac{\tau_2^k}{q^a \overline{q}^b} \, ,
\end{eqaed}
where we recall the nome $q \equiv e^{2\pi i \tau}$, and where $a+b>0$, $a \neq b$. The standard prescription to define the above integral is to cut off the range of $\tau_2$ to $\tau_2 \leq L$ for some sufficiently large (but henceforth fixed) $L$, analogously as how we have done several times above, and then integrate over $\tau_1$ first in the region $\{\tau_2 \geq L\}$. This leaves a finite integral in the remaining region, since the integrand has no Fourier zero-mode by definition. We can extend this approach to integrals including factors of the (non-chiral) reduced internal partition function, here once again denoted $\cZ$ to lighten the notation. Since the worldsheet \ac{CFT} is unitary, no unphysical tachyons can appear in the primary spectrum encoded by $\cZ$; at most, the integral over $\tau_1$ in the region $\tau_2 \geq L$ picks up level-matched contributions obtained combining the tachyonic weights with asymptotically constant weights in $\cZ$ (if any---see \cref{app:additive_towers_anisotropic_limits} for a more thorough discussion on this point). Hence, the integral over the strip $\{\tau_2 \geq L\}$ is (bounded by) $O(1)$ as $t \to \infty$. \\

\noindent \paragraph{Asymptotic analysis of tachyonic terms.} According to the above considerations, we are left with an integral over the cut-off domain $\cF_L$ of the form
\begin{eqaed}\label{eq:tachyonic_integral}
    I_\text{tachyon}(t) = \int_{\cF_L} \td\mu \, \frac{\tau_2^{\frac{c'}{2}}}{q^a \overline{q}^b} \, \cZ \, ,
\end{eqaed}
where we reinstated the precise scaling with $\tau_2$ pertaining to the external sector, as in \cref{eq:behavior_cusp}. Since the tachyonic terms are bounded within $\cF_L$, the integral can be bounded by that of $\cZ$ over $\cF$ (up to possibly adding and subtracting a regulating Eisenstein series before removing the cutoff---cf. \cref{app:eisenstein_series_and_regularization}). Therefore, $I_\text{tachyon}(t) \lesssim t^{c/2}$ is bounded by the familiar volume-like scaling. \\

\noindent In order to examine \ac{KK}-like terms, in principle we could pursue an approach along the lines of \cref{sec:subleading_light_towers}; however, here it proves more convenient to once again define $\cZ = t^{c/2}(A + F(\tau;t))$ and investigate the integral
\begin{eqaed}\label{eq:KK-like_tachyonic_integral}
    I_\text{tachyon}^\text{KK}(t) = t^{\frac{c}{2}}\int_{\cF_L} \td\mu \, \frac{\tau_2^{\frac{c'}{2}}}{q^a \overline{q}^b} \, F\left(\frac{\tau_2}{t}\right) .
\end{eqaed}
The above expression actually makes the lack of $t$-dependence from the integral over $\{\tau_2 \geq L\}$ manifest, since light states in $F$ have worldsheet spin $j=0$ and do not participate in the integral over $\tau_1$. Bounding the $q, \overline{q}$ terms with a constant and using the fact that $F \geq 0$, then changing variables in \cref{eq:KK-like_tachyonic_integral} according to $\tau_2=t u$, as before, one obtains
\begin{eqaed}\label{eq:KK-like_tachyonic_integral_change}
    I_\text{tachyon}^\text{KK}(t) \leq \text{const.} \times t^{\frac{c+c'-2}{2}}\int_{\cF_L(t)} \frac{\td \tau_2 \td u}{u^2} \, u^{\frac{c'}{2}} \, F(u) \, ,
\end{eqaed}
where the rescaled integration domain $\cF_L(t)$ now does not asymptote to the strip $S$, as e.g. in \cref{eq:strip_integral_CF}; rather, it gets squeezed into a vanishingly thin strip. However, bounding it by integrating over a non-vanishing strip, e.g. $\{0 \leq \tau_2 \leq 1\}$ and applying the results of \cref{app:exp_suppression}, we conclude that the integral in \cref{eq:KK-like_tachyonic_integral_change} vanishes as $t \to \infty$. Since the prefactor is precisely the usual \ac{KK}-like term of \cref{eq:wilson_coeff_EFT_expectation}, the upshot is that the \ac{KK}-like terms arising from unphysical tachyons are parametrically smaller than the ones derived in the main text. \\

\noindent In summary, these results imply that the derivations in the main text which bound partition functions in (termwise) absolute value can be applied to the case where unphysical tachyons are present by subtracting the corresponding contributions (or a modular completion thereof---along the lines of \cite{Angelantonj:2010ic, Angelantonj:2011br}---when using modular invariance explicitly) and estimating the remainders.

\bibliographystyle{ytphys}
\baselineskip=.95\baselineskip
\bibliography{refs}

@article{DHoker:2022dxx,
    author = "D'Hoker, Eric and Kaidi, Justin",
    title = "{Lectures on modular forms and strings}",
    eprint = "2208.07242",
    archivePrefix = "arXiv",
    primaryClass = "hep-th",
    month = "8",
    year = "2022",
    journal ="",
    pages ="357",
}

@article{zagier1981rankin,
  title={The Rankin-Selberg method for automorphic functions which are not of rapid decay},
  author={Zagier, Don},
  journal={J. Fac. Sci. Univ. Tokyo Sect. IA Math},
  volume={28},
  number={3},
  pages={415--437},
  year={1981}
}

@article{selberg1940bemerkungen,
  title={Bemerkungen {\"u}ber eine Dirichletsche Reihe, die mit der Theorie der Modulformen nahe verbunden ist},
  author={Selberg, Atle},
  journal={Arch. Math. Naturvid.},
  volume={43},
  pages={47},
  year={1940}
}

@article{Rankin_1939, 
    title={Contributions to the theory of Ramanujan’s function $\tau(n)$ and similar arithmetical functions: II. The order of the Fourier coefficients of integral modular forms}, 
    volume={35}, 
    DOI={10.1017/S0305004100021101}, 
    number={3}, 
    journal={Mathematical Proceedings of the Cambridge Philosophical Society}, 
    author={Rankin, R. A.}, 
    year={1939}, 
    pages={357–372}}

@article{Green:2010wi,
    author = "Green, Michael B. and Russo, Jorge G. and Vanhove, Pierre",
    title = "{Automorphic properties of low energy string amplitudes in various dimensions}",
    eprint = "1001.2535",
    archivePrefix = "arXiv",
    primaryClass = "hep-th",
    reportNumber = "DAMTP-2010-1, IPHT-T-10-001, IHES-P-10-01, ICCUB-10-002",
    doi = "10.1103/PhysRevD.81.086008",
    journal = "Phys. Rev. D",
    volume = "81",
    pages = "086008",
    year = "2010"
}

@article{Angelantonj:2011br,
    author = "Angelantonj, Carlo and Florakis, Ioannis and Pioline, Boris",
    title = "{A new look at one-loop integrals in string theory}",
    eprint = "1110.5318",
    archivePrefix = "arXiv",
    primaryClass = "hep-th",
    reportNumber = "CERN-PH-TH-2011-259, DFTT-23-2011, MPP-2011-11911",
    doi = "10.4310/CNTP.2012.v6.n1.a4",
    journal = "Commun. Num. Theor. Phys.",
    volume = "6",
    pages = "159--201",
    year = "2012"
}

@article{Hartman:2014oaa,
    author = "Hartman, Thomas and Keller, Christoph A. and Stoica, Bogdan",
    title = "{Universal Spectrum of 2d Conformal Field Theory in the Large c Limit}",
    eprint = "1405.5137",
    archivePrefix = "arXiv",
    primaryClass = "hep-th",
    reportNumber = "CALT-68-2889, RUNHETC-2014-07",
    doi = "10.1007/JHEP09(2014)118",
    journal = "JHEP",
    volume = "09",
    pages = "118",
    year = "2014"
}

@article{Afkhami-Jeddi:2020hde,
    author = "Afkhami-Jeddi, Nima and Cohn, Henry and Hartman, Thomas and de Laat, David and Tajdini, Amirhossein",
    title = "{High-dimensional sphere packing and the modular bootstrap}",
    eprint = "2006.02560",
    archivePrefix = "arXiv",
    primaryClass = "hep-th",
    doi = "10.1007/JHEP12(2020)066",
    journal = "JHEP",
    volume = "12",
    pages = "066",
    year = "2020"
}

@article{Maloney:2020nni,
    author = "Maloney, Alexander and Witten, Edward",
    title = "{Averaging over Narain moduli space}",
    eprint = "2006.04855",
    archivePrefix = "arXiv",
    primaryClass = "hep-th",
    doi = "10.1007/JHEP10(2020)187",
    journal = "JHEP",
    volume = "10",
    pages = "187",
    year = "2020"
}

@article{Benjamin:2016fhe,
    author = "Benjamin, Nathan and Dyer, Ethan and Fitzpatrick, A. Liam and Kachru, Shamit",
    title = "{Universal Bounds on Charged States in 2d CFT and 3d Gravity}",
    eprint = "1603.09745",
    archivePrefix = "arXiv",
    primaryClass = "hep-th",
    doi = "10.1007/JHEP08(2016)041",
    journal = "JHEP",
    volume = "08",
    pages = "041",
    year = "2016"
}

@article{Dixon:1990pc,
    author = "Dixon, Lance J. and Kaplunovsky, Vadim and Louis, Jan",
    title = "{Moduli dependence of string loop corrections to gauge coupling constants}",
    reportNumber = "SLAC-PUB-5138, UTTG-36-89",
    doi = "10.1016/0550-3213(91)90490-O",
    journal = "Nucl. Phys. B",
    volume = "355",
    pages = "649--688",
    year = "1991"
}

@article{Antoniadis:1992rq,
    author = "Antoniadis, Ignatios and Gava, E. and Narain, K. S.",
    title = "{Moduli corrections to gauge and gravitational couplings in four-dimensional superstrings}",
    eprint = "hep-th/9204030",
    archivePrefix = "arXiv",
    reportNumber = "CPTH-A162-0392, IC-92-51",
    doi = "10.1016/0550-3213(92)90672-X",
    journal = "Nucl. Phys. B",
    volume = "383",
    pages = "93--109",
    year = "1992"
}

@article{Kaplunovsky:1992vs,
    author = "Kaplunovsky, Vadim S.",
    title = "{One loop threshold effects in string unification}",
    eprint = "hep-th/9205070",
    archivePrefix = "arXiv",
    reportNumber = "ITP-838-STANFORD-REV, ITP-838-STANFORD",
    month = "5",
    year = "1992"
}

@article{Kraus:2006wn,
    author = "Kraus, Per",
    title = "{Lectures on black holes and the AdS(3) / CFT(2) correspondence}",
    eprint = "hep-th/0609074",
    archivePrefix = "arXiv",
    journal = "Lect. Notes Phys.",
    volume = "755",
    pages = "193--247",
    year = "2008"
}

@article{Heidenreich:2024dmr,
    author = "Heidenreich, Ben and Lotito, Matteo",
    title = "{Proving the Weak Gravity Conjecture in perturbative string theory. Part I. The bosonic string}",
    eprint = "2401.14449",
    archivePrefix = "arXiv",
    primaryClass = "hep-th",
    reportNumber = "ACF-T24-01, ACFI-T24-01",
    doi = "10.1007/JHEP05(2025)102",
    journal = "JHEP",
    volume = "05",
    pages = "102",
    year = "2025"
}

@article{Abel:2021tyt,
    author = "Abel, Steven and Dienes, Keith R.",
    title = "{Calculating the Higgs mass in string theory}",
    eprint = "2106.04622",
    archivePrefix = "arXiv",
    primaryClass = "hep-th",
    reportNumber = "IPPP/20/113",
    doi = "10.1103/PhysRevD.104.126032",
    journal = "Phys. Rev. D",
    volume = "104",
    number = "12",
    pages = "126032",
    year = "2021"
}

@article{Kiritsis:1995dx,
    author = "Kiritsis, Elias and Kounnas, Costas",
    editor = "Dijkgraaf, R. and Klebanov, Igor R. and Narain, K. S. and Randjbar-Daemi, S.",
    title = "{One loop calculation of coupling constants in IR regulated string theory}",
    eprint = "hep-th/9509017",
    archivePrefix = "arXiv",
    reportNumber = "CERN-TH-95-222, LPTENS-95-40",
    doi = "10.1016/0920-5632(95)00638-9",
    journal = "Nucl. Phys. B Proc. Suppl.",
    volume = "45BC",
    pages = "207--216",
    year = "1996"
}

@article{Abel:2023hkk,
    author = "Abel, Steven and Dienes, Keith R. and Nutricati, Luca A.",
    title = "{Running of gauge couplings in string theory}",
    eprint = "2303.08534",
    archivePrefix = "arXiv",
    primaryClass = "hep-th",
    reportNumber = "CERN-TH-2023-044, IPPP/23/16",
    doi = "10.1103/PhysRevD.107.126019",
    journal = "Phys. Rev. D",
    volume = "107",
    number = "12",
    pages = "126019",
    year = "2023"
}

@article{Dyer:2017rul,
    author = "Dyer, Ethan and Fitzpatrick, A. Liam and Xin, Yuan",
    title = "{Constraints on Flavored 2d CFT Partition Functions}",
    eprint = "1709.01533",
    archivePrefix = "arXiv",
    primaryClass = "hep-th",
    doi = "10.1007/JHEP02(2018)148",
    journal = "JHEP",
    volume = "02",
    pages = "148",
    year = "2018"
}

@book{Kiritsis:1997hj,
    author = "Kiritsis, Elias",
    title = "{Introduction to superstring theory}",
    eprint = "hep-th/9709062",
    archivePrefix = "arXiv",
    reportNumber = "CERN-TH-97-218",
    isbn = "978-90-6186-894-1",
    publisher = "Leuven U. Press",
    address = "Leuven",
    series = "Leuven notes in mathematical and theoretical physics",
    volume = "B9",
    year = "1998"
}

@inproceedings{Petropoulos:1996xz,
    author = "Petropoulos, P. M.",
    title = "{One loop corrections to coupling constants in string effective field theory}",
    booktitle = "{5th Hellenic School and Workshops on Elementary Particle Physics}",
    eprint = "hep-th/9605012",
    archivePrefix = "arXiv",
    reportNumber = "CERN-TH-96-009, CERN-TH-96-09",
    month = "4",
    year = "1996"
}

@article{Antoniadis:1994sr,
    author = "Antoniadis, Ignatios and Ferrara, S. and Kounnas, C.",
    title = "{Exact supersymmetric string solutions in curved gravitational backgrounds}",
    eprint = "hep-th/9402073",
    archivePrefix = "arXiv",
    reportNumber = "LPTENS-94-04, CPTH-A291-0294, CERN-TH-7148-94, UCLA-94-TEP-5",
    doi = "10.1016/0550-3213(94)90331-X",
    journal = "Nucl. Phys. B",
    volume = "421",
    pages = "343--372",
    year = "1994"
}

@article{Florakis:2016aoi,
    author = "Florakis, Ioannis",
    title = "{Gravitational Threshold Corrections in Non-Supersymmetric Heterotic Strings}",
    eprint = "1611.10323",
    archivePrefix = "arXiv",
    primaryClass = "hep-th",
    doi = "10.1016/j.nuclphysb.2017.01.016",
    journal = "Nucl. Phys. B",
    volume = "916",
    pages = "484--509",
    year = "2017"
}

@article{Aoufia:2024awo,
    author = "Aoufia, Christian and Basile, Ivano and Leone, Giorgio",
    title = "{Species scale, worldsheet CFTs and emergent geometry}",
    eprint = "2405.03683",
    archivePrefix = "arXiv",
    primaryClass = "hep-th",
    doi = "10.1007/JHEP12(2024)111",
    journal = "JHEP",
    volume = "12",
    pages = "111",
    year = "2024"
}

@article{Basile:2024lcz,
    author = "Basile, Ivano and Lust, Dieter",
    title = "{Dark Dimension With (Little) Strings Attached}",
    eprint = "2409.12231",
    archivePrefix = "arXiv",
    primaryClass = "hep-th",
    doi = "10.1002/prop.202400265",
    journal = "Fortsch. Phys.",
    volume = "73",
    number = "4",
    pages = "2400265",
    year = "2025"
}

@article{Montag:1992dm,
    author = "Montag, J. Lee",
    title = "{The one loop five graviton scattering amplitude and its low-energy limit}",
    eprint = "hep-th/9205097",
    archivePrefix = "arXiv",
    reportNumber = "ITP-SB-92-63",
    doi = "10.1016/0550-3213(93)90248-N",
    journal = "Nucl. Phys. B",
    volume = "393",
    pages = "337--360",
    year = "1993"
}

@article{Angelantonj:2002ct,
    author = "Angelantonj, Carlo and Sagnotti, Augusto",
    title = "{Open strings}",
    eprint = "hep-th/0204089",
    archivePrefix = "arXiv",
    reportNumber = "CERN-TH-2002-025, ROM2F-2002-08, LPTENS-02-14, CPHT-RR-020-0202, CPHT-RR-020.0202",
    doi = "10.1016/S0370-1573(02)00273-9",
    journal = "Phys. Rept.",
    volume = "371",
    pages = "1--150",
    year = "2002",
    note = "[Erratum: Phys.Rept. 376, 407 (2003)]"
}

@book{DiFrancesco:1997nk,
    author = "Di Francesco, P. and Mathieu, P. and Senechal, D.",
    title = "{Conformal Field Theory}",
    doi = "10.1007/978-1-4612-2256-9",
    isbn = "978-0-387-94785-3, 978-1-4612-7475-9",
    publisher = "Springer-Verlag",
    address = "New York",
    series = "Graduate Texts in Contemporary Physics",
    year = "1997"
}

@article{Dijkgraaf:1989hb,
    author = "Dijkgraaf, Robbert and Vafa, Cumrun and Verlinde, Erik P. and Verlinde, Herman L.",
    title = "{The Operator Algebra of Orbifold Models}",
    reportNumber = "HUTP-88-A052, THU-88-38",
    doi = "10.1007/BF01238812",
    journal = "Commun. Math. Phys.",
    volume = "123",
    pages = "485",
    year = "1989"
}

@article{Richards:2008jg,
    author = "Richards, David M.",
    title = "{The One-Loop Five-Graviton Amplitude and the Effective Action}",
    eprint = "0807.2421",
    archivePrefix = "arXiv",
    primaryClass = "hep-th",
    reportNumber = "DAMTP-2008-60",
    doi = "10.1088/1126-6708/2008/10/042",
    journal = "JHEP",
    volume = "10",
    pages = "042",
    year = "2008"
}

@article{Berg:2016wux,
    author = "Berg, Marcus and Buchberger, Igor and Schlotterer, Oliver",
    title = "{From maximal to minimal supersymmetry in string loop amplitudes}",
    eprint = "1603.05262",
    archivePrefix = "arXiv",
    primaryClass = "hep-th",
    doi = "10.1007/JHEP04(2017)163",
    journal = "JHEP",
    volume = "04",
    pages = "163",
    year = "2017"
}

@article{Berg:2016fui,
    author = "Berg, Marcus and Buchberger, Igor and Schlotterer, Oliver",
    title = "{String-motivated one-loop amplitudes in gauge theories with half-maximal supersymmetry}",
    eprint = "1611.03459",
    archivePrefix = "arXiv",
    primaryClass = "hep-th",
    doi = "10.1007/JHEP07(2017)138",
    journal = "JHEP",
    volume = "07",
    pages = "138",
    year = "2017"
}

@article{Ellis:1987yx,
    author = "Ellis, John R. and Jetzer, Philippe and Mizrachi, Leah",
    title = "{No Renormalization of the $N=1$ Supergravity Theory Derived From the Heterotic String}",
    reportNumber = "CERN-TH-4739-87, UGVA-DPT-1987-05-538",
    doi = "10.1016/0370-2693(87)90808-2",
    journal = "Phys. Lett. B",
    volume = "196",
    pages = "492--498",
    year = "1987"
}

@article{Ellis:1987dc,
    author = "Ellis, John R. and Jetzer, Philippe and Mizrachi, Leah",
    title = "{One Loop String Corrections to the Effective Field Theory}",
    reportNumber = "CERN-TH-4829/87, UGVA-DPT-1987/08-544",
    doi = "10.1016/0550-3213(88)90214-3",
    journal = "Nucl. Phys. B",
    volume = "303",
    pages = "1--35",
    year = "1988"
}

@article{Anchordoqui:2009mm,
    author = "Anchordoqui, Luis A. and Goldberg, Haim and Lust, Dieter and Nawata, Satoshi and Stieberger, Stephan and Taylor, Tomasz R.",
    title = "{LHC Phenomenology for String Hunters}",
    eprint = "0904.3547",
    archivePrefix = "arXiv",
    primaryClass = "hep-ph",
    reportNumber = "MPP-2009-42, LMU-ASC-18-09",
    doi = "10.1016/j.nuclphysb.2009.06.023",
    journal = "Nucl. Phys. B",
    volume = "821",
    pages = "181--196",
    year = "2009"
}

@article{Banks:1987cy,
    author = "Banks, Tom and Dixon, Lance J. and Friedan, Daniel and Martinec, Emil J.",
    title = "{Phenomenology and Conformal Field Theory Or Can String Theory Predict the Weak Mixing Angle?}",
    reportNumber = "SLAC-PUB-4377",
    doi = "10.1016/0550-3213(88)90551-2",
    journal = "Nucl. Phys. B",
    volume = "299",
    pages = "613--626",
    year = "1988"
}

@article{Banks:1988yz,
    author = "Banks, Tom and Dixon, Lance J.",
    title = "{Constraints on String Vacua with Space-Time Supersymmetry}",
    reportNumber = "PUPT-1086, SCIPP-8805",
    doi = "10.1016/0550-3213(88)90523-8",
    journal = "Nucl. Phys. B",
    volume = "307",
    pages = "93--108",
    year = "1988"
}

@article{Calderon-Infante:2025ldq,
    author = "Calder{\'o}n-Infante, Jos{\'e} and Castellano, Alberto and Herr{\'a}ez, Alvaro",
    title = "{The double EFT expansion in quantum gravity}",
    eprint = "2501.14880",
    archivePrefix = "arXiv",
    primaryClass = "hep-th",
    reportNumber = "CERN-TH-2025-009, EFI-25-1, MPP-2025-6",
    doi = "10.21468/SciPostPhys.19.4.096",
    journal = "SciPost Phys.",
    volume = "19",
    number = "4",
    pages = "096",
    year = "2025"
}

@article{Ooguri:2024ofs,
    author = "Ooguri, Hirosi and Wang, Yifan",
    title = "{Universal bounds on CFT Distance Conjecture}",
    eprint = "2405.00674",
    archivePrefix = "arXiv",
    primaryClass = "hep-th",
    reportNumber = "CALT-TH 2024-015, IPMU 24-0011",
    doi = "10.1007/JHEP12(2024)154",
    journal = "JHEP",
    volume = "12",
    pages = "154",
    year = "2024"
}

@article{Witten:2012ga,
    author = "Witten, Edward",
    title = "{Notes On Super Riemann Surfaces And Their Moduli}",
    eprint = "1209.2459",
    archivePrefix = "arXiv",
    primaryClass = "hep-th",
    doi = "10.4310/PAMQ.2019.v15.n1.a2",
    journal = "Pure Appl. Math. Quart.",
    volume = "15",
    number = "1",
    pages = "57--211",
    year = "2019"
}

@article{Witten:2012bh,
    author = "Witten, Edward",
    title = "{Superstring Perturbation Theory Revisited}",
    eprint = "1209.5461",
    archivePrefix = "arXiv",
    primaryClass = "hep-th",
    month = "9",
    year = "2012"
}

@article{Donagi:2013dua,
    author = "Donagi, Ron and Witten, Edward",
    editor = "Donagi, Ron and Katz, Sheldon and Klemm, Albrecht and Morrison, David R.",
    title = "{Supermoduli Space Is Not Projected}",
    eprint = "1304.7798",
    archivePrefix = "arXiv",
    primaryClass = "hep-th",
    journal = "Proc. Symp. Pure Math.",
    volume = "90",
    pages = "19--72",
    year = "2015"
}

@article{Mirzakhani:2006fta,
    author = "Mirzakhani, Maryam",
    title = "{Simple geodesics and Weil-Petersson volumes of moduli spaces of bordered Riemann surfaces}",
    doi = "10.1007/s00222-006-0013-2",
    journal = "Invent. Math.",
    volume = "167",
    number = "1",
    pages = "179--222",
    year = "2006"
}

@article{Mirzakhani:2006eta,
    author = "Mirzakhani, Maryam",
    title = "{Weil-Petersson volumes and intersection theory on the moduli space of curves}",
    doi = "10.1090/S0894-0347-06-00526-1",
    journal = "J. Am. Math. Soc.",
    volume = "20",
    number = "01",
    pages = "1--24",
    year = "2007"
}

@article{Mirzakhani:2010pla,
    author = "Mirzakhani, Maryam",
    title = "{Growth of Weil-Petersson volumes and random hyperbolic surfaces of large genus}",
    eprint = "1012.2167",
    archivePrefix = "arXiv",
    primaryClass = "math.GN",
    journal = "J. Diff. Geom.",
    volume = "94",
    number = "2",
    pages = "267--300",
    year = "2013"
}

@article{Gross:1988tx,
    author = "Gross, D. J. and Periwal, V.",
    title = "{Gross and Periwal Reply to 'String Perturbation Theory.'}",
    doi = "10.1103/PhysRevLett.61.1517",
    journal = "Phys. Rev. Lett.",
    volume = "61",
    pages = "1517",
    year = "1988"
}

@article{Baker:1988tw,
    author = "Baker, G. A.",
    title = "{STRING PERTURBATION THEORY}",
    doi = "10.1103/PhysRevLett.61.1516",
    journal = "Phys. Rev. Lett.",
    volume = "61",
    pages = "1516",
    year = "1988"
}

@article{Gross:1988ib,
    author = "Gross, David J. and Periwal, Vipul",
    title = "{String Perturbation Theory Diverges}",
    reportNumber = "PUPT-1090",
    doi = "10.1103/PhysRevLett.60.2105",
    journal = "Phys. Rev. Lett.",
    volume = "60",
    pages = "2105",
    year = "1988"
}

@article{Bedroya:2023tch,
    author = "Bedroya, Alek and Raman, Sanjay and Tarazi, Houri-Christina",
    title = "{Non-BPS path to the string lamppost}",
    eprint = "2303.13585",
    archivePrefix = "arXiv",
    primaryClass = "hep-th",
    month = "3",
    year = "2023"
}

@article{Delgado:2024skw,
    author = "Delgado, Matilda and van de Heisteeg, Damian and Raman, Sanjay and Torres, Ethan and Vafa, Cumrun and Xu, Kai",
    title = "{Finiteness and the emergence of dualities}",
    eprint = "2412.03640",
    archivePrefix = "arXiv",
    primaryClass = "hep-th",
    reportNumber = "MPP-2024-224, CERN-TH-2024-204",
    doi = "10.21468/SciPostPhys.19.2.047",
    journal = "SciPost Phys.",
    volume = "19",
    number = "2",
    pages = "047",
    year = "2025"
}

@article{Hellerman:2009bu,
    author = "Hellerman, Simeon",
    title = "{A Universal Inequality for CFT and Quantum Gravity}",
    eprint = "0902.2790",
    archivePrefix = "arXiv",
    primaryClass = "hep-th",
    reportNumber = "IPMU-09-0022, IPMU-09-0022",
    doi = "10.1007/JHEP08(2011)130",
    journal = "JHEP",
    volume = "08",
    pages = "130",
    year = "2011"
}

@article{Hellerman:2010qd,
    author = "Hellerman, Simeon and Schmidt-Colinet, Cornelius",
    title = "{Bounds for State Degeneracies in 2D Conformal Field Theory}",
    eprint = "1007.0756",
    archivePrefix = "arXiv",
    primaryClass = "hep-th",
    reportNumber = "IPMU-10-0104",
    doi = "10.1007/JHEP08(2011)127",
    journal = "JHEP",
    volume = "08",
    pages = "127",
    year = "2011"
}

@article{Kiritsis:1998en,
    author = "Kiritsis, E. and Kounnas, C. and Petropoulos, P. M. and Rizos, J.",
    title = "{String threshold corrections in models with spontaneously broken supersymmetry}",
    eprint = "hep-th/9807067",
    archivePrefix = "arXiv",
    reportNumber = "CERN-TH-97-44, NEIP-97-007, IOA-97-08, LPTENS-97-11, CPTH-S499-0397, CPTH-S499.0397",
    doi = "10.1016/S0550-3213(98)00713-5",
    journal = "Nucl. Phys. B",
    volume = "540",
    pages = "87--148",
    year = "1999"
}

@article{Erbin:2019uiz,
    author = "Erbin, Harold and Maldacena, Juan and Skliros, Dimitri",
    title = "{Two-Point String Amplitudes}",
    eprint = "1906.06051",
    archivePrefix = "arXiv",
    primaryClass = "hep-th",
    doi = "10.1007/JHEP07(2019)139",
    journal = "JHEP",
    volume = "07",
    pages = "139",
    year = "2019"
}

@article{Giribet:2023gub,
    author = "Giribet, Gaston and Labranche, Nicholas and La Madrid, Joan",
    title = "{Remarks on the two-point string amplitudes}",
    eprint = "2303.15658",
    archivePrefix = "arXiv",
    primaryClass = "hep-th",
    doi = "10.1103/PhysRevD.107.106021",
    journal = "Phys. Rev. D",
    volume = "107",
    number = "10",
    pages = "106021",
    year = "2023"
}

@article{Eberhardt:2020bgq,
    author = "Eberhardt, Lorenz",
    title = "{Partition functions of the tensionless string}",
    eprint = "2008.07533",
    archivePrefix = "arXiv",
    primaryClass = "hep-th",
    doi = "10.1007/JHEP03(2021)176",
    journal = "JHEP",
    volume = "03",
    pages = "176",
    year = "2021"
}

@article{Eberhardt:2021jvj,
    author = "Eberhardt, Lorenz",
    title = "{Summing over Geometries in String Theory}",
    eprint = "2102.12355",
    archivePrefix = "arXiv",
    primaryClass = "hep-th",
    doi = "10.1007/JHEP05(2021)233",
    journal = "JHEP",
    volume = "05",
    pages = "233",
    year = "2021"
}

@article{Eberhardt:2021ynh,
    author = "Eberhardt, Lorenz and Pal, Sridip",
    title = "{The disk partition function in string theory}",
    eprint = "2105.08726",
    archivePrefix = "arXiv",
    primaryClass = "hep-th",
    doi = "10.1007/JHEP08(2021)026",
    journal = "JHEP",
    volume = "08",
    pages = "026",
    year = "2021"
}

@article{Ahmadain:2022tew,
    author = "Ahmadain, Amr and Wall, Aron C.",
    title = "{Off-shell strings I: S-matrix and action}",
    eprint = "2211.08607",
    archivePrefix = "arXiv",
    primaryClass = "hep-th",
    doi = "10.21468/SciPostPhys.17.1.005",
    journal = "SciPost Phys.",
    volume = "17",
    number = "1",
    pages = "005",
    year = "2024"
}

@article{Ahmadain:2022eso,
    author = "Ahmadain, Amr and Wall, Aron C.",
    title = "{Off-shell strings II: Black hole entropy}",
    eprint = "2211.16448",
    archivePrefix = "arXiv",
    primaryClass = "hep-th",
    doi = "10.21468/SciPostPhys.17.1.006",
    journal = "SciPost Phys.",
    volume = "17",
    number = "1",
    pages = "006",
    year = "2024"
}

@article{Kounnas:2016gmz,
    author = "Kounnas, Costas and Partouche, Herve",
    title = "{Super no-scale models in string theory}",
    eprint = "1607.01767",
    archivePrefix = "arXiv",
    primaryClass = "hep-th",
    reportNumber = "LPTENS-16-04, CPHT-RR033.062016",
    doi = "10.1016/j.nuclphysb.2016.10.001",
    journal = "Nucl. Phys. B",
    volume = "913",
    pages = "593--626",
    year = "2016"
}

@article{Florakis:2021bws,
    author = "Florakis, Ioannis and Rizos, John and Violaris-Gountonis, Konstantinos",
    title = "{Super no-scale models with Pati-Salam gauge group}",
    eprint = "2110.06752",
    archivePrefix = "arXiv",
    primaryClass = "hep-th",
    doi = "10.1016/j.nuclphysb.2022.115689",
    journal = "Nucl. Phys. B",
    volume = "976",
    pages = "115689",
    year = "2022"
}

@article{Florakis:2022avh,
    author = "Florakis, Ioannis and Rizos, John and Violaris-Gountonis, Konstantinos",
    title = "{Three-generation super no-scale models in heterotic superstrings}",
    eprint = "2206.09732",
    archivePrefix = "arXiv",
    primaryClass = "hep-th",
    doi = "10.1016/j.physletb.2022.137311",
    journal = "Phys. Lett. B",
    volume = "833",
    pages = "137311",
    year = "2022"
}

@article{Abel:2024vov,
    author = "Abel, Steven and Basile, Ivano and Matyas, Viktor G.",
    title = "{Banks-Zaks stabilisation of non-SUSY strings}",
    eprint = "2412.01914",
    archivePrefix = "arXiv",
    primaryClass = "hep-th",
    reportNumber = "LMU-ASC 21/24, IPPP/24/77",
    doi = "10.1007/JHEP04(2025)107",
    journal = "JHEP",
    volume = "04",
    pages = "107",
    year = "2025"
}

@article{Larotonda:2026hxy,
    author = "Larotonda, Vittorio and Montero, Miguel and Tartaglia, Michelangelo",
    title = "{Asymmetric orbifolds with vanishing one-loop vacuum energy}",
    eprint = "2602.07113",
    archivePrefix = "arXiv",
    primaryClass = "hep-th",
    reportNumber = "IFT-26-11",
    month = "2",
    year = "2026"
}

@article{Kachru:1998hd,
    author = "Kachru, Shamit and Kumar, Jason and Silverstein, Eva",
    title = "{Vacuum energy cancellation in a nonsupersymmetric string}",
    eprint = "hep-th/9807076",
    archivePrefix = "arXiv",
    reportNumber = "SLAC-PUB-7875, LBNL-41932, LBL-41932, SU-ITP-98-35, UCB-PTH-98-33",
    doi = "10.1103/PhysRevD.59.106004",
    journal = "Phys. Rev. D",
    volume = "59",
    pages = "106004",
    year = "1999"
}

@article{Angelantonj:2003hr,
    author = "Angelantonj, Carlo and Antoniadis, Ignatios",
    title = "{Suppressing the cosmological constant in nonsupersymmetric type I strings}",
    eprint = "hep-th/0307254",
    archivePrefix = "arXiv",
    reportNumber = "CERN-TH-2003-165",
    doi = "10.1016/j.nuclphysb.2003.09.047",
    journal = "Nucl. Phys. B",
    volume = "676",
    pages = "129--148",
    year = "2004"
}

@article{Dudas:2025yqm,
    author = "Dudas, Emilian and Parameswaran, Susha and Serra, Marco",
    title = "{The Cosmological Constant and Dark Dimensions from Non-Supersymmetric Strings}",
    eprint = "2512.20570",
    archivePrefix = "arXiv",
    primaryClass = "hep-th",
    month = "12",
    year = "2025"
}

@article{Collins:2022nux,
    author = "Collins, Tristan C. and Jafferis, Daniel and Vafa, Cumrun and Xu, Kai and Yau, Shing-Tung",
    title = "{On Upper Bounds in Dimension Gaps of CFT's}",
    eprint = "2201.03660",
    archivePrefix = "arXiv",
    primaryClass = "hep-th",
    month = "1",
    year = "2022"
}

@article{Coudarchet:2023mfs,
    author = "Coudarchet, Thibaut",
    title = "{Hiding the extra dimensions: A review on scale separation in string theory}",
    eprint = "2311.12105",
    archivePrefix = "arXiv",
    primaryClass = "hep-th",
    doi = "10.1016/j.physrep.2024.02.003",
    journal = "Phys. Rept.",
    volume = "1064",
    pages = "1--28",
    year = "2024"
}

@article{Arkani-Hamed:2006emk,
    author = "Arkani-Hamed, Nima and Motl, Lubos and Nicolis, Alberto and Vafa, Cumrun",
    title = "{The String landscape, black holes and gravity as the weakest force}",
    eprint = "hep-th/0601001",
    archivePrefix = "arXiv",
    reportNumber = "HUTP-05-A0057",
    doi = "10.1088/1126-6708/2007/06/060",
    journal = "JHEP",
    volume = "06",
    pages = "060",
    year = "2007"
}

@article{Polchinski:2003bq,
    author = "Polchinski, Joseph",
    editor = "Baer, H. and Belyaev, A.",
    title = "{Monopoles, duality, and string theory}",
    eprint = "hep-th/0304042",
    archivePrefix = "arXiv",
    doi = "10.1142/S0217751X0401866X",
    journal = "Int. J. Mod. Phys. A",
    volume = "19S1",
    pages = "145--156",
    year = "2004"
}

@article{Montero:2022prj,
    author = "Montero, Miguel and Vafa, Cumrun and Valenzuela, Irene",
    title = "{The dark dimension and the Swampland}",
    eprint = "2205.12293",
    archivePrefix = "arXiv",
    primaryClass = "hep-th",
    doi = "10.1007/JHEP02(2023)022",
    journal = "JHEP",
    volume = "02",
    pages = "022",
    year = "2023"
}

@article{Anchordoqui:2025nmb,
    author = "Anchordoqui, Luis A. and Antoniadis, Ignatios and Lust, Dieter",
    title = "{Two Micron-Size Dark Dimensions}",
    eprint = "2501.11690",
    archivePrefix = "arXiv",
    primaryClass = "hep-th",
    reportNumber = "MPP-2025-5, LMU-ASC 02/25",
    doi = "10.1002/prop.70015",
    journal = "Fortsch. Phys.",
    volume = "73",
    number = "8",
    pages = "e70015",
    year = "2025"
}

@article{Andriot:2025cyi,
    author = "Andriot, David and Cribiori, Niccol{\`o} and Van Riet, Thomas",
    title = "{Scale separation, rolling solutions, and entropy bounds}",
    eprint = "2504.08634",
    archivePrefix = "arXiv",
    primaryClass = "hep-th",
    doi = "10.1103/5rkw-5qfk",
    journal = "Phys. Rev. D",
    volume = "112",
    number = "2",
    pages = "026028",
    year = "2025"
}

@article{Cribiori:2025oek,
    author = "Cribiori, Niccol{\`o} and Tonioni, Flavio",
    title = "{Cosmological constraints from UV/IR mixing}",
    eprint = "2507.02738",
    archivePrefix = "arXiv",
    primaryClass = "hep-th",
    doi = "10.1007/JHEP02(2026)035",
    journal = "JHEP",
    volume = "02",
    pages = "035",
    year = "2026"
}

@article{Cohen:1998zx,
    author = "Cohen, Andrew G. and Kaplan, David B. and Nelson, Ann E.",
    title = "{Effective field theory, black holes, and the cosmological constant}",
    eprint = "hep-th/9803132",
    archivePrefix = "arXiv",
    reportNumber = "BUHEP-98-7, DOE-ER-40561-358, INT-98-00-6, UW-PT-97-24",
    doi = "10.1103/PhysRevLett.82.4971",
    journal = "Phys. Rev. Lett.",
    volume = "82",
    pages = "4971--4974",
    year = "1999"
}

@article{Bousso:1999xy,
    author = "Bousso, Raphael",
    title = "{A Covariant entropy conjecture}",
    eprint = "hep-th/9905177",
    archivePrefix = "arXiv",
    reportNumber = "SU-ITP-99-23",
    doi = "10.1088/1126-6708/1999/07/004",
    journal = "JHEP",
    volume = "07",
    pages = "004",
    year = "1999"
}

@article{Bedroya:2025fwh,
    author = "Bedroya, Alek and Obied, Georges and Vafa, Cumrun and Wu, David H.",
    title = "{Evolving Dark Sector and the Dark Dimension Scenario}",
    eprint = "2507.03090",
    archivePrefix = "arXiv",
    primaryClass = "astro-ph.CO",
    month = "7",
    year = "2025"
}

@article{Heckman:2024trz,
    author = "Heckman, Jonathan J. and Vafa, Cumrun and Weigand, Timo and Xu, Fengjun",
    title = "{Dark dimension and the grand unification of forces}",
    eprint = "2409.01405",
    archivePrefix = "arXiv",
    primaryClass = "hep-th",
    reportNumber = "ZMP-HH/24-19",
    doi = "10.1103/PhysRevD.111.046014",
    journal = "Phys. Rev. D",
    volume = "111",
    number = "4",
    pages = "046014",
    year = "2025"
}

@article{Gonzalo:2022jac,
    author = "Gonzalo, Eduardo and Montero, Miguel and Obied, Georges and Vafa, Cumrun",
    title = "{Dark dimension gravitons as dark matter}",
    eprint = "2209.09249",
    archivePrefix = "arXiv",
    primaryClass = "hep-ph",
    doi = "10.1007/JHEP11(2023)109",
    journal = "JHEP",
    volume = "11",
    pages = "109",
    year = "2023"
}

@article{Anchordoqui:2022txe,
    author = "Anchordoqui, Luis A. and Antoniadis, Ignatios and Lust, Dieter",
    title = "{Dark dimension, the swampland, and the dark matter fraction composed of primordial black holes}",
    eprint = "2206.07071",
    archivePrefix = "arXiv",
    primaryClass = "hep-th",
    reportNumber = "MPP-2022-60, LMU-ASC 24/22",
    doi = "10.1103/PhysRevD.106.086001",
    journal = "Phys. Rev. D",
    volume = "106",
    number = "8",
    pages = "086001",
    year = "2022"
}

@article{Anchordoqui:2024dxu,
    author = "Anchordoqui, Luis A. and Antoniadis, Ignatios and Lust, Dieter",
    title = "{More on black holes perceiving the dark dimension}",
    eprint = "2403.19604",
    archivePrefix = "arXiv",
    primaryClass = "hep-th",
    reportNumber = "MPP-2024-67; LMU-ASC 04/24",
    doi = "10.1103/PhysRevD.110.015004",
    journal = "Phys. Rev. D",
    volume = "110",
    number = "1",
    pages = "015004",
    year = "2024"
}

@article{Benjamin:2021ygh,
    author = "Benjamin, Nathan and Collier, Scott and Fitzpatrick, A. Liam and Maloney, Alexander and Perlmutter, Eric",
    title = "{Harmonic analysis of 2d CFT partition functions}",
    eprint = "2107.10744",
    archivePrefix = "arXiv",
    primaryClass = "hep-th",
    doi = "10.1007/JHEP09(2021)174",
    journal = "JHEP",
    volume = "09",
    pages = "174",
    year = "2021"
}

@article{Lee:2019wij,
    author = "Lee, Seung-Joo and Lerche, Wolfgang and Weigand, Timo",
    title = "{Emergent strings from infinite distance limits}",
    eprint = "1910.01135",
    archivePrefix = "arXiv",
    primaryClass = "hep-th",
    reportNumber = "CERN-TH-2019-159",
    doi = "10.1007/JHEP02(2022)190",
    journal = "JHEP",
    volume = "02",
    pages = "190",
    year = "2022"
}

@article{Ooguri:2006in,
    author = "Ooguri, Hirosi and Vafa, Cumrun",
    title = "{On the Geometry of the String Landscape and the Swampland}",
    eprint = "hep-th/0605264",
    archivePrefix = "arXiv",
    reportNumber = "CALT-68-2600, HUTP-06-A017",
    doi = "10.1016/j.nuclphysb.2006.10.033",
    journal = "Nucl. Phys. B",
    volume = "766",
    pages = "21--33",
    year = "2007"
}

@article{Stout:2022phm,
    author = "Stout, John",
    title = "{Infinite Distances and Factorization}",
    eprint = "2208.08444",
    archivePrefix = "arXiv",
    primaryClass = "hep-th",
    month = "8",
    year = "2022"
}

@article{Herraez:2024kux,
    author = {Herr{\'a}ez, Alvaro and L{\"u}st, Dieter and Masias, Joaquin and Scalisi, Marco},
    title = "{On the Origin of Species Thermodynamics and the Black Hole - Tower Correspondence}",
    eprint = "2406.17851",
    archivePrefix = "arXiv",
    primaryClass = "hep-th",
    reportNumber = "MPP-2024-123, LMU-ASC 08/24",
    doi = "10.21468/SciPostPhys.18.3.083",
    journal = "SciPost Phys.",
    volume = "18",
    year = "2025",
    pages= "61"
}

@article{Basile:2023blg,
    author = {Basile, Ivano and L{\"u}st, Dieter and Montella, Carmine},
    title = "{Shedding black hole light on the emergent string conjecture}",
    eprint = "2311.12113",
    archivePrefix = "arXiv",
    primaryClass = "hep-th",
    reportNumber = "LMU-ASC 35/23, MPP-2023-262",
    doi = "10.1007/JHEP07(2024)208",
    journal = "JHEP",
    volume = "07",
    pages = "208",
    year = "2024"
}

@article{Basile:2024dqq,
    author = "Basile, Ivano and Cribiori, Niccol{\`o} and Lust, Dieter and Montella, Carmine",
    title = "{Minimal black holes and species thermodynamics}",
    eprint = "2401.06851",
    archivePrefix = "arXiv",
    primaryClass = "hep-th",
    reportNumber = "LMU-ASC 02/24, MPP-2024-6",
    doi = "10.1007/JHEP06(2024)127",
    journal = "JHEP",
    volume = "06",
    pages = "127",
    year = "2024"
}

@article{Herraez:2025clp,
    author = {Herr{\'a}ez, Alvaro and L{\"u}st, Dieter and Masias, Joaquin and Montella, Carmine},
    title = "{A short overview on the Black Hole-Tower Correspondence and Species Thermodynamics}",
    eprint = "2506.02335",
    archivePrefix = "arXiv",
    primaryClass = "hep-th",
    reportNumber = "MPP-2025-114",
    doi = "10.22323/1.490.0161",
    journal = "PoS",
    volume = "CORFU2024",
    pages = "161",
    year = "2025"
}

@article{Bedroya:2024ubj,
    author = "Bedroya, Alek and Mishra, Rashmish K. and Wiesner, Max",
    title = "{Density of states, black holes and the Emergent String Conjecture}",
    eprint = "2405.00083",
    archivePrefix = "arXiv",
    primaryClass = "hep-th",
    doi = "10.1007/JHEP01(2025)144",
    journal = "JHEP",
    volume = "01",
    pages = "144",
    year = "2025"
}

@article{McNamara:2020uza,
    author = "McNamara, Jacob and Vafa, Cumrun",
    title = "{Baby Universes, Holography, and the Swampland}",
    eprint = "2004.06738",
    archivePrefix = "arXiv",
    primaryClass = "hep-th",
    month = "4",
    year = "2020"
}

@phdthesis{McNamaraThesis,
	author = "McNamara, Jacob",
	title = "{The Kinematics of Quantum Gravity}",
	eprint = {https://nrs.harvard.edu/URN-3:HUL.INSTREPOS:37372201},
	school = "Harvard University",
	year = "2022"
}

@article{Green:1999pv,
    author = "Green, Michael B. and Vanhove, Pierre",
    title = "{The Low-energy expansion of the one loop type II superstring amplitude}",
    eprint = "hep-th/9910056",
    archivePrefix = "arXiv",
    reportNumber = "DAMTP-1999-124, CERN-TH-99-200, SACLAY-SPH-T-99-071",
    doi = "10.1103/PhysRevD.61.104011",
    journal = "Phys. Rev. D",
    volume = "61",
    pages = "104011",
    year = "2000"
}

@article{Green:2008uj,
    author = "Green, Michael B. and Russo, Jorge G. and Vanhove, Pierre",
    title = "{Low energy expansion of the four-particle genus-one amplitude in type II superstring theory}",
    eprint = "0801.0322",
    archivePrefix = "arXiv",
    primaryClass = "hep-th",
    reportNumber = "DAMTP-2007-96, SPHT-T-07-126, UB-ECM-PF-07-29",
    doi = "10.1088/1126-6708/2008/02/020",
    journal = "JHEP",
    volume = "02",
    pages = "020",
    year = "2008"
}

@article{Green:1997tv,
    author = "Green, Michael B. and Gutperle, Michael",
    title = "{Effects of D instantons}",
    eprint = "hep-th/9701093",
    archivePrefix = "arXiv",
    reportNumber = "DAMTP-96-104",
    doi = "10.1016/S0550-3213(97)00269-1",
    journal = "Nucl. Phys. B",
    volume = "498",
    pages = "195--227",
    year = "1997"
}

@article{Green:1997di,
    author = "Green, Michael B. and Vanhove, Pierre",
    title = "{D instantons, strings and M theory}",
    eprint = "hep-th/9704145",
    archivePrefix = "arXiv",
    reportNumber = "DAMTP-97-31, CPTH-S501-0497",
    doi = "10.1016/S0370-2693(97)00785-5",
    journal = "Phys. Lett. B",
    volume = "408",
    pages = "122--134",
    year = "1997"
}

@article{Obers:1998fb,
    author = "Obers, N. A. and Pioline, B.",
    title = "{U duality and M theory}",
    eprint = "hep-th/9809039",
    archivePrefix = "arXiv",
    reportNumber = "CERN-TH-98-282, CPHT-S639-0898",
    doi = "10.1016/S0370-1573(99)00004-6",
    journal = "Phys. Rept.",
    volume = "318",
    pages = "113--225",
    year = "1999"
}

@article{Obers:1999um,
    author = "Obers, N. A. and Pioline, B.",
    title = "{Eisenstein series and string thresholds}",
    eprint = "hep-th/9903113",
    archivePrefix = "arXiv",
    reportNumber = "NORDITA-1999-18-HE, NBI-HE-99-06, CPHT-S710-0299",
    doi = "10.1007/s002200050022",
    journal = "Commun. Math. Phys.",
    volume = "209",
    pages = "275--324",
    year = "2000"
}

@article{Blumenhagen:2024ydy,
    author = "Blumenhagen, Ralph and Cribiori, Niccol{\`o} and Gligovic, Aleksandar and Paraskevopoulou, Antonia",
    title = "{Emergence of R$^{4}$-terms in M-theory}",
    eprint = "2404.01371",
    archivePrefix = "arXiv",
    primaryClass = "hep-th",
    reportNumber = "MPP-2024-72",
    doi = "10.1007/JHEP07(2024)018",
    journal = "JHEP",
    volume = "07",
    pages = "018",
    year = "2024"
}

@article{Blumenhagen:2024lmo,
    author = "Blumenhagen, Ralph and Cribiori, Niccol{\`o} and Gligovic, Aleksandar and Paraskevopoulou, Antonia",
    title = "{Reflections on an M-theoretic Emergence Proposal}",
    eprint = "2404.05801",
    archivePrefix = "arXiv",
    primaryClass = "hep-th",
    reportNumber = "MPP-2024-73",
    doi = "10.22323/1.463.0238",
    journal = "PoS",
    volume = "CORFU2023",
    pages = "238",
    year = "2024"
}

@article{vandeHeisteeg:2023dlw,
    author = "van de Heisteeg, Damian and Vafa, Cumrun and Wiesner, Max and Wu, David H.",
    title = "{Species scale in diverse dimensions}",
    eprint = "2310.07213",
    archivePrefix = "arXiv",
    primaryClass = "hep-th",
    doi = "10.1007/JHEP05(2024)112",
    journal = "JHEP",
    volume = "05",
    pages = "112",
    year = "2024"
}

@article{Aoufia:2025ppe,
    author = "Aoufia, Christian and Castellano, Alberto and Ib{\'a}{\~n}ez, Luis",
    title = "{Laplacians in Various Dimensions and the Swampland}",
    eprint = "2506.03253",
    archivePrefix = "arXiv",
    primaryClass = "hep-th",
    reportNumber = "IFT-UAM/CSIC-25-61, EFI-25-8",
    month = "6",
    year = "2025"
}

@article{Castellano:2025ljk,
    author = "Castellano, Alberto and Zatti, Matteo",
    title = "{Black hole entropy, quantum corrections and EFT transitions}",
    eprint = "2502.02655",
    archivePrefix = "arXiv",
    primaryClass = "hep-th",
    reportNumber = "EFI-25-2, MPP-2025-13",
    doi = "10.1007/JHEP08(2025)112",
    journal = "JHEP",
    volume = "08",
    pages = "112",
    year = "2025"
}

@article{Castellano:2025yur,
    author = {Castellano, Alberto and L{\"u}st, Dieter and Montella, Carmine and Zatti, Matteo},
    title = "{Quantum Calabi-Yau Black Holes and Non-Perturbative D0-brane Effects}",
    eprint = "2505.15920",
    archivePrefix = "arXiv",
    primaryClass = "hep-th",
    reportNumber = "EFI-25-06,MPP-2025-107,LMU-ASC 13/25",
    month = "5",
    year = "2025"
}

@article{Cecotti:1991me,
    author = "Cecotti, Sergio and Vafa, Cumrun",
    title = "{Topological antitopological fusion}",
    reportNumber = "HUTP-91-A031, SISSA-69-91-EP",
    doi = "10.1016/0550-3213(91)90021-O",
    journal = "Nucl. Phys. B",
    volume = "367",
    pages = "359--461",
    year = "1991"
}

@article{Bershadsky:1993ta,
    author = "Bershadsky, M. and Cecotti, S. and Ooguri, H. and Vafa, C.",
    editor = "Greene, B. and Yau, Shing-Tung",
    title = "{Holomorphic anomalies in topological field theories}",
    eprint = "hep-th/9302103",
    archivePrefix = "arXiv",
    reportNumber = "HUTP-93-A008, RIMS-915",
    doi = "10.1016/0550-3213(93)90548-4",
    journal = "Nucl. Phys. B",
    volume = "405",
    pages = "279--304",
    year = "1993"
}

@article{Bershadsky:1993cx,
    author = "Bershadsky, M. and Cecotti, S. and Ooguri, H. and Vafa, C.",
    title = "{Kodaira-Spencer theory of gravity and exact results for quantum string amplitudes}",
    eprint = "hep-th/9309140",
    archivePrefix = "arXiv",
    reportNumber = "HUTP-93-A025, RIMS-946, SISSA-142-93-EP",
    doi = "10.1007/BF02099774",
    journal = "Commun. Math. Phys.",
    volume = "165",
    pages = "311--428",
    year = "1994"
}

@article{Aganagic:2002qg,
    author = "Aganagic, Mina and Marino, Marcos and Vafa, Cumrun",
    title = "{All loop topological string amplitudes from Chern-Simons theory}",
    eprint = "hep-th/0206164",
    archivePrefix = "arXiv",
    reportNumber = "HUTP-02-A024",
    doi = "10.1007/s00220-004-1067-x",
    journal = "Commun. Math. Phys.",
    volume = "247",
    pages = "467--512",
    year = "2004"
}

@article{Grimm:2007tm,
    author = "Grimm, Thomas W. and Klemm, Albrecht and Marino, Marcos and Weiss, Marlene",
    title = "{Direct Integration of the Topological String}",
    eprint = "hep-th/0702187",
    archivePrefix = "arXiv",
    reportNumber = "CERN-PH-TH-2007-039, MAD-TH-07-04",
    doi = "10.1088/1126-6708/2007/08/058",
    journal = "JHEP",
    volume = "08",
    pages = "058",
    year = "2007"
}

@article{Iwaki:2023cek,
    author = "Iwaki, Kohei and Marino, Marcos",
    title = "{Resurgent Structure of the Topological String and the First Painlev{\'e} Equation}",
    eprint = "2307.02080",
    archivePrefix = "arXiv",
    primaryClass = "hep-th",
    doi = "10.3842/SIGMA.2024.028",
    journal = "SIGMA",
    volume = "20",
    pages = "028",
    year = "2024"
}

@article{Gu:2023mgf,
    author = "Gu, Jie and Kashani-Poor, Amir-Kian and Klemm, Albrecht and Marino, Marcos",
    title = "{Non-perturbative topological string theory on compact Calabi-Yau 3-folds}",
    eprint = "2305.19916",
    archivePrefix = "arXiv",
    primaryClass = "hep-th",
    doi = "10.21468/SciPostPhys.16.3.079",
    journal = "SciPost Phys.",
    volume = "16",
    number = "3",
    pages = "079",
    year = "2024"
}

@article{Marino:2024tbx,
    author = "Marino, Marcos",
    title = "{Les Houches lectures on non-perturbative topological strings}",
    eprint = "2411.16211",
    archivePrefix = "arXiv",
    primaryClass = "hep-th",
    doi = "10.21468/SciPostPhysLectNotes.112",
    month = "11",
    year = "2024"
}

@article{Hattab:2024ewk,
    author = "Hattab, Jarod and Palti, Eran",
    title = "{Non-perturbative topological string theory on compact Calabi-Yau manifolds from M-theory}",
    eprint = "2408.09255",
    archivePrefix = "arXiv",
    primaryClass = "hep-th",
    doi = "10.1007/JHEP04(2025)017",
    journal = "JHEP",
    volume = "04",
    pages = "017",
    year = "2025"
}

@article{Hattab:2024chf,
    author = "Hattab, Jarod and Palti, Eran",
    title = "{Emergent potentials and non-perturbative open topological strings}",
    eprint = "2408.12302",
    archivePrefix = "arXiv",
    primaryClass = "hep-th",
    doi = "10.1007/JHEP10(2024)195",
    journal = "JHEP",
    volume = "10",
    pages = "195",
    year = "2024"
}

@article{Hattab:2024yol,
    author = "Hattab, Jarod and Palti, Eran",
    title = "{On Calabi-Yau Manifolds at Strong Topological String Coupling}",
    eprint = "2409.01721",
    archivePrefix = "arXiv",
    primaryClass = "hep-th",
    doi = "10.1002/prop.202400199",
    journal = "Fortsch. Phys.",
    volume = "72",
    number = "12",
    pages = "2400199",
    year = "2024"
}

@article{Antoniadis:1993ze,
    author = "Antoniadis, Ignatios and Gava, E. and Narain, K. S. and Taylor, T. R.",
    title = "{Topological amplitudes in string theory}",
    eprint = "hep-th/9307158",
    archivePrefix = "arXiv",
    reportNumber = "NUB-3071, IC-93-202, CPTH-A258-0793",
    doi = "10.1016/0550-3213(94)90617-3",
    journal = "Nucl. Phys. B",
    volume = "413",
    pages = "162--184",
    year = "1994"
}

@article{Antoniadis:2007ta,
    author = "Antoniadis, I. and Hohenegger, S.",
    editor = "Baulieu, Laurent and de Boer, Jan and Douglas, Michael R. and Rabinovici, Eliezer and Vanhove, Pierre and Windey, Paul",
    title = "{Topological amplitudes and physical couplings in string theory}",
    eprint = "hep-th/0701290",
    archivePrefix = "arXiv",
    reportNumber = "CERN-PH-TH-2007-024",
    doi = "10.1016/j.nuclphysbps.2007.06.011",
    journal = "Nucl. Phys. B Proc. Suppl.",
    volume = "171",
    pages = "176--195",
    year = "2007"
}

@article{Ooguri:1995cp,
    author = "Ooguri, Hirosi and Vafa, Cumrun",
    title = "{All loop N=2 string amplitudes}",
    eprint = "hep-th/9505183",
    archivePrefix = "arXiv",
    reportNumber = "HUTP-95-A017, LBL-37271, UCB-PTH-95-16",
    doi = "10.1016/0550-3213(95)00365-Y",
    journal = "Nucl. Phys. B",
    volume = "451",
    pages = "121--161",
    year = "1995"
}

@article{Ooguri:2004zv,
    author = "Ooguri, Hirosi and Strominger, Andrew and Vafa, Cumrun",
    title = "{Black hole attractors and the topological string}",
    eprint = "hep-th/0405146",
    archivePrefix = "arXiv",
    reportNumber = "HUTP-04-A020, CALT-68-2501",
    doi = "10.1103/PhysRevD.70.106007",
    journal = "Phys. Rev. D",
    volume = "70",
    pages = "106007",
    year = "2004"
}

@article{Dabholkar:2004yr,
    author = "Dabholkar, Atish",
    title = "{Exact counting of black hole microstates}",
    eprint = "hep-th/0409148",
    archivePrefix = "arXiv",
    reportNumber = "SLAC-PUB-10717, SU-ITP-04-36, TIFR-TH-04-23",
    doi = "10.1103/PhysRevLett.94.241301",
    journal = "Phys. Rev. Lett.",
    volume = "94",
    pages = "241301",
    year = "2005"
}

@article{Cribiori:2022nke,
    author = {Cribiori, Niccol{\`o} and L{\"u}st, Dieter and Staudt, Georgina},
    title = "{Black hole entropy and moduli-dependent species scale}",
    eprint = "2212.10286",
    archivePrefix = "arXiv",
    primaryClass = "hep-th",
    reportNumber = "LMU-ASC 56/22, MPP-2022-289",
    doi = "10.1016/j.physletb.2023.138113",
    journal = "Phys. Lett. B",
    volume = "844",
    pages = "138113",
    year = "2023"
}

@article{Cribiori:2023ffn,
    author = "Cribiori, Niccol{\`o} and Lust, Dieter and Montella, Carmine",
    title = "{Species entropy and thermodynamics}",
    eprint = "2305.10489",
    archivePrefix = "arXiv",
    primaryClass = "hep-th",
    reportNumber = "LMU-ASC 18/23, MPP-2023-97",
    doi = "10.1007/JHEP10(2023)059",
    journal = "JHEP",
    volume = "10",
    pages = "059",
    year = "2023"
}

@article{vandeHeisteeg:2022btw,
    author = "van de Heisteeg, Damian and Vafa, Cumrun and Wiesner, Max and Wu, David H.",
    title = "{Moduli-dependent species scale}",
    eprint = "2212.06841",
    archivePrefix = "arXiv",
    primaryClass = "hep-th",
    doi = "10.4310/bpam.2024.v1.n1.a1",
    journal = "Beijing J. Pure Appl. Math.",
    volume = "1",
    number = "1",
    pages = "1--41",
    year = "2024"
}

@article{Castellano:2023stg,
    author = "Castellano, Alberto and Ruiz, Ignacio and Valenzuela, Irene",
    title = "{Universal Pattern in Quantum Gravity at Infinite Distance}",
    eprint = "2311.01501",
    archivePrefix = "arXiv",
    primaryClass = "hep-th",
    reportNumber = "CERN-TH-2023-203",
    doi = "10.1103/PhysRevLett.132.181601",
    journal = "Phys. Rev. Lett.",
    volume = "132",
    number = "18",
    pages = "181601",
    year = "2024"
}

@article{Castellano:2023jjt,
    author = "Castellano, Alberto and Ruiz, Ignacio and Valenzuela, Irene",
    title = "{Stringy evidence for a universal pattern at infinite distance}",
    eprint = "2311.01536",
    archivePrefix = "arXiv",
    primaryClass = "hep-th",
    reportNumber = "CERN-TH-2023-204",
    doi = "10.1007/JHEP06(2024)037",
    journal = "JHEP",
    volume = "06",
    pages = "037",
    year = "2024"
}

@article{Basile:2025bql,
    author = "Basile, Ivano and Staudt, Georgina",
    title = "{Corrections to the CRV pattern}",
    eprint = "2503.00107",
    archivePrefix = "arXiv",
    primaryClass = "hep-th",
    month = "2",
    year = "2025"
}

@article{Cvetic:1989ii,
    author = "Cvetic, Mirjam and Ovrut, Burt A. and Louis, Jan",
    title = "{The Zamolodchikov Metric and Effective Lagrangians in String Theory}",
    reportNumber = "UPR-0380T",
    doi = "10.1103/PhysRevD.40.684",
    journal = "Phys. Rev. D",
    volume = "40",
    pages = "684",
    year = "1989"
}

@article{Gerchkovitz:2014gta,
    author = "Gerchkovitz, Efrat and Gomis, Jaume and Komargodski, Zohar",
    title = "{Sphere Partition Functions and the Zamolodchikov Metric}",
    eprint = "1405.7271",
    archivePrefix = "arXiv",
    primaryClass = "hep-th",
    doi = "10.1007/JHEP11(2014)001",
    journal = "JHEP",
    volume = "11",
    pages = "001",
    year = "2014"
}

@article{Zamolodchikov:1986gt,
    author = "Zamolodchikov, A. B.",
    title = "{Irreversibility of the Flux of the Renormalization Group in a 2D Field Theory}",
    journal = "JETP Lett.",
    volume = "43",
    pages = "730--732",
    year = "1986"
}

@article{Angelantonj:2023egh,
    author = "Angelantonj, Carlo and Florakis, Ioannis and Leone, Giorgio",
    title = "{Tachyons and misaligned supersymmetry in closed string vacua}",
    eprint = "2301.13702",
    archivePrefix = "arXiv",
    primaryClass = "hep-th",
    doi = "10.1007/JHEP06(2023)174",
    journal = "JHEP",
    volume = "06",
    pages = "174",
    year = "2023"
}

@article{Leone:2023qfd,
    author = "Leone, Giorgio",
    title = "{Tachyons and Misaligned Supersymmetry in orientifold vacua}",
    eprint = "2308.09757",
    archivePrefix = "arXiv",
    primaryClass = "hep-th",
    doi = "10.1007/JHEP11(2023)066",
    journal = "JHEP",
    volume = "11",
    pages = "066",
    year = "2023"
}

@phdthesis{Leone:2024xae,
    author = "Leone, Giorgio",
    title = "{Aspects of Stability, Rigidity and Unitarity in String Vacua}",
    eprint = "2408.00132",
    archivePrefix = "arXiv",
    primaryClass = "hep-th",
    school = "Universit{\`a} degli Studi di Torino, Italy, Turin U.",
    year = "2024"
}

@article{Leone:2025mwo,
    author = "Leone, Giorgio and Raucci, Salvatore",
    title = "{Aspects of strings without spacetime supersymmetry}",
    eprint = "2509.24703",
    archivePrefix = "arXiv",
    primaryClass = "hep-th",
    reportNumber = "IFT-UAM/CSIC-25-100",
    month = "9",
    year = "2025"
}

@article{Angelantonj:2022dsx,
    author = "Angelantonj, Carlo and Antoniadis, Ignatios and Florakis, Ioannis and Jiang, Hongliang",
    title = "{Refined topological amplitudes from the {\ensuremath{\Omega}}-background in string theory}",
    eprint = "2202.13205",
    archivePrefix = "arXiv",
    primaryClass = "hep-th",
    reportNumber = "QMUL-PH-22-07",
    doi = "10.1007/JHEP05(2022)143",
    journal = "JHEP",
    volume = "05",
    pages = "143",
    year = "2022"
}

@article{Angelantonj:2019qfw,
    author = "Angelantonj, Carlo and Antoniadis, Ignatios",
    title = "{The String Geometry Behind Topological Amplitudes}",
    eprint = "1910.03347",
    archivePrefix = "arXiv",
    primaryClass = "hep-th",
    doi = "10.1007/JHEP01(2020)005",
    journal = "JHEP",
    volume = "01",
    pages = "005",
    year = "2020"
}

@article{Dvali:2001gx,
    author = "Dvali, G. R. and Gabadadze, Gregory and Kolanovic, Marko and Nitti, F.",
    title = "{Scales of gravity}",
    eprint = "hep-th/0106058",
    archivePrefix = "arXiv",
    reportNumber = "NYU-TH-01-06-03, TPI-MINN-01-26",
    doi = "10.1103/PhysRevD.65.024031",
    journal = "Phys. Rev. D",
    volume = "65",
    pages = "024031",
    year = "2002"
}

@article{Veneziano:2001ah,
    author = "Veneziano, G.",
    title = "{Large N bounds on, and compositeness limit of, gauge and gravitational interactions}",
    eprint = "hep-th/0110129",
    archivePrefix = "arXiv",
    reportNumber = "CERN-TH-2001-278",
    doi = "10.1088/1126-6708/2002/06/051",
    journal = "JHEP",
    volume = "06",
    pages = "051",
    year = "2002"
}

@article{Dvali:2007hz,
    author = "Dvali, Gia",
    title = "{Black Holes and Large N Species Solution to the Hierarchy Problem}",
    eprint = "0706.2050",
    archivePrefix = "arXiv",
    primaryClass = "hep-th",
    doi = "10.1002/prop.201000009",
    journal = "Fortsch. Phys.",
    volume = "58",
    pages = "528--536",
    year = "2010"
}

@article{Dvali:2007wp,
    author = "Dvali, Gia and Redi, Michele",
    title = "{Black Hole Bound on the Number of Species and Quantum Gravity at LHC}",
    eprint = "0710.4344",
    archivePrefix = "arXiv",
    primaryClass = "hep-th",
    doi = "10.1103/PhysRevD.77.045027",
    journal = "Phys. Rev. D",
    volume = "77",
    pages = "045027",
    year = "2008"
}

@article{Dvali:2009ks,
    author = "Dvali, Gia and Lust, Dieter",
    title = "{Evaporation of Microscopic Black Holes in String Theory and the Bound on Species}",
    eprint = "0912.3167",
    archivePrefix = "arXiv",
    primaryClass = "hep-th",
    reportNumber = "CERN-PH-TH-2009-243, MPP-2009-205, LMU-ASC-56-09",
    doi = "10.1002/prop.201000008",
    journal = "Fortsch. Phys.",
    volume = "58",
    pages = "505--527",
    year = "2010"
}

@article{Dvali:2010vm,
    author = "Dvali, Gia and Gomez, Cesar",
    title = "{Species and Strings}",
    eprint = "1004.3744",
    archivePrefix = "arXiv",
    primaryClass = "hep-th",
    reportNumber = "CERN-PH-TH-2010-069, IFT-UAM-CSIC-10-25",
    month = "4",
    year = "2010"
}

@article{Dvali:2012uq,
    author = "Dvali, Gia and Gomez, Cesar and Lust, Dieter",
    title = "{Black Hole Quantum Mechanics in the Presence of Species}",
    eprint = "1206.2365",
    archivePrefix = "arXiv",
    primaryClass = "hep-th",
    reportNumber = "MPP-2012-102, LMU-ASC-38-12, MPP--2012--102",
    doi = "10.1002/prop.201300002",
    journal = "Fortsch. Phys.",
    volume = "61",
    pages = "768--778",
    year = "2013"
}

@article{ValeixoBento:2025iqu,
    author = "Valeixo Bento, Bruno and Melo, Jo\~ao",
    title = "{EFT \& Species Scale: Friends or foes?}",
    eprint = "2501.08230",
    archivePrefix = "arXiv",
    primaryClass = "hep-th",
    month = "1",
    year = "2025"
}

@article{Caron-Huot:2024lbf,
    author = "Caron-Huot, Simon and Li, Yue-Zhou",
    title = "{Gravity and a universal cutoff for field theory}",
    eprint = "2408.06440",
    archivePrefix = "arXiv",
    primaryClass = "hep-th",
    doi = "10.1007/JHEP02(2025)115",
    journal = "JHEP",
    volume = "02",
    pages = "115",
    year = "2025"
}

@article{Castellano:2022bvr,
    author = "Castellano, Alberto and Herr\'aez, Alvaro and Ib\'a\~nez, Luis E.",
    title = "{The emergence proposal in quantum gravity and the species scale}",
    eprint = "2212.03908",
    archivePrefix = "arXiv",
    primaryClass = "hep-th",
    reportNumber = "IFT-UAM/CSIC-22-149",
    doi = "10.1007/JHEP06(2023)047",
    journal = "JHEP",
    volume = "06",
    pages = "047",
    year = "2023"
}

@article{Blumenhagen:2023yws,
    author = "Blumenhagen, Ralph and Gligovic, Aleksandar and Paraskevopoulou, Antonia",
    title = "{The emergence proposal and the emergent string}",
    eprint = "2305.10490",
    archivePrefix = "arXiv",
    primaryClass = "hep-th",
    reportNumber = "MPP-2023-96",
    doi = "10.1007/JHEP10(2023)145",
    journal = "JHEP",
    volume = "10",
    pages = "145",
    year = "2023"
}

@article{vandeHeisteeg:2023ubh,
    author = "van de Heisteeg, Damian and Vafa, Cumrun and Wiesner, Max",
    title = "{Bounds on Species Scale and the Distance Conjecture}",
    eprint = "2303.13580",
    archivePrefix = "arXiv",
    primaryClass = "hep-th",
    doi = "10.1002/prop.202300143",
    journal = "Fortsch. Phys.",
    volume = "71",
    number = "10-11",
    pages = "2300143",
    year = "2023"
}

@article{Castellano:2023aum,
    author = "Castellano, Alberto and Herr\'aez, Alvaro and Ib\'a\~nez, Luis E.",
    title = "{On the species scale, modular invariance and the gravitational EFT expansion}",
    eprint = "2310.07708",
    archivePrefix = "arXiv",
    primaryClass = "hep-th",
    doi = "10.1007/JHEP12(2024)019",
    journal = "JHEP",
    volume = "12",
    pages = "019",
    year = "2024"
}

@phdthesis{Castellano:2024bna,
    author = "Castellano, Alberto",
    title = "{The Quantum Gravity Scale and the Swampland}",
    eprint = "2409.10003",
    archivePrefix = "arXiv",
    primaryClass = "hep-th",
    school = "U. Autonoma, Madrid (main)",
    year = "2024"
}

@article{Cribiori:2023sch,
    author = {Cribiori, Niccol\`o and L\"ust, Dieter},
    title = "{A Note on Modular Invariant Species Scale and Potentials}",
    eprint = "2306.08673",
    archivePrefix = "arXiv",
    primaryClass = "hep-th",
    reportNumber = "LMU-ASC 20/23, MPP-2023-126",
    doi = "10.1002/prop.202300150",
    journal = "Fortsch. Phys.",
    volume = "71",
    number = "10-11",
    pages = "2300150",
    year = "2023"
}

@article{Cribiori:2024qsv,
    author = "Cribiori, Niccol\`o and Lust, Dieter",
    title = "{String dualities and modular symmetries in supergravity: a review}",
    eprint = "2411.06516",
    archivePrefix = "arXiv",
    primaryClass = "hep-th",
    reportNumber = "LMU-ASC 18/24, MPP-2024-206",
    month = "11",
    year = "2024"
}

@article{Camanho:2014apa,
    author = "Camanho, Xian O. and Edelstein, Jose D. and Maldacena, Juan and Zhiboedov, Alexander",
    title = "{Causality Constraints on Corrections to the Graviton Three-Point Coupling}",
    eprint = "1407.5597",
    archivePrefix = "arXiv",
    primaryClass = "hep-th",
    doi = "10.1007/JHEP02(2016)020",
    journal = "JHEP",
    volume = "02",
    pages = "020",
    year = "2016"
}

@article{Cheung:2016wjt,
    author = "Cheung, Clifford and Remmen, Grant N.",
    title = "{Positivity of Curvature-Squared Corrections in Gravity}",
    eprint = "1608.02942",
    archivePrefix = "arXiv",
    primaryClass = "hep-th",
    reportNumber = "CALT-TH-2016-018",
    doi = "10.1103/PhysRevLett.118.051601",
    journal = "Phys. Rev. Lett.",
    volume = "118",
    number = "5",
    pages = "051601",
    year = "2017"
}

@article{Caron-Huot:2022ugt,
    author = "Caron-Huot, Simon and Li, Yue-Zhou and Parra-Martinez, Julio and Simmons-Duffin, David",
    title = "{Causality constraints on corrections to Einstein gravity}",
    eprint = "2201.06602",
    archivePrefix = "arXiv",
    primaryClass = "hep-th",
    reportNumber = "CALT-TH 2021-003",
    doi = "10.1007/JHEP05(2023)122",
    journal = "JHEP",
    volume = "05",
    pages = "122",
    year = "2023"
}

@article{Caron-Huot:2022jli,
    author = "Caron-Huot, Simon and Li, Yue-Zhou and Parra-Martinez, Julio and Simmons-Duffin, David",
    title = "{Graviton partial waves and causality in higher dimensions}",
    eprint = "2205.01495",
    archivePrefix = "arXiv",
    primaryClass = "hep-th",
    reportNumber = "CALT-TH 2022-17",
    doi = "10.1103/PhysRevD.108.026007",
    journal = "Phys. Rev. D",
    volume = "108",
    number = "2",
    pages = "026007",
    year = "2023"
}

@article{Albert:2024yap,
    author = "Albert, Jan and Knop, Waltraut and Rastelli, Leonardo",
    title = "{Where is tree-level string theory?}",
    eprint = "2406.12959",
    archivePrefix = "arXiv",
    primaryClass = "hep-th",
    reportNumber = "YITP-SB-2024-12",
    doi = "10.1007/JHEP02(2025)157",
    journal = "JHEP",
    volume = "02",
    pages = "157",
    year = "2025"
}

@article{Cheung:2025nhw,
    author = "Cheung, Clifford and Remmen, Grant N.",
    title = "{Multipositivity bounds for scattering amplitudes}",
    eprint = "2505.05553",
    archivePrefix = "arXiv",
    primaryClass = "hep-th",
    reportNumber = "CALT-TH 2025-010",
    doi = "10.1103/wt4x-2149",
    journal = "Phys. Rev. D",
    volume = "112",
    number = "1",
    pages = "016017",
    year = "2025"
}

@article{Cheung:2025tbr,
    author = "Cheung, Clifford and Remmen, Grant N. and Sciotti, Francesco and Tarquini, Michele",
    title = "{Strings from Almost Nothing}",
    eprint = "2508.09246",
    archivePrefix = "arXiv",
    primaryClass = "hep-th",
    reportNumber = "CALT-TH 2025-027",
    month = "8",
    year = "2025"
}

@article{Calisto:2025tjo,
    author = "Calisto, Francesco and Cheung, Clifford and Remmen, Grant N. and Sciotti, Francesco and Tarquini, Michele",
    title = "{Completeness from Gravitational Scattering}",
    eprint = "2512.11955",
    archivePrefix = "arXiv",
    primaryClass = "hep-th",
    reportNumber = "CALT-TH 2025-040",
    month = "12",
    year = "2025"
}

@article{Cheung:2024obl,
    author = "Cheung, Clifford and Hillman, Aaron and Remmen, Grant N.",
    title = "{Uniqueness criteria for the Virasoro-Shapiro amplitude}",
    eprint = "2408.03362",
    archivePrefix = "arXiv",
    primaryClass = "hep-th",
    reportNumber = "CALT-TH 2024-030",
    doi = "10.1103/PhysRevD.111.086034",
    journal = "Phys. Rev. D",
    volume = "111",
    number = "8",
    pages = "086034",
    year = "2025"
}

@article{Cheung:2024uhn,
    author = "Cheung, Clifford and Hillman, Aaron and Remmen, Grant N.",
    title = "{Bootstrap Principle for the Spectrum and Scattering of Strings}",
    eprint = "2406.02665",
    archivePrefix = "arXiv",
    primaryClass = "hep-th",
    reportNumber = "CALT-TH 2024-022",
    doi = "10.1103/PhysRevLett.133.251601",
    journal = "Phys. Rev. Lett.",
    volume = "133",
    number = "25",
    pages = "251601",
    year = "2024"
}

@article{Bachu:2022gof,
    author = "Bachu, Brad and Hillman, Aaron",
    title = "{Stringy Completions of the Standard Model from the Bottom Up}",
    eprint = "2212.03871",
    archivePrefix = "arXiv",
    primaryClass = "hep-th",
    month = "12",
    year = "2022"
}

@article{Arkani-Hamed:2023jwn,
    author = "Arkani-Hamed, Nima and Cheung, Clifford and Figueiredo, Carolina and Remmen, Grant N.",
    title = "{Multiparticle Factorization and the Rigidity of String Theory}",
    eprint = "2312.07652",
    archivePrefix = "arXiv",
    primaryClass = "hep-th",
    reportNumber = "CALT-TH 2023-051",
    doi = "10.1103/PhysRevLett.132.091601",
    journal = "Phys. Rev. Lett.",
    volume = "132",
    number = "9",
    pages = "091601",
    year = "2024"
}

@article{Arkani-Hamed:2020blm,
    author = "Arkani-Hamed, Nima and Huang, Tzu-Chen and Huang, Yu-tin",
    title = "{The EFT-Hedron}",
    eprint = "2012.15849",
    archivePrefix = "arXiv",
    primaryClass = "hep-th",
    reportNumber = "NCTS-TH/2014, CALT-TH 2020-061",
    doi = "10.1007/JHEP05(2021)259",
    journal = "JHEP",
    volume = "05",
    pages = "259",
    year = "2021"
}

@article{Geiser:2022exp,
    author = "Geiser, Nicholas and Lindwasser, Lukas W.",
    title = "{Generalized Veneziano and Virasoro amplitudes}",
    eprint = "2210.14920",
    archivePrefix = "arXiv",
    primaryClass = "hep-th",
    doi = "10.1007/JHEP04(2023)031",
    journal = "JHEP",
    volume = "04",
    pages = "031",
    year = "2023"
}

@article{Guerrieri:2021ivu,
    author = "Guerrieri, Andrea and Penedones, Joao and Vieira, Pedro",
    title = "{Where Is String Theory in the Space of Scattering Amplitudes?}",
    eprint = "2102.02847",
    archivePrefix = "arXiv",
    primaryClass = "hep-th",
    doi = "10.1103/PhysRevLett.127.081601",
    journal = "Phys. Rev. Lett.",
    volume = "127",
    number = "8",
    pages = "081601",
    year = "2021"
}

@article{Guerrieri:2022sod,
    author = "Guerrieri, Andrea and Murali, Harish and Penedones, Joao and Vieira, Pedro",
    title = "{Where is M-theory in the space of scattering amplitudes?}",
    eprint = "2212.00151",
    archivePrefix = "arXiv",
    primaryClass = "hep-th",
    doi = "10.1007/JHEP06(2023)064",
    journal = "JHEP",
    volume = "06",
    pages = "064",
    year = "2023"
}

@article{Montero:2020icj,
    author = "Montero, Miguel and Vafa, Cumrun",
    title = "{Cobordism Conjecture, Anomalies, and the String Lamppost Principle}",
    eprint = "2008.11729",
    archivePrefix = "arXiv",
    primaryClass = "hep-th",
    doi = "10.1007/JHEP01(2021)063",
    journal = "JHEP",
    volume = "01",
    pages = "063",
    year = "2021"
}

@article{Bedroya:2021fbu,
    author = "Bedroya, Alek and Hamada, Yuta and Montero, Miguel and Vafa, Cumrun",
    title = "{Compactness of brane moduli and the String Lamppost Principle in d {\ensuremath{>}} 6}",
    eprint = "2110.10157",
    archivePrefix = "arXiv",
    primaryClass = "hep-th",
    doi = "10.1007/JHEP02(2022)082",
    journal = "JHEP",
    volume = "02",
    pages = "082",
    year = "2022"
}

@article{Hamada:2021bbz,
    author = "Hamada, Yuta and Vafa, Cumrun",
    title = "{8d supergravity, reconstruction of internal geometry and the Swampland}",
    eprint = "2104.05724",
    archivePrefix = "arXiv",
    primaryClass = "hep-th",
    doi = "10.1007/JHEP06(2021)178",
    journal = "JHEP",
    volume = "06",
    pages = "178",
    year = "2021"
}

@article{Kim:2019ths,
    author = "Kim, Hee-Cheol and Tarazi, Houri-Christina and Vafa, Cumrun",
    title = "{Four-dimensional $\mathbf{\mathcal{N}=4}$ SYM theory and the swampland}",
    eprint = "1912.06144",
    archivePrefix = "arXiv",
    primaryClass = "hep-th",
    doi = "10.1103/PhysRevD.102.026003",
    journal = "Phys. Rev. D",
    volume = "102",
    number = "2",
    pages = "026003",
    year = "2020"
}

@article{Kim:2019vuc,
    author = "Kim, Hee-Cheol and Shiu, Gary and Vafa, Cumrun",
    title = "{Branes and the Swampland}",
    eprint = "1905.08261",
    archivePrefix = "arXiv",
    primaryClass = "hep-th",
    doi = "10.1103/PhysRevD.100.066006",
    journal = "Phys. Rev. D",
    volume = "100",
    number = "6",
    pages = "066006",
    year = "2019"
}

@article{Baykara:2025gcc,
    author = "Baykara, Zihni Kaan and Dierigl, Markus and Kim, Hee-Cheol and Vafa, Cumrun and Xu, Kai",
    title = "{Bounds on Discrete Gauge Symmetries in Supergravity}",
    eprint = "2511.09613",
    archivePrefix = "arXiv",
    primaryClass = "hep-th",
    reportNumber = "CERN-TH-2025-233",
    month = "11",
    year = "2025"
}

@article{Kim:2024eoa,
    author = "Kim, Hee-Cheol and Vafa, Cumrun and Xu, Kai",
    title = "{Finite Landscape of 6d N=(1,0) Supergravity}",
    eprint = "2411.19155",
    archivePrefix = "arXiv",
    primaryClass = "hep-th",
    doi = "10.21468/SciPostPhys.20.1.016",
    journal = "SciPost Phys.",
    volume = "20",
    pages = "016",
    year = "2026"
}

@article{Arkani-Hamed:2007ryu,
    author = "Arkani-Hamed, Nima and Dubovsky, Sergei and Nicolis, Alberto and Villadoro, Giovanni",
    title = "{Quantum Horizons of the Standard Model Landscape}",
    eprint = "hep-th/0703067",
    archivePrefix = "arXiv",
    doi = "10.1088/1126-6708/2007/06/078",
    journal = "JHEP",
    volume = "06",
    pages = "078",
    year = "2007"
}

@article{Marchesano:2024gul,
    author = "Marchesano, Fernando and Shiu, Gary and Weigand, Timo",
    title = "{The Standard Model from String Theory: What Have We Learned?}",
    eprint = "2401.01939",
    archivePrefix = "arXiv",
    primaryClass = "hep-th",
    reportNumber = "IFT-UAM/CSIC-24-01, ZMP-HH/24-01",
    doi = "10.1146/annurev-nucl-102622-012235",
    journal = "Ann. Rev. Nucl. Part. Sci.",
    volume = "74",
    pages = "113--140",
    year = "2024"
}

@article{McAllister:2025qwq,
    author = "McAllister, Liam and Schachner, Andreas",
    title = "{TASI Lectures on de Sitter Vacua}",
    eprint = "2512.17095",
    archivePrefix = "arXiv",
    primaryClass = "hep-th",
    month = "12",
    year = "2025"
}

@article{Basile:2024oms,
    author = "Basile, Ivano and Buoninfante, Luca and Di Filippo, Francesco and Knorr, Benjamin and Platania, Alessia and Tokareva, Anna",
    title = "{Lectures in quantum gravity}",
    eprint = "2412.08690",
    archivePrefix = "arXiv",
    primaryClass = "hep-th",
    doi = "10.21468/SciPostPhysLectNotes.98",
    journal = "SciPost Phys. Lect. Notes",
    volume = "98",
    pages = "1",
    year = "2025"
}

@article{Lee:2018urn,
    author = "Lee, Seung-Joo and Lerche, Wolfgang and Weigand, Timo",
    title = "{Tensionless Strings and the Weak Gravity Conjecture}",
    eprint = "1808.05958",
    archivePrefix = "arXiv",
    primaryClass = "hep-th",
    reportNumber = "CERN-TH-2018-190",
    doi = "10.1007/JHEP10(2018)164",
    journal = "JHEP",
    volume = "10",
    pages = "164",
    year = "2018"
}

@article{Klaewer:2020lfg,
    author = "Klaewer, Daniel and Lee, Seung-Joo and Weigand, Timo and Wiesner, Max",
    title = "{Quantum corrections in 4d $N$ = 1 infinite distance limits and the weak gravity conjecture}",
    eprint = "2011.00024",
    archivePrefix = "arXiv",
    primaryClass = "hep-th",
    reportNumber = "CTPU-PTC-20-24, IFT-UAM/CSIC-20-148, MITP/20-064, ZMP-HH/20-21",
    doi = "10.1007/JHEP03(2021)252",
    journal = "JHEP",
    volume = "03",
    pages = "252",
    year = "2021"
}

@article{Alvarez-Garcia:2021pxo,
    author = {{\'A}lvarez-Garc{\'\i}a, Rafael and Kl{\"a}wer, Daniel and Weigand, Timo},
    title = "{Membrane limits in quantum gravity}",
    eprint = "2112.09136",
    archivePrefix = "arXiv",
    primaryClass = "hep-th",
    reportNumber = "ZMP-HH/21-25",
    doi = "10.1103/PhysRevD.105.066024",
    journal = "Phys. Rev. D",
    volume = "105",
    number = "6",
    pages = "066024",
    year = "2022"
}

@article{Hassfeld:2025uoy,
    author = "Hassfeld, Bjoern and Monnee, Jeroen and Weigand, Timo and Wiesner, Max",
    title = "{Emergent strings in Type IIB Calabi-Yau compactifications}",
    eprint = "2504.01066",
    archivePrefix = "arXiv",
    primaryClass = "hep-th",
    doi = "10.1007/JHEP01(2026)140",
    journal = "JHEP",
    volume = "01",
    pages = "140",
    year = "2026"
}

@article{Monnee:2025ynn,
    author = "Monnee, Jeroen and Weigand, Timo and Wiesner, Max",
    title = "{Physics and Geometry of Complex Structure Limits in Type IIB Calabi-Yau Compactifications}",
    eprint = "2509.07056",
    archivePrefix = "arXiv",
    primaryClass = "hep-th",
    month = "9",
    year = "2025"
}

@article{Kaufmann:2024gqo,
    author = "Kaufmann, Lukas and Lanza, Stefano and Weigand, Timo",
    title = "{Asymptotics of 5d supergravity theories and the emergent string conjecture}",
    eprint = "2412.12251",
    archivePrefix = "arXiv",
    primaryClass = "hep-th",
    doi = "10.1007/JHEP06(2025)230",
    journal = "JHEP",
    volume = "06",
    pages = "230",
    year = "2025"
}

@article{Monnee:2025msf,
    author = "Monnee, Jeroen and Weigand, Timo and Wiesner, Max",
    title = "{K-Points and Type IIB/Heterotic Duality with NS5-Branes}",
    eprint = "2510.02435",
    archivePrefix = "arXiv",
    primaryClass = "hep-th",
    reportNumber = "ZMP-HH-25/18",
    month = "10",
    year = "2025"
}

@article{Cummings:2026giw,
    author = "Cummings, Charlie and Heckman, Jonathan J.",
    title = "{The Topological Equivalence Principle: On Decoupling TFTs from Gravity}",
    eprint = "2601.09781",
    archivePrefix = "arXiv",
    primaryClass = "hep-th",
    month = "1",
    year = "2026"
}

@article{Montero:2024qtz,
    author = "Montero, Miguel and Valenzuela, Irene",
    title = "{Quantum corrections to DGKT and the Weak Gravity Conjecture}",
    eprint = "2412.00189",
    archivePrefix = "arXiv",
    primaryClass = "hep-th",
    reportNumber = "IFT-24-155, CERN-TH-2024-190",
    doi = "10.1007/JHEP07(2025)057",
    journal = "JHEP",
    volume = "07",
    pages = "057",
    year = "2025"
}

@article{Bedroya:2025ltj,
    author = "Bedroya, Alek and Steinhardt, Paul J.",
    title = "{Holography vs. Scale Separation}",
    eprint = "2509.25313",
    archivePrefix = "arXiv",
    primaryClass = "hep-th",
    month = "9",
    year = "2025"
}

@article{Bedroya:2025fie,
    author = "Bedroya, Alek and Steinhardt, Paul J.",
    title = "{Holographic Constraints on the String Landscape}",
    eprint = "2511.15784",
    archivePrefix = "arXiv",
    primaryClass = "hep-th",
    month = "11",
    year = "2025"
}

@article{Bobev:2025yxp,
    author = "Bobev, Nikolay and Paul, Hynek and Revello, Filippo",
    title = "{A Holographic Constraint on Scale Separation}",
    eprint = "2512.11031",
    archivePrefix = "arXiv",
    primaryClass = "hep-th",
    month = "12",
    year = "2025"
}

@article{Apers:2026lgi,
    author = "Apers, Fien",
    title = "{On the DGKT brane dual and its decoupling}",
    eprint = "2601.15093",
    archivePrefix = "arXiv",
    primaryClass = "hep-th",
    month = "1",
    year = "2026"
}

@article{Tachikawa:2021mby,
    author = "Tachikawa, Yuji and Yamashita, Mayuko",
    title = "{Topological Modular Forms and the Absence of All Heterotic Global Anomalies}",
    eprint = "2108.13542",
    archivePrefix = "arXiv",
    primaryClass = "hep-th",
    doi = "10.1007/s00220-023-04761-2",
    journal = "Commun. Math. Phys.",
    volume = "402",
    number = "2",
    pages = "1585--1620",
    year = "2023",
    note = "[Erratum: Commun.Math.Phys. 402, 2131 (2023)]"
}

@article{Tachikawa:2021mvw,
    author = "Tachikawa, Yuji",
    title = "{Topological modular forms and the absence of a heterotic global anomaly}",
    eprint = "2103.12211",
    archivePrefix = "arXiv",
    primaryClass = "hep-th",
    doi = "10.1093/ptep/ptab060",
    journal = "PTEP",
    volume = "2022",
    number = "4",
    pages = "04A107",
    year = "2022"
}

@article{Cho:2023mhw,
    author = "Cho, Minjae and Kim, Manki",
    title = "{A worldsheet description of flux compactifications}",
    eprint = "2311.04959",
    archivePrefix = "arXiv",
    primaryClass = "hep-th",
    reportNumber = "MIT-CTP/5623",
    doi = "10.1007/JHEP05(2024)247",
    journal = "JHEP",
    volume = "05",
    pages = "247",
    year = "2024"
}

@article{Kim:2024dnw,
    author = "Kim, Manki",
    title = "{String perturbation theory of Klebanov-Strassler throat}",
    eprint = "2409.19048",
    archivePrefix = "arXiv",
    primaryClass = "hep-th",
    doi = "10.1007/JHEP05(2025)234",
    journal = "JHEP",
    volume = "05",
    pages = "234",
    year = "2025"
}

@article{Frenkel:2025wko,
    author = "Frenkel, Alexander and Kim, Manki",
    title = "{Non-linear sigma model in string field theory}",
    eprint = "2509.20527",
    archivePrefix = "arXiv",
    primaryClass = "hep-th",
    month = "9",
    year = "2025"
}

@article{Kim:2026kex,
    author = "Kim, Manki",
    title = "{On string loops in Calabi-Yau orientifolds in large volume}",
    eprint = "2601.08973",
    archivePrefix = "arXiv",
    primaryClass = "hep-th",
    month = "1",
    year = "2026"
}

@article{Cano:2021tfs,
    author = "Cano, Pablo A. and Murcia, {\'A}ngel",
    title = "{Duality-invariant extensions of Einstein-Maxwell theory}",
    eprint = "2104.07674",
    archivePrefix = "arXiv",
    primaryClass = "hep-th",
    doi = "10.1007/JHEP08(2021)042",
    journal = "JHEP",
    volume = "08",
    pages = "042",
    year = "2021"
}

@article{Kiritsis:1994ta,
    author = "Kiritsis, Elias and Kounnas, Costas",
    title = "{Infrared regularization of superstring theory and the one loop calculation of coupling constants}",
    eprint = "hep-th/9501020",
    archivePrefix = "arXiv",
    reportNumber = "CERN-TH-7472-94, LPTENS-94-36",
    doi = "10.1016/0550-3213(95)00156-M",
    journal = "Nucl. Phys. B",
    volume = "442",
    pages = "472--493",
    year = "1995"
}

@article{Kiritsis:1994yv,
    author = "Kiritsis, Elias and Kounnas, Costas",
    title = "{Curved four-dimensional space-times as infrared regulator in superstring theories}",
    eprint = "hep-th/9410212",
    archivePrefix = "arXiv",
    reportNumber = "CERN-TH-7471-94, LPTENS-94-29",
    doi = "10.1016/0920-5632(95)00441-B",
    journal = "Nucl. Phys. B Proc. Suppl.",
    volume = "41",
    pages = "331--340",
    year = "1995"
}

@article{Kutasov:1990sv,
    author = "Kutasov, David and Seiberg, Nathan",
    title = "{Number of degrees of freedom, density of states and tachyons in string theory and CFT}",
    reportNumber = "PUPT-1221, RU-90-60",
    doi = "10.1016/0550-3213(91)90426-X",
    journal = "Nucl. Phys. B",
    volume = "358",
    pages = "600--618",
    year = "1991"
}

@article{Kiritsis:1996dn,
    author = "Kiritsis, E. and Kounnas, C. and Petropoulos, P. M. and Rizos, J.",
    title = "{Universality properties of N=2 and N=1 heterotic threshold corrections}",
    eprint = "hep-th/9608034",
    archivePrefix = "arXiv",
    reportNumber = "CERN-TH-96-90, SISSA-87-96-EP, LPTENS-96-26",
    doi = "10.1016/S0550-3213(96)00550-0",
    journal = "Nucl. Phys. B",
    volume = "483",
    pages = "141--171",
    year = "1997"
}

@article{Petropoulos:1996rr,
    author = "Petropoulos, P. M. and Rizos, J.",
    title = "{Universal moduli dependent string thresholds in Z(2) x Z(2) orbifolds}",
    eprint = "hep-th/9601037",
    archivePrefix = "arXiv",
    reportNumber = "CERN-TH-95-284, SISSA-170-95-EP",
    doi = "10.1016/0370-2693(96)00230-4",
    journal = "Phys. Lett. B",
    volume = "374",
    pages = "49--56",
    year = "1996"
}

@inproceedings{Kiritsis:1996yb,
    author = "Kiritsis, E. and Kounnas, C. and Petropoulos, P. M. and Rizos, J.",
    title = "{On the heterotic effective action at one loop gauge couplings and the gravitational sector}",
    booktitle = "{5th Hellenic School and Workshops on Elementary Particle Physics}",
    eprint = "hep-th/9605011",
    archivePrefix = "arXiv",
    reportNumber = "CERN-TH-96-091, CERN-TH-96-91, SISSA-60-96-EP",
    month = "4",
    year = "1996"
}

@inproceedings{Kiritsis:1995ga,
    author = "Kiritsis, Elias and Kounnas, Costas",
    title = "{Infrared regulated string theory and loop corrections to coupling constants}",
    booktitle = "{STRINGS 95: Future Perspectives in String Theory}",
    eprint = "hep-th/9507051",
    archivePrefix = "arXiv",
    reportNumber = "CERN-TH-95-172, LPTENS-95-29",
    pages = "287--308",
    month = "7",
    year = "1995"
}

@article{Angelantonj:2016gkz,
    author = "Angelantonj, Carlo and Israel, Dan and Sarkis, Matthieu",
    title = "{Threshold corrections in heterotic flux compactifications}",
    eprint = "1611.09442",
    archivePrefix = "arXiv",
    primaryClass = "hep-th",
    doi = "10.1007/JHEP08(2017)032",
    journal = "JHEP",
    volume = "08",
    pages = "032",
    year = "2017"
}

@article{Gross:1986iv,
    author = "Gross, David J. and Witten, Edward",
    title = "{Superstring Modifications of Einstein's Equations}",
    reportNumber = "Print-86-0250 (PRINCETON)",
    doi = "10.1016/0550-3213(86)90429-3",
    journal = "Nucl. Phys. B",
    volume = "277",
    pages = "1",
    year = "1986"
}

@article{Arkani-Hamed:1998jmv,
    author = "Arkani-Hamed, Nima and Dimopoulos, Savas and Dvali, G. R.",
    title = "{The Hierarchy problem and new dimensions at a millimeter}",
    eprint = "hep-ph/9803315",
    archivePrefix = "arXiv",
    reportNumber = "SLAC-PUB-7769, SU-ITP-98-13",
    doi = "10.1016/S0370-2693(98)00466-3",
    journal = "Phys. Lett. B",
    volume = "429",
    pages = "263--272",
    year = "1998"
}

@article{Antoniadis:2001sw,
    author = "Antoniadis, Ignatios and Dimopoulos, Savas and Giveon, Amit",
    title = "{Little string theory at a TeV}",
    eprint = "hep-th/0103033",
    archivePrefix = "arXiv",
    reportNumber = "CERN-TH-2001-066, RI-07-00, ITP-01-04",
    doi = "10.1088/1126-6708/2001/05/055",
    journal = "JHEP",
    volume = "05",
    pages = "055",
    year = "2001"
}

@article{Lust:2008qc,
    author = "Lust, Dieter and Stieberger, Stephan and Taylor, Tomasz R.",
    title = "{The LHC String Hunter's Companion}",
    eprint = "0807.3333",
    archivePrefix = "arXiv",
    primaryClass = "hep-th",
    reportNumber = "MPP-2008-05, LMU-ASC-38-08",
    doi = "10.1016/j.nuclphysb.2008.09.012",
    journal = "Nucl. Phys. B",
    volume = "808",
    pages = "1--52",
    year = "2009"
}

@article{Lerche:1987qk,
    author = "Lerche, W. and Nilsson, B. E. W. and Schellekens, A. N. and Warner, N. P.",
    title = "{Anomaly Cancelling Terms From the Elliptic Genus}",
    reportNumber = "CERN-TH-4765/87",
    doi = "10.1016/0550-3213(88)90468-3",
    journal = "Nucl. Phys. B",
    volume = "299",
    pages = "91--116",
    year = "1988"
}

@article{Schellekens:1986xh,
    author = "Schellekens, A. N. and Warner, N. P.",
    title = "{Anomalies, Characters and Strings}",
    reportNumber = "CERN-TH-4529/86",
    doi = "10.1016/0550-3213(87)90108-8",
    journal = "Nucl. Phys. B",
    volume = "287",
    pages = "317",
    year = "1987"
}

@article{Castellano:2021mmx,
    author = "Castellano, Alberto and Herr{\'a}ez, Alvaro and Ib{\'a}{\~n}ez, Luis E.",
    title = "{IR/UV mixing, towers of species and swampland conjectures}",
    eprint = "2112.10796",
    archivePrefix = "arXiv",
    primaryClass = "hep-th",
    doi = "10.1007/JHEP08(2022)217",
    journal = "JHEP",
    volume = "08",
    pages = "217",
    year = "2022"
}

@article{Angelantonj:2010ic,
    author = "Angelantonj, Carlo and Cardella, Matteo and Elitzur, Shmuel and Rabinovici, Eliezer",
    title = "{Vacuum stability, string density of states and the Riemann zeta function}",
    eprint = "1012.5091",
    archivePrefix = "arXiv",
    primaryClass = "hep-th",
    reportNumber = "DFTT-26-2010",
    doi = "10.1007/JHEP02(2011)024",
    journal = "JHEP",
    volume = "02",
    pages = "024",
    year = "2011"
}

@article{Abel:2015oxa,
    author = "Abel, Steven and Dienes, Keith R. and Mavroudi, Eirini",
    title = "{Towards a nonsupersymmetric string phenomenology}",
    eprint = "1502.03087",
    archivePrefix = "arXiv",
    primaryClass = "hep-th",
    doi = "10.1103/PhysRevD.91.126014",
    journal = "Phys. Rev. D",
    volume = "91",
    number = "12",
    pages = "126014",
    year = "2015"
}

@article{Dienes:1990ij,
    author = "Dienes, Keith R.",
    title = "{New string partition functions with vanishing cosmological constant}",
    reportNumber = "CLNS-89/966",
    doi = "10.1103/PhysRevLett.65.1979",
    journal = "Phys. Rev. Lett.",
    volume = "65",
    pages = "1979--1982",
    year = "1990"
}

@article{Witten:2013tpa,
    author = "Witten, Edward",
    title = "{Notes On Holomorphic String And Superstring Theory Measures Of Low Genus}",
    eprint = "1306.3621",
    archivePrefix = "arXiv",
    primaryClass = "hep-th",
    month = "6",
    year = "2013"
}

@article{Donagi:2024bbe,
    author = "Donagi, Ron and Ott, Nadia",
    title = "{A measure on the moduli space of super Riemann surfaces with Ramond punctures}",
    eprint = "2410.12912",
    archivePrefix = "arXiv",
    primaryClass = "math.AG",
    month = "10",
    year = "2024"
}

@article{Wang:2022aem,
    author = "Wang, Charles and Yin, Xi",
    title = "{Supermoduli and PCOs at Genus Two}",
    eprint = "2205.10377",
    archivePrefix = "arXiv",
    primaryClass = "hep-th",
    doi = "10.1007/JHEP01(2023)144",
    journal = "JHEP",
    volume = "01",
    pages = "144",
    year = "2023"
}

@article{Sonoda:1988mf,
    author = "Sonoda, Hidenori",
    title = "{SEWING CONFORMAL FIELD THEORIES}",
    reportNumber = "LBL-25140",
    doi = "10.1016/0550-3213(88)90066-1",
    journal = "Nucl. Phys. B",
    volume = "311",
    pages = "401--416",
    year = "1988"
}

@article{Sonoda:1988fq,
    author = "Sonoda, Hidenori",
    title = "{SEWING CONFORMAL FIELD THEORIES. 2.}",
    reportNumber = "LBL-25316",
    doi = "10.1016/0550-3213(88)90067-3",
    journal = "Nucl. Phys. B",
    volume = "311",
    pages = "417--432",
    year = "1988"
}

@article{Cribiori:2026btb,
    author = "Cribiori, Niccol{\`o} and Paraskevopoulou, Antonia and Van Riet, Thomas",
    title = "{A domain wall bound on anti-de Sitter vacua}",
    eprint = "2603.08779",
    archivePrefix = "arXiv",
    primaryClass = "hep-th",
    reportNumber = "MPP-2026-25",
    month = "3",
    year = "2026"
}

@article{Friedan:2012hi,
    author = "Friedan, Daniel and Konechny, Anatoly",
    title = "{Curvature formula for the space of 2-d conformal field theories}",
    eprint = "1206.1749",
    archivePrefix = "arXiv",
    primaryClass = "hep-th",
    doi = "10.1007/JHEP09(2012)113",
    journal = "JHEP",
    volume = "09",
    pages = "113",
    year = "2012"
}

@article{Demirtas:2021nlu,
    author = "Demirtas, Mehmet and Kim, Manki and McAllister, Liam and Moritz, Jakob and Rios-Tascon, Andres",
    title = "{Small cosmological constants in string theory}",
    eprint = "2107.09064",
    archivePrefix = "arXiv",
    primaryClass = "hep-th",
    doi = "10.1007/JHEP12(2021)136",
    journal = "JHEP",
    volume = "12",
    pages = "136",
    year = "2021"
}

@article{Demirtas:2021ote,
    author = "Demirtas, Mehmet and Kim, Manki and McAllister, Liam and Moritz, Jakob and Rios-Tascon, Andres",
    title = "{Exponentially Small Cosmological Constant in String Theory}",
    eprint = "2107.09065",
    archivePrefix = "arXiv",
    primaryClass = "hep-th",
    doi = "10.1103/PhysRevLett.128.011602",
    journal = "Phys. Rev. Lett.",
    volume = "128",
    number = "1",
    pages = "011602",
    year = "2022"
}

@article{Lust:2022lfc,
    author = {L{\"u}st, Severin and Vafa, Cumrun and Wiesner, Max and Xu, Kai},
    title = "{Holography and the KKLT scenario}",
    eprint = "2204.07171",
    archivePrefix = "arXiv",
    primaryClass = "hep-th",
    doi = "10.1007/JHEP10(2022)188",
    journal = "JHEP",
    volume = "10",
    pages = "188",
    year = "2022"
}

@article{Bena:2024are,
    author = {Bena, Iosif and Li, Yixuan and L{\"u}st, Severin},
    title = "{KKLT compactifications ex nihilo}",
    eprint = "2410.22400",
    archivePrefix = "arXiv",
    primaryClass = "hep-th",
    doi = "10.1103/gcyw-pc83",
    journal = "Phys. Rev. D",
    volume = "113",
    number = "4",
    pages = "046013",
    year = "2026"
}

@article{Perlmutter:2015iya,
    author = "Perlmutter, Eric",
    title = "{Virasoro conformal blocks in closed form}",
    eprint = "1502.07742",
    archivePrefix = "arXiv",
    primaryClass = "hep-th",
    doi = "10.1007/JHEP08(2015)088",
    journal = "JHEP",
    volume = "08",
    pages = "088",
    year = "2015"
}

@article{Antoniadis:1999ge,
    author = "Antoniadis, Ignatios and Bachas, C. and Dudas, E.",
    title = "{Gauge couplings in four-dimensional type I string orbifolds}",
    eprint = "hep-th/9906039",
    archivePrefix = "arXiv",
    reportNumber = "CPHT-S714-0399, LPT-ORSAY-99-12, CPHT-S714.0399",
    doi = "10.1016/S0550-3213(99)00452-6",
    journal = "Nucl. Phys. B",
    volume = "560",
    pages = "93--134",
    year = "1999"
}

@article{Komatsu:2025cai,
    author = "Komatsu, Shota and Kusuki, Yuya and Meineri, Marco and Ooguri, Hirosi",
    title = "{Continuous Family of Conformal Field Theories and Exactly Marginal Operators}",
    eprint = "2512.11045",
    archivePrefix = "arXiv",
    primaryClass = "hep-th",
    reportNumber = "CALT 2025-041, IPMU 25-0054, KYUSHU-HET-333, RIKEN-iTHEMS-Report-25",
    month = "12",
    year = "2025"
}

\end{document}